\documentclass[11pt]{article}
\usepackage{etex}
\usepackage{youngtab}
\usepackage{amscd,amsmath,amssymb,amsfonts,xspace,mathrsfs,mathtools,amsthm}
\usepackage{silence}
\usepackage{xr-hyper}
\usepackage{jheppub}
\usepackage[dvipsnames]{xcolor}
\usepackage[utf8]{inputenc}
\usepackage{bbm}
\usepackage{graphicx}
\usepackage{tabu} 
\usepackage[color,matrix,arrow]{xy}
\usepackage{wasysym} 
\usepackage{stmaryrd}
\usepackage[shortlabels]{enumitem}
\usepackage{array,amsmath}
\usepackage{slashed}
\usepackage{tensor}
\usepackage{caption}
\usepackage{subcaption}
\usepackage{enumitem}
\usepackage[textsize=tiny]{todonotes}
\usepackage{cellspace, hhline}
\usepackage{lineno}
\usepackage{multirow}
\usepackage{svg}
\usepackage{booktabs, cellspace, hhline}
\usepackage{longtable}
\usepackage{comment}
\usepackage{placeins} 
\usepackage{lipsum}

\usepackage{tikz}
\usepackage{tikz-cd}
\usepackage{diagbox}
\usetikzlibrary{positioning}
\usetikzlibrary{calc}
\usetikzlibrary{decorations.pathreplacing,calligraphy}

\usepackage{xstring}
\usetikzlibrary{decorations.pathmorphing} 
\usetikzlibrary{decorations.markings} 
\usetikzlibrary{arrows} 
\usetikzlibrary{shapes} 
\usetikzlibrary{matrix} 
\usetikzlibrary{positioning} 
\usepackage[english]{babel} 
\usepackage[autostyle]{csquotes}
\usetikzlibrary{positioning,fit}
\usepackage[dvipsnames]{xcolor}
\tikzset{->-/.style={decoration={
  markings,
  mark=at position #1 with {\arrow{>}}},postaction={decorate}}}

\theoremstyle{definition}

\newtheorem{example}{Example}

\DeclarePairedDelimiterX\braket[2]{\langle}{\rangle}{#1\delimsize\vert\mathopen{}#2}%

\newcommand{\diag}{\text{diag}}

\newcommand{\rk}{\text{rk}}

\newcommand{\be}{\begin{equation}} 
\newcommand{\ee}{\end{equation}} 
\newcommand{\bea}{\begin{equation} \begin{aligned}} \newcommand{\eea}{\end{aligned} \end{equation}}
\newcommand{\bes}{\begin{equation*}}
\newcommand{\ees}{\end{equation*}}

\newcommand{\ov}{\over}

\newcommand{\kk}{\mathsf{k}}

\newcommand{\CN}{\mathcal{N}}
\newcommand{\CO}{\mathcal{O}} 

\newcommand{\CQ}{\mathcal{Q}}
\newcommand{\CR}{\mathcal{R}}
\newcommand{\CS}{\mathcal{S}}

\newcommand{\SL}{\text{SL}(2,\BZ)}

\renewcommand{\t}{\widetilde }
\newcommand{\Z}{\mathbb{Z}}
\newcommand{\C}{\mathbb{C}}
\newcommand{\K}{\mathbb{K}}
\newcommand{\R}{\mathbb{R}}

\renewcommand{\SL}{{\mathscr L}}

\newcommand{\qcoh}{\mathsf{q}}
\newcommand{\qk}{q}
\newcommand{\bbC}{\mathbb{C}}
\renewcommand{\l}{\mathsf{l}}
\usetikzlibrary{fit,backgrounds}
\numberwithin{equation}{section}  
\allowdisplaybreaks  
\numberwithin{table}{section}

\usepackage{lscape}

\usepackage[normalem]{ulem}

 \title{
 Schubert line defects in 3d GLSMs, part II: 
Partial flag manifolds and parabolic quantum polynomials}

\abstract{We construct Schubert line defects in the 3d $\CN=2$ supersymmetric gauged linear sigma model (GLSM) with target space a partial flag manifold $X={\rm Fl}({\boldsymbol{k}};n)$, generalizing our construction for complete flag manifolds given in a companion paper (part I)~\cite{Closset:2025cfm}. In the context of the 3d GLSM/quantum K-theory correspondence, the Schubert line defects are constructed as 1d $\mathcal{N}=2$ supersymmetric gauge theories coupled to the 3d field theory, and they flow to objects supported on Schubert varieties $X_w \subseteq X$ in the quantum K-theory.  The flavored Witten index of the 1d defect is expected to compute the Chern character of $[\CO_w]$ --- more precisely, it gives us a polynomial representative of the Schubert class in the quantum K-theory ring. We give strong evidence for this claim by showing in examples that the Witten indices of Schubert defects indeed reproduce a recently-defined set of polynomials that represent the Schubert classes in the Whitney presentation, which we call the parabolic Whitney polynomials. Moreover, upon using the quantum ring relations, we can convert these polynomials into seemingly new polynomials in the Toda presentation, which we call the parabolic quantum Grothendieck polynomials. These new polynomials specialize to known polynomials in various limits, including to the quantum Grothendieck polynomials in the case of the complete flag.  
In the 2d limit, our construction also realizes the Schubert classes $[X_w]$ in the quantum cohomology ring of the partial flag manifold, and the parabolic quantum Grothendieck polynomials then reduce to previously known parabolic quantum Schubert polynomials.
}

\author[a]{Cyril Closset,}
\affiliation[a]{ School of Mathematics, University of Birmingham,\\ 
 Watson Building, Edgbaston, Birmingham B15 2TT, UK}
\emailAdd{c.closset@bham.ac.uk}

\author[b]{Wei Gu,}
\affiliation[b]{Zhejiang Institute of Modern Physics, School of Physics, Zhejiang University,\\
Hangzhou, Zhejiang 310058, China}
\emailAdd{guwei2875@zju.edu.cn}

\author[a,c]{Osama Khlaif,}
\emailAdd{osama.khlaif@phys.ens.fr}
\affiliation[c]{Philippe Meyer Institute, Physics Department, École Normale Supérieure (ENS), Université PSL, 24 rue Lhomond, F-75231 Paris, France}

\author[d]{Eric Sharpe,}
\affiliation[d]{Physics Department, 850 West Campus Drive, Virginia Tech,\\
Blacksburg, VA 24061, US}
\emailAdd{ersharpe@vt.edu}

\author[d]{Hao Zhang,}
\emailAdd{hzhang96@vt.edu}

\author[e,f]{Hao Zou}
\affiliation[e]{Center for Mathematics and Interdisciplinary Sciences, Fudan University, Shanghai 200433, China}
\affiliation[f]{Shanghai Institute for Mathematics and Interdisciplinary Sciences, Shanghai 200433, China}
\emailAdd{haozou@fudan.edu.cn}

\begin{document}

\maketitle

\section{Introduction}

The 3d gauged linear sigma model (GLSM)/quantum K-theory correspondence~\cite{Bullimore:2014awa, Jockers:2018sfl, Jockers:2019wjh, Jockers:2021omw, Ueda:2019qhg, Koroteev:2017nab,Bullimore:2018jlp,Bullimore:2019qnt,Bullimore:2020nhv, Gu:2020zpg,Dedushenko:2021mds, Gu:2022yvj,Dedushenko:2023qjq, Gu:2023tcv, Gu:2023fpw, Gu:2025abc, Gu:2021yek, Gu:2021beo,Sharpe:2024ujm,Closset:2023bdr,Closset:2023izb, Huq-Kuruvilla:2024tsg, Huq-Kuruvilla:2025nlf, ahkmox} relates the quantum K-theory of the GLSM target space $X$ to the half-BPS lines of the 3d gauge theory. While earlier work e.g.~\cite{Closset:2016arn, Jockers:2019lwe, Gu:2022yvj} focused on Wilson lines, which correspond to locally-free sheaves ({\it i.e.} vector bundles) over $X$, more recently there has been renewed interest in the GLSM realization of non-locally-free sheaves supported on non-trivial subvarieties of $X$~\cite{Closset:2023bdr, Gu:2025tda, Closset:2025cfm}. The general expectation is that they correspond to defect lines, which are concretely realized as a 1d $\CN=2$ supersymmetric gauge theory --- {\it i.e.} a gauged supersymmetric quantum mechanics (SQM) --- coupled to the 3d bulk.

This paper is meant to be read together with our companion paper~\cite{Closset:2025cfm}, henceforth called `part I,' where we defined and studied the {\it Schubert line defects} $\SL_w$ that realize Schubert classes in the quantum K-theory of the complete flag manifold. In this work, we generalize this construction and define Schubert line defects in any {\it partial flag manifold}. Recall that the partial flag manifold 
\be
X\,=\, {\rm Fl}(\boldsymbol{k}; n)\,=\, {\rm Fl}(k_1,  \cdots, k_s; n)
\ee
is the set of all flags $V_\bullet$ with dimension vector $\boldsymbol{k}$:
\be
 {\rm Fl}(\boldsymbol{k}; n)\,:=\,\{V_\bullet \,=\, (0\,\subset\, V_1\,\subset\, V_2\,\subset\, \cdots\,\subset \,V_{s}\,\subset\,\C^n)~|~\dim(V_\ell) \,=\, k_\ell\}~.
\ee
The GLSM for a partial flag manifold (see \cite{Donagi:2007hi}) takes the form of a linear quiver gauge theory with unitary gauge group $G=U(k_1)\times \cdots \times U(k_s)$. The complexified gauge groups $GL(k_\ell)= U(k_\ell)_\C$ correspond to the freedom in choosing a basis in the vector space $V_\ell$ that is part of a flag $V_\bullet$.

\medskip
\noindent
{\bf Schubert varieties in partial flag manifolds.}
In order to explain some of our terminology, let us recall here that the partial flag manifold can alternatively be described as a quotient of complex groups:
\be\label{Fl as Un ov P}
 {\rm Fl}(\boldsymbol{k}; n)\,\cong\, GL(n)/P~, \qquad P \,\supset\, U(k_1)\,\times\,U(k_2-k_1)\,\times\,\cdots\,\times\,U(k_s-k_{s-1})\,\times\,U(n-k_s)~,
\ee
where $P$ is a parabolic subgroup with the Levi subgroup as indicated. For the complete flag, $P=B$ is the Borel subgroup whose compact subgroup is the maximal torus $U(1)^n$, while a general Levi has a non-trivial Weyl group denoted by ${\rm W}_P$.

The Schubert variety $X_w$ in ${\rm Fl}(\boldsymbol{k}; n)$ is defined as the subset of flags that intersect some reference complete flag of $\C^n$ in certain codimensions which are determined by a `parabolic' permutation of $n$ elements (more precisely, a minimal coset representative):
\be
w\, \in\, {\rm W}^P \, :=\, S_n/{\rm W}_P~,
\ee
where $S_n$ is the Weyl group of $GL(n)$. The permutation $w$ determines the relative position of the flags $V_\bullet \in X_w$ with respect to the reference complete flag or, equivalently, the $B$-orbit $B\cdot wP/P$ (the Schubert cell) whose closure gives us $X_w$.%
\footnote{To be more precise, the Schubert varieties $X_w$ we will study are isomorphic to the $B^-$ orbits (with $B^-$ the Borel of lower-triangular matrices), see~\protect\cite[appendix D.3]{Anderson_Fulton_2023}.}

\begin{table}[t]
\renewcommand{\arraystretch}{1.1}
\centering
\begin{tabular}{|c||c|c|c|}
\hline
Notation  & Name of polynomial & Geometric interpretation & References \\
\hline\hline

\multirow{2}{*}{$\mathfrak{S}_w(\sigma)$}
 & \multirow{2}{*}{(Double) Schubert polynomial}
 & \multirow{2}{*}{$[X_w]\in{\rm H}^\bullet({\rm Fl}(n))$}
 & \cite{MacDonald1991NotesOS, lascoux2006symmetry, LS82b, Fulton92} \\
 & & & \cite[(A.8)]{Closset:2025cfm} \\
\hline

\multirow{2}{*}{$\mathfrak{S}_w^{(\qcoh)}(\sigma)$}
 & Quantum (double)
 & \multirow{2}{*}{$[X_w]\in{\rm QH}^\bullet({\rm Fl}(n))$}
 & \cite{fomin1997quantum, kirillov2000quantum, LamShimozono} \\
 & Schubert polynomial
 & & \cite[(A.18)]{Closset:2025cfm} \\
\hline

\multirow{2}{*}{$\mathfrak{S}_w{\scriptsize\begin{bmatrix}\qcoh\\\boldsymbol{k}\end{bmatrix}}(\sigma)$}
 & Parabolic quantum
 & \multirow{2}{*}{$[X_w]\in{\rm QH}^\bullet({\rm Fl}(\boldsymbol{k};n))$}
 & \cite{LamShimozono} \\
 & (double) Schubert polynomial
 & & \eqref{par-q-elementary-polys-defn} \\
\hline

\multirow{2}{*}{$\mathfrak{Z}_w^{(\boldsymbol{k};n)}(\widetilde{\sigma})$}
 & Cohomological parabolic
 & \multirow{2}{*}{$[X_w]\in{\rm QH}^\bullet({\rm Fl}(\boldsymbol{k};n))$}
 & \multirow{2}{*}{\eqref{k-poly-defn}} \\
 & (double) Whitney polynomial
 & & \\
\hline\hline

\multirow{2}{*}{$\mathfrak{G}_w(x)$}
 & (Double)
 & \multirow{2}{*}{$[\mathcal{O}_w]\in{\rm K}({\rm Fl}(n))$}
 & \cite{Buch02, lascoux1982structure, Las90,
 10.1215/S0012-7094-94-07627-8} \\
 & Grothendieck polynomial
 & & \cite[(A.26)]{Closset:2025cfm} \\
\hline

\multirow{2}{*}{$\mathfrak{G}_w^{(\qk)}(x)$}
 & Quantum (double)
 & \multirow{2}{*}{$[\mathcal{O}_w]\in{\rm QK}({\rm Fl}(n))$}
 & \cite{lenart2006quantum, MNS23} \\
 & Grothendieck polynomial
 & & \cite[(A.32)]{Closset:2025cfm} \\
\hline

\multirow{2}{*}{$\mathfrak{G}_w{\scriptsize\begin{bmatrix}\qk\\\boldsymbol{k}\end{bmatrix}}(x)$}
 & {\bf Proposal:} Parabolic quantum
 & \multirow{2}{*}{$[\mathcal{O}_w]\in{\rm QK}({\rm Fl}(\boldsymbol{k};n))$}
 & \multirow{2}{*}{\eqref{qpdgp}} \\
 & (double) Grothendieck polynomial
 & & \\
\hline

\multirow{2}{*}{$\mathfrak{W}_w^{(\boldsymbol{k};n)}(x^{(\bullet)})$}
 & K-theoretic parabolic
 & \multirow{2}{*}{$[\mathcal{O}_w]\in{\rm QK}({\rm Fl}(\boldsymbol{k};n))$}
 & \cite[Thm.~5.9]{ahkmox} \\
 & (double) Whitney polynomial
 & & \eqref{qpw poly def} \\
\hline

\end{tabular}
\caption{List of the polynomials we consider in this work and in part I~\protect\cite{Closset:2025cfm}. The (quantum) Schubert polynomials represent the (quantum) cohomology classes dual to the Schubert varieties inside the partial flag manifold Fl$(\boldsymbol{k};n)$. The (quantum) Grothendieck polynomials represent classes of the structure sheaves $\CO_w$ in the (quantum) K-theory ring of Fl$(\boldsymbol{k};n)$. In the last column, we list references for each one of these polynomials: the first line gives references to the mathematics literature. 
    The second line is a reference to the equation where they are defined in this paper or in part I. Note that all these polynomials are naturally defined equivariantly (this is the meaning of the `double' in the name, with the equivariant variables $m$ (cohomology) or $y$ (K-theory) kept implicit here). The non-equivariant polynomials are obtained as a non-equivariant limit ($m=0$ or $y=1$, in our notation).} 
    \label{tab:polys and refs}
\end{table}

\medskip
\noindent
{\bf Schubert defect lines as 1d quivers.} Generalizing part I~\cite{Closset:2025cfm}, we construct the Schubert line defects $\SL_w$ as 1d $\CN=2$ quivers coupled to the 3d $\CN=2$ quiver defining the 3d GLSM with target space $X$. The gauge group of the 1d quiver is determined by $w\in {\rm W}^P$. This construction also generalizes an earlier construction of Schubert line defects (also called Grothendieck lines) for the Grassmannian manifold ${\rm Gr}(k,n)= {\rm Fl}(k;n)$~\cite{Closset:2023bdr,Gu:2025tda} --- essentially, while Schubert line defects on the Grassmannian are linear quivers, Schubert line defects on partial flag manifolds are {\it rectangular quivers.} We will show that the worldline theory of the coupled 1d-3d system localizes on the Schubert variety $X_w$, by the same logic as in part I (and conjecturally realizes
a geometric resolution $\t X_w$ of the Schubert variety $X_w$, as happens in the complete flags cases of part I). It is then natural to identify the Schubert line defects with the Schubert classes in the (equivariant, quantum) K-theory:
\be\label{def Ow intro}
[\CO_w] \,:= \,[\CO_{X_w}] \,\cong\, [\CO_{\t X_w}]\, \in\, {\rm QK}_T({\rm Fl}(\boldsymbol{k}; n))~. 
\ee
The fact that these objects naturally live in {\it quantum} K-theory is the key statement of the 3d GLSM/QK correspondence: the twisted chiral ring of half-BPS line operators of the 3d GLSM is isomorphic to the quantum K-theory ring of the target space. The quantum parameters $\qk_\ell$ ($\ell=1, \cdots, s$) are the exponentiated FI parameters of the 3d unitary gauge theory.

\medskip
\noindent
{\bf Parabolic Whitney polynomials and parabolic quantum Grothendieck polynomials: a proposal.} Quantum cohomology or K-theory rings are often usefully represented as rings of special polynomials modulo a $q$-dependent ideal. The formal variables of those polynomials are certain (K-theoretic) Chern roots. In the present case, we have a minimal standard presentation:
\be\label{QK ring explicit intro}
{\rm QK}_T({\rm Fl}(\boldsymbol{k};n)) \,\cong\, \K[x_1, \cdots, x_n]^{{\rm W}_P} / (\t {\rm QI})~.
\ee
 Here, the variables $x_i$ denote exponentiated Chern roots of the quotient vector bundles%
\footnote{Here we have $\CQ_\ell$ for $\ell=1, \cdots, s+1$, with $\CS_0:=1$ and $\CS_{s+1}:=\C^n$.} 
$\CQ_\ell = \CS_\ell/\CS_{\ell-1}$, where $\CS_\ell$ denote the tautological vector bundle whose fiber over $V_\bullet$ is $V_\ell$, and $\t {\rm QI}$ denotes a particular ideal. The ring coefficients  $\K\equiv \Z(y_1, \cdots, y_n, q_1, \cdots, q_s)$ include the equivariant parameters  $y_i$ for the $SL(n)\subset GL(n)$ isometry%
\footnote{In our physics conventions, the non-equivariant limit corresponds to $y_i=1$ ($\forall i$).} as well as the quantum parameters. 
Picking any particular basis of K-theory classes that generate the QK-theory ring, it is natural to ask for polynomial representatives in~\eqref{QK ring explicit intro}. Exactly as in part I, we will achieve this for the Schubert classes~\eqref{def Ow intro} by a two-step computation in the 3d GLSM:
\begin{enumerate}
    \item We first consider polynomials in a different set of variables, denoted by $x^{(\bullet)} = x^{(\ell)}_{a_\ell}$, which are the K-theoretic Chern roots for the vector bundles $\CS_\ell$. These polynomials were constructed very recently in the mathematical literature~\cite{ahkmox}. They are known to represent the quantum K-theory classes $[\CO_w]$ in the Whitney presentation of the QK ring, which extends~\eqref{QK ring explicit intro} to a larger set of variables including all $x^{(\bullet)}$ variables. We will  call these polynomials the {\it parabolic Whitney polynomials} and denote them by $\mathfrak{W}_w^{({\boldsymbol{k}}; n)}$. Importantly, they only depend on the variables $x^{(\ell)}$ and not on $x_i$ nor on the quantum parameters $\qk$.
     The Chern roots $x^{(\ell)}$  arise most naturally from physics, since they span the semi-classical Coulomb branch of the 3d gauge theory. Physically, our main result in this paper is that the flavored Witten indices of the 1d $\CN=2$ supersymmetric quiver gauge theories defining the Schubert line defects are {\it exactly equal} to the
     parabolic Whitney polynomials:
    \be\label{tG for xbullet intro}
    \mathcal{I}^{({\rm 1d})}[\SL_w] \,= \,\mathfrak{W}_w^{({\boldsymbol{k}}; n)} \,=\, {\rm ch}(\CO_w)~.
    \ee
   The Witten index can be computed explicitly as a residue integral~\cite{Hori:2014tda} while the Whitney polynomial has a combinatorial definition~\cite{ahkmox}, and the two computations are rather different. Hence, from the mathematical perspective, the first equality in~\eqref{tG for xbullet intro}, stating that `Witten = Whitney,' is a non-trivial conjecture. 
    Note that the 3d gauge parameters $x^{(\bullet)}$ enter as flavor parameters from the perspective of the 1d defect theory. The Witten index should then be interpreted as a Chern character of the (equivariant) K-theory class corresponding to the line defect $\SL_w$, which is the second equality in~\eqref{tG for xbullet intro}, properly interpreted.%
    \footnote{Note that there is no quantum generalization of the classical homomorphism ${\rm ch}\, :  {\rm K}(X)\rightarrow {\rm H}^\bullet(X)$. This second equality in~\protect\eqref{tG for xbullet intro} can be interpreted literally in classical K-theory, or as a schematic way of saying that the polynomial represents the quantum K-theory class.}

    \item We then use the 3d twisted chiral ring relations in the Whitney formulation, which contains both types of Chern roots $x$ and $x^{(\bullet)}$~\cite{Witten:1993xi,Gu:2023tcv}, to fully eliminate the $x^{(\bullet)}$ variables in terms of the $x$ variables. This variable elimination procedure can be performed systematically using standard computational algebraic geometry methods (see~\cite{Closset:2023vos} for a physics discussion which directly applies in the present context) or by direct computation, as we will demonstrate. Since the 3d twisted chiral ring of the 3d GLSM is identified with the quantum K-theory ring of the partial flag manifold, we find:
    \be\label{pqgp from GP intro}
 {{\mathfrak{G}_w{\scriptsize\begin{bmatrix}
        \qk\\\boldsymbol{k}
    \end{bmatrix}}(x,y)}   \,\cong \,      \mathfrak{W}_w^{({\boldsymbol{k}}; n)} (x^{(\bullet)}, y)~, \qquad\text{with}\quad
{\mathfrak{G}_w{\scriptsize\begin{bmatrix}
        \qk\\\boldsymbol{k}
    \end{bmatrix}}}\,\in\, \K[x_1, \cdots, x_n]^{{\rm W}_P}~,}
    \ee
where the equality holds inside the QK-theory ring. These latter polynomials must exactly represent $[\CO_w]$ in the QK-theory ring in the simpler Toda presentation~\eqref{QK ring explicit intro}. Note that they do explicitly depend on the quantum parameters --- this comes about because the twisted chiral ring relations themselves depend on $\qk_\ell$. 
\end{enumerate}

\begin{figure}[t!]
    \centering
    \includegraphics[width=1\linewidth]{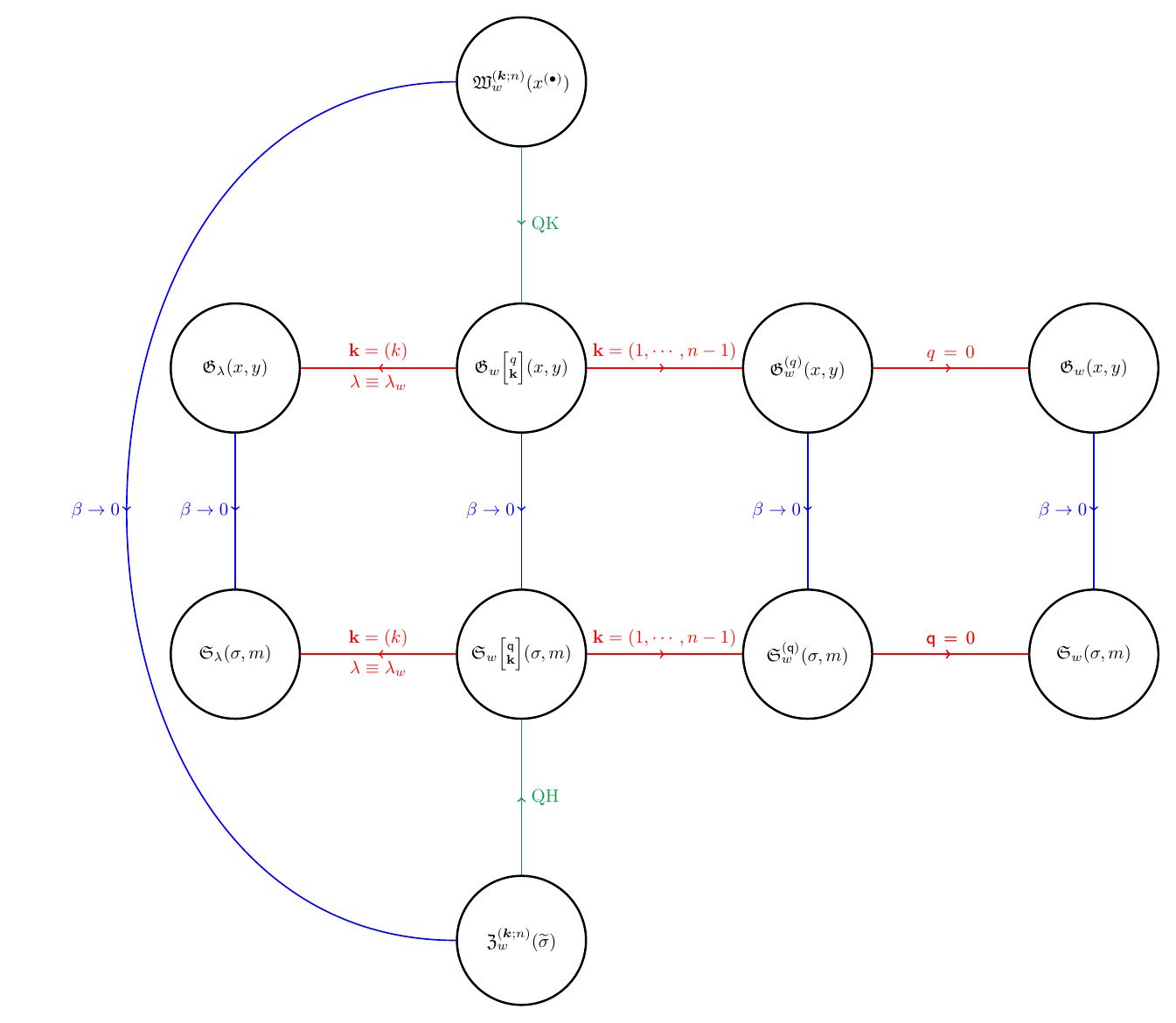}
    \caption{Limits connecting all the (quantum and classical) polynomials that we consider in this work (including part I). One can go from the parabolic Whitney polynomials to the parabolic quantum Grothendieck ones using the quantum K-theory ring relations. Similarly, one can go from the parabolic cohomological Whitney polynomials to their parabolic quantum Schubert counterparts via the quantum cohomology ring relations. Starting with the parabolic quantum double Grothendieck polynomial, one obtains all the other Toda polynomials in various limits. See table~\ref{tab:polys and refs} for terminology and references. For the special case of the Grassmannian manifold (on the left of the figure), the partition $\lambda=\lambda_w$ is defined in~\protect\eqref{defn par k}.}
    \label{fig:network of polys}
\end{figure}

\noindent 
The polynomials $\mathfrak{G}_w$ representing $[\CO_w]$ in the quantum K-theory ring~\eqref{QK ring explicit intro} for general partial flag manifolds (that is, given any $n$ and $\boldsymbol{k}$) have not been explicitly worked out in the mathematical literature, to the best of our knowledge.  The natural mathematical offshoot of our physical approach is therefore an explicit construction of such polynomials, which we call the {\it parabolic quantum (double) Grothendieck polynomials}. We give an explicit combinatorial formula for the polynomials on the left-hand-side of~\eqref{pqgp from GP intro} that naturally generalizes and subsumes all known quantum (and classical) polynomials for Schubert classes of flag manifolds, as detailed in table~\ref{tab:polys and refs}. This is presented in section~\ref{sec: quantum polys}, which can be read independently of the rest of the paper --- see subsection~\ref{subsec:pqgp def} for the explicit definitions of the parabolic quantum Grothendieck polynomials. The latter specializes to the quantum Grothendieck polynomials of Lenart and Maeno~\cite{lenart2006quantum,MNS23} in the case of the complete flag manifold, and to the ordinary Grothendieck polynomials in the case of the Grassmannian manifold, exactly as one would expect. In the cohomological limit (the 3d to 2d dimensional reduction in physics), our new quantum polynomials reduce to the previously-known parabolic quantum Schubert polynomials of Lam and Shimozono~\cite{LamShimozono}, which are known to represent the cohomology classes $[X_w]$ in the quantum cohomology ring.

In this paper, we check the proposed equality between the parabolic Whitney polynomials (and parabolic quantum Grothendieck polynomials) as defined in subsection~\ref{sec: quantum polys}, on the one hand, and the physical computation of the 1d Witten indices, on the other hand, in all examples of partial flag manifolds with $n\leq 4$. We leave a general proof of the first equality in~\eqref{tG for xbullet intro} as an interesting challenge for future work.

\medskip 
\noindent
\textbf{Plan of the paper.} The paper is structured as follows. In section \ref{sec: GLSM to partial flag}, we give a lightning review of the 2d and 3d GLSMs with partial flag manifolds as their target spaces. We review a particularly useful presentation of the quantum cohomology ring in the 2d GLSM, and of the quantum K-theory ring in the 3d GLSM, which both arise as twisted chiral rings. In section \ref{sec:Schubert in partial}, we review the definition of the Schubert variety in partial flag manifolds, setting up our notations and conventions. 

In section~\ref{sec: quantum polys}, we discuss the Whitney polynomials, and we explicitly use the ring relations to derive the parabolic quantum Grothendieck polynomials. In section \ref{sect:proposal defect}, we detail our proposal for the Schubert line defect as 1d rectangular quivers. We discuss the calculation of the corresponding 1d flavored Witten indices and show in examples how these give us back the expected Chern characters expressed as quantum polynomials. We also discuss the special case of the Grassmannian manifold and show how our proposal reduces to the Grothendieck lines defined in~\cite{Closset:2023bdr} in that case. 

Finally, taking the 2d limit of the 3d GLSM and of our construction, we discuss in section~\ref{sec:0d2dsystem} the Schubert point defects that correspond to quantum cohomology classes $[X_w]$. We compute their 0d partition functions and show that they reproduce the Schubert classes in the quantum cohomology ring of the partial flag, in examples. We end with some conclusions and remarks in section \ref{sec:conclusion}. Various explicit results are listed in appendices \ref{app:124}, \ref{app:134}, and \ref{app:234}.

\medskip 
\noindent
\textbf{Mathematica notebook for parabolic quantum polynomials.} For the convenience of the reader, this paper is accompanied by a \textsc{Mathematica}~\cite{Wolfram} notebook implementing all the polynomials defined in section~\ref{sec: quantum polys}. The notebook computes efficiently all polynomials for any partial flag manifold with $n\leq 5$ in a reasonable running time (seconds or minutes on a laptop). It also computes Hasse diagrams corresponding to embeddings of Schubert varieties.

\section{Review of GLSMs for partial flag manifolds}\label{sec: GLSM to partial flag}

In this section, we review aspects of the 2d and 3d GLSMs with partial flag manifolds as targets,
following \cite{Witten:1993xi, Donagi:2007hi}. Recall that, for $0<k_1<k_2<\cdots<k_s<n$, the partial flag manifold is defined as:
\begin{equation}\label{partial-flag-defn}
	{\rm Fl}(k_1, \cdots, k_s;n) := \left\{\,V_\bullet \equiv (0\subset V_1\subset V_2\subset\cdots\subset V_s\subset V_{s+1}\equiv\C^n)~\mid~\dim(V_\ell) = k_\ell\,\right\}~.
\end{equation}
We will also use the notation Fl$({\boldsymbol{k}};n)$ whenever convenient. We will focus on describing the quantum cohomology and quantum K-theory rings of ${\rm Fl}(k_1, \cdots, k_s;n)$, which arise as the twisted chiral rings of the 2d and 3d GLSM, respectively.

\subsection{2d GLSMs and quantum cohomology of partial flag manifolds}

\begin{figure}[t]
    \centering
    \includegraphics[scale=1.1]{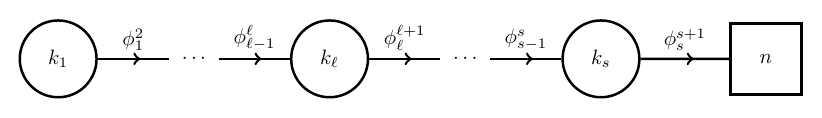}
    \caption{The 2d quiver gauge theory of interest in this paper. The circle nodes denote vector multiplets for the gauge groups $U(k_\ell)$ and the square node stands for the flavor symmetry group $SU(n)$. The ranks of the gauge groups are such that $1\leq k_1<k_2<\cdots<k_s\leq n-1$. The arrows denote bifundamental chiral multiplets --- here the fields $\phi_{\ell}^{\ell+1}$ denote the corresponding bifundamental complex scalars. The classical Higgs branch of this theory is the partial flag manifold Fl$(k_1, \cdots, k_s;n)$.}
    \label{fig:partial Flag quiver}
\end{figure}
Consider a 2d $\mathcal{N}=(2,2)$ gauge theory with the unitary gauge group:
\begin{equation}\label{par flag gauge group}
		G \,=\, U(k_1)\,\times\,U(k_2)\,\times\,\cdots\,\times\,U(k_s)~.
\end{equation}
To each pair of consecutive gauge groups $U(k_\ell)\times U(k_{\ell+1})$, we couple  a chiral multiplet $\Phi_\ell^{\ell+1} = (\phi_\ell^{\ell+1}, \cdots)$ in the bifundamental representation $(\square_{k_\ell}, \overline{\square}_{k_{\ell+1}})$. In addition, the last gauge node $U(k_s)$ also supports $n$ chiral multiplets $\Phi_s^{s+1} = (\phi_s^{s+1},\cdots)$ in the fundamental representation. This data is best summarized by the quiver diagram displayed in figure~\ref{fig:partial Flag quiver}.

The semi-classical supersymmetric vacua of this 2d GLSM are the solutions to the equations:
\begin{align}\label{vac-eq-par-flag}
    \begin{split}
    &\left(\,\widetilde{\sigma}^{(\ell)}_{a_\ell} \,- \,\widetilde{\sigma}_{a_{\ell+1}}^{(\ell+1)}\,\right)\, {\left(\phi_\ell^{\ell+1}\right)^{a_\ell}}_{a_{\ell+1}} \,=\, 0 \qquad \text{(no summation)~,}
    \\
    &\phi_\ell^{\ell+1}\,{\phi_\ell^{\ell+1\;\dagger}} \,-\,\phi_{\ell-1}^{\ell\;\dagger}\,{\phi_{\ell-1}^{\ell}} \,=\, \frac{\widetilde{\zeta}_\ell}{2\pi} \,\mathbb{I}_{k_\ell}~, 
    \end{split}
\end{align}
for $\ell=1, \cdots, s$, up to a $G$ gauge transformation. In the first equation, $\widetilde{\sigma}^{(\ell)}_\bullet$ are the diagonal components of the adjoint complex scalar fields $\widetilde{\sigma}^{(\ell)}$, which are the lowest components of the $U(k_\ell)$ vector multiplets. Here it is understood that $\widetilde{\sigma}^{(s+1)}_{i} \equiv {m}_{i}$ are the complex twisted masses associated with the flavor symmetry group $SU(n)$ --- these are constrained by the traceless condition $\sum_{i=1}^n {m}_i = 0$, but we often find it convenient to keep a full $U(n)$ symmetry manifest.   The second set of equations are the $D$-term equations,  $\widetilde{\zeta}_\ell$ denotes the 2d real FI parameter associated with the overall $U(1)\subseteq U(k_\ell)$, and $\mathbb{I}_{k_\ell}$ denotes the identity matrix of size $k_\ell\times k_\ell$.

\subsubsection{Partial flags as GIT quotients}
The Higgs branch of the theory is obtained by taking the VEVs of the scalars $\widetilde{\sigma}^{(\bullet)}$ to be trivial and all FI parameters $\widetilde{\zeta}_\ell$ to be positive. In this case, the resulting equations in \eqref{vac-eq-par-flag} are none other than the symplectic quotient definition of the partial flag manifold~\eqref{partial-flag-defn} --- see~\cite[Section 2]{Donagi:2007hi} for more details. As a GIT quotient, we can write:
\begin{equation}
		{\rm Fl}(\boldsymbol{k};n) \,=\, \left(\,\bigtimes_{\ell=1}^{s} \,\C^{k_\ell k_{\ell+1}} \,\right)\sslash_{\underline{\widetilde{\zeta}}\,>\,0}\, G_\C~,
\end{equation}
with, $\underline{\widetilde{\zeta}} := (\widetilde{\zeta}_1, \cdots, \widetilde{\zeta}_s)$, and $G_\C \,:=\, GL(k_1)\,\times\,\cdots\,\times\,GL(k_s)$ is the complexification of the gauge group \eqref{par flag gauge group}. The complex dimension and Euler characteristic of this complex manifold are given by:
\begin{equation}\label{dim and chi par flag}
\begin{split}
		&\dim({\rm Fl}(\boldsymbol{k};n)) \, =\, \sum_{\ell=1}^s\, k_\ell\,(\,k_{\ell+1}\,-\,k_\ell\,)~, \qquad \chi({\rm Fl}(\boldsymbol{k};n) ) \,=\, \frac{n!}{\prod_{\ell=0}^{s} \,(\,k_{\ell+1}\,-\,k_\ell\,)!}~,\end{split}
\end{equation}
respectively, with the understanding that $k_0 \equiv 0$   and $k_{s+1} \equiv n$.

\begin{table}[t]
\renewcommand{\arraystretch}{1.4}
\centering
\begin{equation*}
\begin{array}{|c||c|c|c|}
\hline
&{\rm Fl}(k_1, \cdots, k_s;n)& \dim & \chi\\
\hline\hline
\text{Grassmannian manifold} & {\rm Gr}(k,n) \,\equiv\, {\rm Fl}(k;n)&\, k\,(\,n\,-\,k\,) \,& \binom{n}{k}\\ 
\hline
\text{Incidence flag manifold} & {\rm Fl}(1, n-1;n) & 2\,n\,-\,3&\, n\,(\,n\,-\,1\,)\, \\
\hline
\text{Complete flag manifold} & \; {\rm Fl}(n) \,\equiv\,  {\rm Fl}(1, 2, \cdots, n-1;n)  \,&\frac{n\,(\,n\,-\,1\,)}{2} \;& n!\\
\hline
\end{array}
\end{equation*}
\caption{Three special instances of partial flag manifolds: the complex Grassmannian, the incidence flag manifold, and the complete flag manifold. Their complex dimension and Euler characteristic are given in the last two columns.}
\label{tab:par flag examples}
\end{table}

\medskip
\noindent
\textbf{Special cases.} In table~\ref{tab:par flag examples}, we list three special instances of the partial flag manifold Fl$(\boldsymbol{k};n)$. In this paper, we will look in some detail at the Grassmannian and at the incidence flag cases. The complete flag manifold is discussed in our companion paper~\cite{Closset:2025cfm}.

\medskip
\noindent
\textbf{Twisted masses and equivariant parameters.} We note that the complex twisted masses ${m}_i$ that we introduced earlier are interpreted as the equivariant parameters along the maximal torus $T\,\equiv\,U(1)^{n-1}\subset\,SL(n)$ of the isometry group of the partial flag variety.

\medskip
\noindent
\textbf{Universal tautological and quotient bundles.} Associated with the partial flag manifold \eqref{partial-flag-defn}, we have the tautological vector bundles $\{\mathcal{S}_1, \cdots, \mathcal{S}_s\}$ where, at a particular point $V_\bullet \in {\rm Fl}(\boldsymbol{k};n)$, the fiber of the rank-$k_\ell$ bundle $\mathcal{S}_\ell$ is precisely the vector space $V_\ell$. From the perspective of the 2d GLSM, the Chern roots of $\mathcal{S}_\ell$ correspond to the diagonal components of the complex scalar field $\widetilde{\sigma}^{(\ell)}$~\cite{Witten:1993xi, Donagi:2007hi,Gu:2023tcv}. 
The partial flag manifold also comes equipped with the universal quotient bundles $\mathcal{Q}_\ell$ defined as:\footnote{Note we use a different numbering convention for $\CQ_\ell$ compared to~\protect\cite{Gu:2023tcv}, where they defined $\mathcal{Q}_1  = \mathcal{S}_2/\mathcal{S}_1$.}
\begin{equation}
		\mathcal{Q}_\ell\,\equiv\, \begin{cases}
				\mathcal{S}_1~, \qquad & \ell\,=\,1~,\\
				\mathcal{S}_\ell/\mathcal{S}_{\ell-1}~, \qquad &\ell\,=\,2, \cdots, s~,\\
				\C^n/\mathcal{S}_{s}~, \qquad &\ell\, =\, s+1~.
		\end{cases}
\end{equation}
These quotient bundles have ranks:
\begin{equation}
		\rk(\mathcal{Q}_\ell) \,=\,k_\ell\,-\,k_{\ell-1}~, 
\end{equation}
with the understanding that $k_0\equiv 0$ and $k_{s+1}\equiv n$. We have a total of $n$ Chern roots for the quotient bundles, which we denote by $\{\sigma_1, \cdots, \sigma_n\}$. For instance, the Chern roots of $\mathcal{Q}_\ell$ are:
\begin{equation}\label{roots of Qell}
   \widehat{\sigma}^{(\ell)}\,\equiv\,\{\,\sigma_{k_{\ell-1}+1}\,,\, \cdots\,,\, \sigma_{k_\ell}\,\}~.
\end{equation}
In the classical cohomology ring of the partial flag manifold, ${\rm H}^\bullet({\rm Fl}(\boldsymbol{k};n))$, the Chern roots of the quotient bundles can be related to those of the tautological bundles via the exact sequences:
\begin{equation}\label{SES-ell}
    0 \,\longrightarrow\, \mathcal{S}_{\ell}\,\longrightarrow\, \mathcal{S}_{\ell+1}\, \longrightarrow\, \mathcal{Q}_{\ell+1}\, \longrightarrow\, 0~,
\end{equation}
which imply that:
\begin{equation}\label{classical coh par}
	c(\mathcal{S}_{\ell}) \,\cup\,c(\mathcal{Q}_{\ell+1})\,=\, c(\mathcal{S}_{\ell+1})~, 
\end{equation}
where, $c(E) = 1+ c_1(E) + \cdots $ denotes the total Chern class of the vector bundle $E$.

\subsubsection{The 2d A-model for the partial flag manifold} 
Recall that the GLSMs we are considering here have $\mathcal{N}=(2,2)$ supersymmetry, where we denote the four supercharges by $Q_{\pm}$, $\overline{Q}_{\pm}$. We are interested in the twisted chiral ring $\mathcal{R}_{\rm 2d}$. It is captured by the A-model subsector of the full theory, corresponding to the (equivariant) cohomology of the supercharge $Q_{\rm A}= \overline{Q}_+ +{Q}_-$ defining the A-model~\cite{Witten:1988xj}. In the present case, the twisted chiral ring is isomorphic to the (equivariant) quantum cohomology ring of the target space:
\begin{equation}\label{2d R and QH}
    \mathcal{R}_{\rm 2d}\,\cong\,{\rm QH}_T^\bullet({\rm Fl}(\boldsymbol{k};n))~.
\end{equation}
Viewing $Q_{\rm A}$ as a scalar supercharge in the $A$-model allows us to define the 2d GLSM along any closed Riemann surface $\Sigma$. In particular, this A-model computes genus-zero Gromov--Witten (GW) invariants~\cite{Closset:2015rna}.

\medskip
\noindent
\textbf{Ring relations and genus-0 Gromov--Witten invariants.} Let us take $\{\Omega_w\,:\,w\,\in\, W\}$ to be a basis for the cohomology ring of the partial flag manifold Fl$(\boldsymbol{k};n)$ with $W$ being some indexing set --- we will focus on a particular basis shortly. Let us denote the corresponding point operators by $O_w\in \CR_{\rm 2d}$. One can work out the twisted chiral ring product in this basis by computing the 2-point and 3-point functions on the Riemann sphere $\mathbb{P}^1$~\cite{Witten:1993xi, Closset:2015rna}:
\begin{equation}
    \eta_{w\,w'}\,:=\,\langle\, O_w\,O_{w'}\,\rangle_{\mathbb{P}^1}~,\qquad C_{w\,w'\,w''}\,:=\,\langle\,O_w\,O_{w'}\,O_{w''}\,\rangle_{\mathbb{P}^1}~.
\end{equation}
The two-point functions determine the topological metric. We then have the quantum cohomology ring product:
\begin{equation}\label{OPE 2d}
    O_w\,\star\,O_{w'}\,=\,\,{C_{w\,w'}}^{w''}\,O_{w''}~,
\end{equation}
where repeated indices are summed over. Here, we defined
\begin{equation}
    {C_{w\,w'}}^{w''}\,:=\,\eta^{w''\,v}\,C_{w\, w'\, v}~, \qquad \eta^{w\, v}\,\eta_{v\,w'}\,=\,\delta_{w'}^w~.
\end{equation}
From the point of the target space, $\eta_{w\,w'}$ and $C_{w\,w'\,w''}$ are genus-0 GW invariants of Fl$(\boldsymbol{k};n)$. The fusion relations \eqref{OPE 2d} are interpreted as a quantum deformation of the cup product:
\begin{equation}
    \Omega_w\,\cup_\qcoh\,\Omega_{w'}\,=\,{C_{w,w'}}^{w''}\,\Omega_{w''}\,=\,\Omega_w\,\cup\,\Omega_{w'}\,+\,\qcoh\,(\cdots)~.
\end{equation}

Any cohomology class $\Omega_w$ can be represented as a polynomial in the Chern roots of the tautological and quotient bundles. For our purposes,  it will be most useful to have an explicit presentation of the quantum ring QH$_T^\bullet({\rm Fl}(\boldsymbol{k};n))$~\cite{Astashkevich:1993ks} in terms of relations between these Chern roots, generalizing the classical relations~\eqref{classical coh par}. Such relations can be derived from the effective twisted superpotential of the 2d gauge theory~\cite{Witten:1993xi, Morrison:1994fr} --- or, equivalently, from the Bethe ansatz equations (BAEs) of the model~\cite{Nekrasov:2009uh}. For the partial flag manifold, this was worked out by some of the authors in~\cite[equation (4.68)]{Gu:2023tcv}:
\begin{equation}\label{quantum coh par}
    c(\mathcal{S}_{\ell})\,\cup_\qcoh \,c(\mathcal{Q}_{\ell+1}) \,= \,c(\mathcal{S}_{\ell+1}) \,+\,(-1)^{k_{\ell}-k_{\ell-1}} \,\qcoh_{\ell} \,c(\mathcal{S}_{\ell-1})~,
\end{equation}
where $\qcoh_\ell \sim e^{-2\pi \, \widetilde{\zeta}_\ell}$ are the quantum parameters defined in terms of the 2d (complexified) FI parameters $\widetilde{\zeta}_\ell$.

\medskip
\noindent
{\bf Explicit form of the QH ring relations.}
 At the practical level, for $\ell = 0,1, \cdots, s$, the quantum relations~\eqref{quantum coh par} can be expanded as follows:
\begin{equation}\label{QH par flag explicit}
    \begin{split}
        \sum_{a+b= c} e_{a}\,(\widetilde{\sigma}^{(\ell)})\,e_b(\widehat{\sigma}^{(\ell+1)}) \,=\, e_c(\widetilde{\sigma}^{(\ell+1)})\,+\,(-1)^{k_\ell-k_{\ell-1}}\,\qcoh_\ell\,e_{c+k_{\ell-1}-k_{\ell+1}}(\widetilde{\sigma}^{(\ell-1)})~,
        \end{split}
\end{equation}
for $c = 1, \cdots, k_{\ell+1}$. Here, $e_a(\sigma)$ denotes elementary symmetric polynomials (in the appropriate number of variables), and $\widehat{\sigma}^{(\ell+1)}$ denotes the Chern roots~\eqref{roots of Qell} for the quotient bundle $\mathcal{Q}_{\ell+1}$. For instance, the relations for $\ell = 0$ can be rewritten as:
\begin{equation}\label{sigma11=sigma1}
    \sigma_c\, =\, \widetilde{\sigma}_c^{(1)}~, \qquad c \,=\, 1, \cdots, k_1~,
\end{equation}
which is simply the statement that $\CQ_1=\CS_1$ by definition. 
Note that setting $k_{\ell}= \ell$ in~\eqref{QH par flag explicit} reproduces the complete flag relations~\cite[equation~(3.12)]{Closset:2025cfm}.

\subsubsection{Explicit examples}

Let us look at the quantum cohomology relations \eqref{QH par flag explicit} in detail for a class of examples to be revisited in later sections. (For the complete flag manifold Fl$(n)$, see subsection 3.1 of \cite{Closset:2025cfm}.) More precisely, for our purposes, we only write down the relations for $\ell=0,\cdots, s-1$ in~\eqref{QH par flag explicit}, which allow us to express the variables $\t\sigma^{(\ell)}$ fully in terms of the variable $\sigma_i$. Plugging this back into~\eqref{QH par flag explicit} with $\ell=s$ would then give us the quantum cohomology ring relations expressed in the $\sigma_i$ variables only --- we will elaborate on this point in section~\ref{sec: quantum polys}.

\begin{example} {\bf The incidence variety Fl$(1,n-1;n)$.} In this case, the quantum relations \eqref{QH par flag explicit} give us $\sigma_1 = \widetilde{\sigma}_1^{(1)}$ and 
\begin{equation}\label{QCoH of Fl(1,n-1;n)}
    e_a(\widetilde{\sigma}^{(2)})\,=\,e_a^{n-1}(\sigma)\,+\,\qcoh_1\,\delta_{a,n-1}~, \qquad a\,=\,1, \cdots,n-1~. 
\end{equation}
Here and in the following, we are using the notation $e_i^j(\sigma)\equiv e_i(\sigma_1, \cdots, \sigma_j)$. 
\end{example}

\begin{example} \textbf{The manifold Fl$(1,2;4)$.} In this case, one finds:
\begin{equation}\label{QCoH of Fl(1,2,4)}
    \begin{split}
        \widetilde{\sigma}_1^{(1)}\,=\,\sigma_1~, \qquad e_1(\widetilde{\sigma}^{(2)})\,=\, e_1^2(\sigma)~,\qquad e_2(\widetilde{\sigma}^{(2)})\,=\,e_2^2(\sigma)\,+\,\qcoh_1~.
    \end{split}
\end{equation}
\end{example}

\begin{example} {\bf The manifold Fl$(2,3;4)$.} In this case, taking $\ell=0$ in \eqref{QH par flag explicit} gives us:
\begin{equation}\label{QCoH of Fl(2,3,4) I}
    \sigma_1 \,= \,\widetilde{\sigma}_1^{(1)}~, \qquad\sigma_2 \,= \,\widetilde{\sigma}_2^{(1)}~,
\end{equation}
and plugging this into the $\ell=1$ relations in \eqref{QH par flag explicit} gives us:
\begin{equation}\label{QCoH of Fl(2,3,4) II}
   e_a(\widetilde{\sigma}^{(2)})\, =\,  e_a^{3}(\sigma)\,-\,\qcoh_1\,\delta_{a,3}~, \qquad a= 1,2,3~.
\end{equation}
\end{example}

\subsection{3d GLSMs and quantum K-theory of partial flag manifolds}\label{subsec: 3d GLSM and QK}
The three-dimensional uplift of the above GLSM is obtained by considering a 3d $\mathcal{N}=2$ supersymmetric quiver gauge theory with the same gauge groups and field content as in figure~\ref{fig:partial Flag quiver}, together with specific 3d Chern--Simons levels to be discussed momentarily. We then consider this 3d field theory on $\mathbb{R}^2\times S^1_\beta$.  Along $\mathbb{R}^2$, we retain a 2d $\mathcal{N}=(2,2)$ supersymmetric gauge theory of Kaluza--Klein (KK) type, with the KK scale $\beta^{-1}$ being the inverse radius of the compactification circle.

The semi-classical supersymmetric vacua of the 3d theory in $\R^3$ are solutions to the following set of equations: 
\begin{align}\label{3d-vac-eq-par-flag}
    \begin{split}
    &\left(\sigma^{(\ell)}_{a_\ell} - \sigma_{a_{\ell+1}}^{(\ell+1)}\right)\; {\left(\phi_\ell^{\ell+1}\right)^{a_\ell}}_{a_{\ell+1}} = 0~, \qquad (\text{no summation})~,\\
    &\phi_\ell^{\ell+1}\,{\phi_\ell^{\ell+1\;\dagger}} -\phi_{\ell-1}^{\ell\;\dagger}\,{\phi_{\ell-1}^{\ell}} = \frac{1}{2\pi} {\sf F}^{(\ell)}(\sigma, \zeta)~,
    \end{split}
\end{align}
up to a $G$ gauge transformation \eqref{par flag gauge group}. Here, $\sigma^{(\ell)}_\bullet$ are the diagonal elements of the 3d real adjoint scalar in the 3d $\mathcal{N}=2$ $U(k_\ell)$ gauge vector bundle. Meanwhile, $\phi_\ell^{\ell+1}$ is a 3d bifundamental complex scalar that is defined similarly as in 2d above.\footnote{We denote the bifundamental complex scalars by $\phi$ in 2d and in 3d. This should cause no confusion.} The new ingredient in~\eqref{3d-vac-eq-par-flag} compared to~\eqref{vac-eq-par-flag} is the matrix ${\sf F}^{(\ell)}(\sigma, \zeta)$. It is a diagonal matrix of size $k_\ell\times k_\ell$ which encodes the effective contributions of the 3d gauge Chern--Simons terms. More explicitly, we write its diagonal entries as \cite{Intriligator:2013lca, Closset:2023jiq}:
\begin{multline}
	\mathsf{F}_{a_\ell}^{(\ell)} (\sigma, \zeta)\, \equiv\, \zeta_\ell\, +\, \kk_\ell \,\sigma_{a_\ell}^{(\ell)}\, +\, {\l}_\ell \,\sum_{b_\ell=1}^{k_\ell}\, \sigma^{(\ell)}_{b_\ell} \,-\,\frac{1}{2}\,\sum_{a_{\ell-1}=1}^{k_{\ell-1}} \,\left|\,-\,\sigma^{(\ell)}_{a_\ell} \,+\, \sigma^{(\ell-1)}_{a_{\ell-1}}\,\right| \\+\, \frac{1}{2}\,\sum_{a_{\ell+1}=1}^{k_{\ell+1}}\, \left|\,\sigma^{(\ell)}_{a_\ell}\, -\, \sigma^{(\ell+1)}_{a_{\ell+1}}\,\right|~,
\end{multline}
for $a_\ell  = 1, \cdots, k_\ell$. Here, $\zeta_\ell$ is the 3d real FI parameter associated with $U(1)\subseteq U(k_\ell)$. Here, $\kk_\ell$ and $\l_\ell$ denote the Chern--Simons levels associated with the $\ell$-th gauge group:
\begin{equation}
		U(k_\ell)_{\kk_\ell, \,\kk_\ell + k_\ell \l_\ell} \cong \frac{SU(k_\ell)_{\kk_\ell}\,\times\,U(1)_{k_\ell(\kk_\ell + k_\ell \l_\ell)}}{\Z_{k_{\ell}}}~.
\end{equation}
For more details on the conventions we are following here for the Chern--Simons levels, see subsection 2.2 of \cite{Closset:2023vos}. 

\medskip
\noindent
\textbf{Types of 3d SUSY vacua.} Following the analysis that was done in \cite{Closset:2023jiq} for 3d $\mathcal{N}=2$ SQCD ({\it i.e.} the quiver gauge theory with $s=1$), we expect that the solutions to the 3d vacuum equations \eqref{3d-vac-eq-par-flag} come in the following four possible families:
\begin{itemize}
		\item \textbf{Higgs vacua:} This is the case where the real scalars $\sigma^{(\ell)}$ obtain vanishing VEVs. Then, the VEVs of the complex scalar fields $\phi_\ell^{\ell+1}$ span the flag manifold Fl$(k_1, \cdots, k_s;n)$ as long as all 3d FI parameters are large and positive, matching the target space of the 2d GLSM discussed above. 
		\item	\textbf{Topological vacua:} In this case, the real scalars obtain non-trivial VEVs giving masses to the complex scalars. Integrating out the massive bifundamental chiral matter multiplets, we obtain pure 3d $\mathcal{N}=2$ CS theory with gauge group given by $G$ in \eqref{par flag gauge group} which is a 3d TQFT; hence the name of the associated type of vacua. 
        
		\item \textbf{Hybrid vacua:} These are obtained by taking a mixture of the above two cases. That is, we take part of the fields $\sigma^{(\ell)}$ to have a trivial VEV and the other part to have non-vanishing ones. The earlier sector gives rise to a particular flag manifold; meanwhile, the latter sector leads to topological vacua. 
        
		\item \textbf{Strongly-coupled vacua:} Near the origin of the classical Coulomb branch of the theory, where the semi-classical approximation breaks down, there is a possibility for new (quantum) vacua to appear for some specific CS levels.
\end{itemize}
While topological and hybrid vacua are expected for generic CS levels, we expect that there exists a `geometric window' among all possible CS levels such that the only 3d vacua are Higgs vacua. In such a case, further compactifying $\R^3$ to $\R^2\times S^1_\beta$ gives us a 3d GLSM, wherein the 3d Higgs branch becomes the target space discussed above.

The full analysis of the solutions to \eqref{3d-vac-eq-par-flag} is beyond the scope of this work --- see~\cite{Closset:2023jiq} for the full result in the Grassmannian case. Instead of determining the full geometric window for partial flag manifolds, we will simply focus on a `standard' choice of CS levels such that the twisted chiral ring of the 3d GLSM becomes isomorphic to the (standard) quantum K-theory ring of the partial flag manifold.  See for example \cite{Jockers:2019lwe,Closset:2023bdr, Closset:2023izb,Huq-Kuruvilla:2025nlf} for discussions of more general Chern-Simons levels and their relation to twistings of quantum K theory \cite{coates-givental}, including the Ruan-Zhang levels \cite{ruan2018level,rwz1} as special cases.
In this work, we focus on the standard choice defined below, leaving the exploration of the other cases for future work.

\medskip
\noindent
\textbf{The standard CS levels.} We fix the `standard' CS levels for the 3d GLSM with target Fl$(\boldsymbol{k};n)$ as follows. Start with the 3d $\mathcal{N}=4$ quiver gauge theory described by figure \ref{fig:partial Flag quiver}, where each gauge node is now associated with an $\mathcal{N}=4$ vector multiplet while the arrows denote $\mathcal{N}=4$ hypermultiplets. In the $\mathcal{N}=2$ language, this means that we have an extra adjoint matter multiplet associated with each gauge node, and an extra bifundamental chiral multiplet between every two consecutive nodes going in the opposite direction.

The resulting hyperk\"ahler Higgs branch is an example of a Nakajima quiver variety~\cite{nakajima1998quiver}. In this case, it is the cotangent bundle of the partial flag manifold, $T^*{\rm Fl}(k_1, \cdots, k_s;n)$, and it is known that its twisted chiral ring corresponds to the quantum K-theory ring --- see~{\it e.g.}~\cite{Koroteev:2017nab}. To go back to our 3d $\mathcal{N}=2$ quiver gauge theory, one needs to softly break supersymmetry by giving masses to the extra adjoint and anti-chiral bifundamental multiplets, therefore `integrating out' the cotangent bundle fibers.  Integrating out the massive multiplets generates non-trivial 3d $\CN=2$ CS levels for the $U(k_\ell)$ gauge groups.%
\footnote{We refer to section 2 of \protect\cite{Closset:2023vos} for a review of this well-known fact, and for our CS level conventions.} In this way, we obtain the following CS levels (see also section 3 of \cite{Gu:2023tcv}):
\begin{equation}\label{standard levels}
		\kk_\ell \,=\, k_\ell\, -\, \frac{k_{\ell-1} + k_{\ell+1}}{2}~,\qquad \mathsf{l}_\ell \,=\, -1~, \qquad \forall \ell = 1, \cdots, s~,
\end{equation}
keeping in mind that $k_0\equiv 0 $ and $k_{s+1} \equiv n$. In addition to~\eqref{standard levels},  we also have mixed CS levels between the $U(1)$ factors of each two consecutive gauge groups:
\begin{equation}\label{standard mixed level}
    \kk_{\ell, \ell+1} = \frac{1}{2}~:\qquad \underbrace{U(k_\ell)\times U(k_{\ell+1}}_{\frac{1}{2}})~,\qquad i = 1, \cdots, s-1~.
\end{equation}

\subsubsection{The 3d A-model of the partial flag manifold}
Let us now consider the 3d theory on $\Sigma \times S^1_\beta$ with the topological A-twist along $\Sigma=\mathbb{P}^1$. The cohomology of the supercharge $Q_{\rm A}$ consists of half-BPS line operators $\SL$ wrapping the circle fibre at some point on $\Sigma$. These line operators are in one-to-one correspondence with the K-theory classes of coherent sheaves $[\mathcal{E}_{\SL}]$ on the partial flag manifold -- see part~I~\cite[section 2]{Closset:2025cfm} for a review. Analogously to \eqref{2d R and QH}, the 3d twisted chiral ring of these line operators is isomorphic to the (equivariant) quantum K-theory ring of Fl$(\boldsymbol{k};n)$:
\begin{equation}\label{3d R QK}
    \mathcal{R}_{\rm 3d}\,\cong\,{\rm QK}_T({\rm Fl}(\boldsymbol{k};n))~.
\end{equation}
The ring product corresponds to the fusion of parallel half-BPS lines in the $Q_{\rm A}$-cohomology.

\medskip
\noindent
\textbf{Ring relations and genus-0 K-theoretic GW invariants.} Taking a basis $\{\SL_w\,:\,w\in W\}$ of $\mathcal{R}_{\rm 3d}$, the fusion relations can be written as:
\begin{equation}\label{3d OPE relations}
    \SL_w\,\star\,\SL_{w'}\,=\,{\mathcal{N}_{w\,w'}}^{w''}\,\SL_{w''}~,
\end{equation}
where the fusion coefficients appearing on the RHS are defined in terms of the 2-point and the 3-point functions, as in 2d:
\bea
  &  g_{w\,w'}\,:=\,\langle\,\SL_w\,\SL_{w'}\,\rangle_{\mathbb{P}^1\times S^1_\beta}~, \qquad &&\mathcal{N}_{w\,w'\,w''}\,:=\,\langle\,\SL_w\,\SL_{w'}\,\SL_{w''}\,\rangle_{\mathbb{P}^1\times S^1_\beta}~,\\
&    {\mathcal{N}_{w\,w'}}^{w''}\,:=\,g^{w''\,v}\, {\mathcal{N}_{w\,w'\,v}}~, \qquad&& g^{w\,v}\,g_{v\,w'}\,=\,\delta_{w'}^w~.
\eea
The topological metric and structure constants defined above are the genus-0 K-theoretic GW invariants, and the products \eqref{3d OPE relations} should match the quantum K-theory of the flag manifold:
\begin{equation}\label{QK relations}
    [\mathcal{E}_w]\,\otimes_q\,[\mathcal{E}_{w'}]\,=\, {\mathcal{N}_{w\,w'}}^{w''}\,[\mathcal{E}_{w''}]\,=\,[\mathcal{E}_w]\,\otimes\,[\mathcal{E}_{w'}]\,+\,\qk(\cdots)~.
\end{equation}
Here, $\otimes$ is the usual tensor product operation in the K$({\rm Fl}(\boldsymbol{k};n))$ and $\otimes_q$ denotes the quantum product in QK$({\rm Fl}(\boldsymbol{k};n))$. Note that $\qk_\ell = e^{-\beta\, \zeta_\ell}$ is defined in terms of the 3d real FI parameters (complexified by a flavor holonomy along $S^1_\beta$  for the topological symmetries).

In what follows, we will be interested in a particular characteristic of the coherent sheaves on Fl$(\boldsymbol{k};n)$; namely, the Chern character:
\begin{equation}
    {\rm ch}\,:\,{\rm K}({\rm Fl}(\boldsymbol{k};n))\,\longrightarrow\,{\rm H}^\bullet({\rm Fl}(\boldsymbol{k};n))~.
\end{equation}
These Chern characters are polynomials in terms of the K-theoretic Chern roots of the tautological bundles $x_{a_\ell}^{(\ell)} = e^{-\beta \t\sigma_{a_\ell}^{(\ell)}}$ and those of the quotient bundles $x_i= e^{-\beta \sigma_i}$. Here, the complex parameters $\widetilde{\sigma}_{a_\ell}^{(\ell)}$ are the complexification of the 3d real parameters $\sigma_{a_\ell}^{(\ell)}$ by the holonomy of the corresponding gauge field along $S^1_\beta$. Therefore, one needs to rewrite the quantum K-theory ring relations \eqref{QK relations} in terms of these Chern roots. 
Following the discussion in \cite[section 4]{Gu:2023tcv}, the Bethe ansatz equations (BAEs) of our 3d A-model can be rewritten elegantly in terms of the Hirzebruch $\lambda$-classes of the universal tautological and quotient bundles as follows \cite[equation (4.42)]{Gu:2023tcv}:
\begin{equation}\label{QK lambda relations}
    \lambda_t(\mathcal{S}_\ell)\,\otimes_q\,\lambda_t(\mathcal{Q}_{\ell+1}) = \lambda_t(\mathcal{S}_{\ell+1})\,-\,t^{k_{\ell+1}-k_\ell} \,\frac{\qk_\ell}{1-\qk_\ell}\,\det(\mathcal{Q}_{\ell+1})\,\otimes_q\, (\lambda_t(\mathcal{S}_{\ell})-\lambda_t(\mathcal{S}_{\ell-1}))~.
\end{equation}
Here, the variable $t$ is a formal variable that enters into the definition of the $\lambda$ class of the vector bundles:
\begin{equation}
    \lambda_t(E) \,:= \,1 \,+ \, t\,[E]\, + \,\cdots\, + \, t^{\rk(E)}\,[\wedge^{\rk(E)}E] ~,
\end{equation}
where $[\cdot]$ denotes the corresponding K-theory class. 
The relations \eqref{QK lambda relations} can also be rewritten in terms of the  Chern characters of the $\lambda$ classes, using the relations: 
\begin{equation}\label{explicit lambda classes}
    \begin{split}
{\rm ch}(\lambda_t(\mathcal{S}_{\ell}))\,=\,\prod_{a_{\ell}=1}^{k_\ell} \left(1 \, + \, t\,x_{a_\ell}^{(\ell)}\right)~,\qquad {\rm ch}(\lambda_t(\mathcal{Q}_{\ell+1}) )\,=\,\prod_{i=k_{\ell}+1}^{k_{\ell+1}}\left(1\,+\,t\,x_i\right)~.
    \end{split}
\end{equation}
In particular, we have ${\rm ch}(\lambda_{t}(\mathcal{S}_{s+1})) = {\rm ch}(\lambda_t(\bbC^n)) = \prod_{i=1}^n(1+t\,y_i)$.\footnote{The $GL(n)$-equivariant parameters $y_i$ correspond to fugacities of the 3d flavor symmetry group $SU(n)$ and they are related to the 2d twisted masses via $y_i\equiv e^{-\beta m_i}$.} Moreover, note that ${\rm ch}(\det(\mathcal{Q}_{\ell+1})) = x_{k_\ell+1}\,\cdots\,x_{k_{\ell+1}} $.

\medskip
\noindent
{\bf Explicit form of the QK ring relations.} Using the characters \eqref{explicit lambda classes}, one can rewrite the quantum K-theory ring relations \eqref{QK lambda relations} in terms of elementary symmetric polynomials in the K-theoretic Chern roots. Namely, for $\ell=0, \cdots, s$, we have:
\begin{multline}\label{explicit QI rels for QK}
\sum_{a\,+\,b\,=\,c}\,e_a(x^{(\ell)})\,\,e_b(\widehat{x}^{(\ell+1)})\,=\,e_c(x^{(\ell+1)})\\\,-\,\frac{\qk_\ell}{1\,-\,\qk_\ell}\,x_{k_\ell+1}\,\cdots\,x_{k_{\ell+1}}\left(\,e_{c-k_{\ell+1}+k_\ell}(x^{(\ell)})\,-\,e_{c-k_{\ell+1}+k_\ell}(x^{(\ell-1)})\,\right)~,
\end{multline}
for $c = 1, \cdots, k_{\ell+1}$.
Here, similarly to \eqref{roots of Qell}, we denote by
\begin{equation}\label{Ktheory roots of Qell}
   \widehat{x}^{(\ell)}\,\equiv\,\{\,x_{k_{\ell-1}+1}\,,\, \cdots\,,\, x_{k_\ell}\,\}
\end{equation}
the set of K-theoretic Chern roots of the quotient bundle $\mathcal{Q}_{\ell}$.

\medskip
\noindent
\textbf{2d limit: back to quantum cohomology.} The strict dimensional reduction of the 3d GLSM to the ordinary 2d GLSM corresponds to setting $\beta\rightarrow 0$, so that:
\be\label{xy 2d limit}
x\, =\, e^{-\beta \sigma}\, \rightarrow \,1\,-\, \beta\, \sigma\, +\, \cdots~,\qquad\quad
y \,=\, e^{-\beta m}\, \rightarrow \,1\,-\, \beta\, m\, +\, \cdots~,
\ee
for all K-theoretic Chern roots. The quantum parameters $\qk_\ell$ and $\qcoh_\ell$ are related via:
\begin{equation}\label{qk to qcoh}
    \qk_\ell\,\equiv \,(-\beta)^{k_{\ell+1}-k_{\ell-1}}\,\qcoh_\ell ~,
\end{equation}
Up to a convention-dependent sign, this follows from the RG flow of the 2d FI parameters: while $\qk_\ell$ is naturally dimensionless, the 2d quantum parameters $\qcoh_\ell$ actually denote non-trivial RG-invariant scales set by the 1-loop beta functions for the FI parameters in 2d, hence the above relations when we match the RG and KK scales upon dimensional reduction.  
It is straightforward to see that, in this 2d limit, the quantum K-theory relations \eqref{explicit QI rels for QK} reduce to the quantum cohomology relations~\eqref{QH par flag explicit}.

\subsubsection{Some explicit examples}\label{examples eliminationQK}
Let us now discuss some examples that will be of interest to us in this work. As in the cohomology case, we focus on the equations for $\ell=0, \cdots, s-1$ in~\eqref{explicit QI rels for QK}, allowing us to express the $x^{(\bullet)}$ variables in terms of the $x$ variables. For the complete flag Fl$(n)$ case, see equations (3.18) of \cite{Closset:2025cfm} and the examples below them.

\begin{example} \textbf{The incidence variety Fl$(1,n-1;n)$.} In this case, we get that $x_1 = x_1^{(1)}$ and, 
\begin{equation}\label{QK of Fl(1,n-1;n)}
    e_a(x^{(2)}) \,=\, e_a^{n-1}(x)\,+\,\frac{\qk_1}{1\,-\,\qk_1}\,e^{n-1}_{n-1}(x)\,\delta_{n-1,a}~, \qquad a = 1, \cdots, n-1~.
\end{equation}
Note that these equations reduce to equations (2.20) in \cite{Closset:2025cfm} for the case $n=3$, and, to \eqref{QCoH of Fl(1,n-1;n)} in the 2d limit discussed above.
\end{example}

\begin{example}\textbf{The manifold Fl$(1,2;4)$.} For this case, we get the following ring relations:
\begin{equation}\label{QK of Fl(1,2,4)}
    \begin{split}
        x_1^{(1)}\,=\,x_1~, \qquad e_1(x^{(2)})\,=\,e_1^2(x)~, \qquad e_2(x^{(2)})\,=\, \frac{e_2^2(x)}{1\,-\,q_1}~.
    \end{split}
\end{equation}
\end{example}

\begin{example} \textbf{The manifold Fl$(2,3;4)$.} In this case, we have the relations:
\begin{equation}\label{QK of Fl(2,3,4) I}
    x_1 \,= \,x_1^{(1)}~, \qquad x_2 \,= \,x_2^{(1)}~, 
\end{equation}
for $\ell = 0$. And, 
\begin{equation}\label{QK of Fl(2,3,4) II}
\begin{split}
    &e_1(x^{(2)}) \,=\, x_1 \,+\, x_2 \,+\, x_3~,\\
    &e_2(x^{(2)}) \,=\, x_1\, x_2\, +\, \frac{x_1\, x_3}{1\,-\,\qk_1}\, +\, \frac{x_2\, x_3}{1\,-\,\qk_1}~,\\
    &e_3(x^{(2)}) \,=\, \frac{e_3^3(x)}{1\,-\,\qk_1}~,
\end{split}
\end{equation}
which indeed reduce to \eqref{QCoH of Fl(2,3,4) I} and \eqref{QCoH of Fl(2,3,4) II} in the 2d limit.
\end{example}
\section{Schubert varieties in partial flag manifolds}\label{sec:Schubert in partial}
The purpose of this section is to review the definitions of the Schubert classes of the partial flag manifold. In cohomology, these classes are given in terms of the Schubert varieties; meanwhile, in K-theory, they are given by the K-theory class of the structure sheaf of the Schubert variety.

In the partial flag manifold Fl$(\boldsymbol{k};n)$, the Schubert classes are indexed by a particular subset of the permutation group of $n$ characters $S_n$. The elements of this set depend on the rank vector $\boldsymbol{k}$. Before we discuss this in more detail, and to set the notation, let us start by reviewing some definitions and properties of permutations. We follow the conventions of Anderson and Fulton's book \cite{Anderson_Fulton_2023}. 

\subsection{Definitions and conventions}
We present an element  $w\in S_{n}$ of the permutation group as follows:\footnote{In what follows, we use either this notation for a permutation $w$ or $(w_1 \cdots w_{n})$, where the relation between the two is understood; namely, $w_i\equiv w(i)$.}
\begin{equation}
    w \,=\, \begin{pmatrix}
        1 \;&\;2\;&\cdots\;&\; n\\
        w(1)\;&\;w(2)&\cdots \;&\; w(n)
    \end{pmatrix} \,\equiv\, (w(1)\;w(2)\;\cdots\;w(n))~.
\end{equation}
(As in the companion paper \cite{Closset:2025cfm}, we denote the permutations in the `window' notation rather than the `cycle' notation, in which, for example, the identity is $(1\, 2\, 3\, \cdots\, n)$.  This convention is also used
in, for example, \cite{billey2000singular}.)
The permutation
$w$ is said to have a \textit{descent} at position $i$ if $w(i) > w(i+1)$. The length of $w$, denoted by $\ell(w)$, is defined as:
\begin{equation}\label{leng-perm}
    \ell(w)\, \equiv\, \# \{\,(i,j)\; :\;
    i\,< \,j \quad  \text{and}\quad w(i)\,>\,w(j)\, \}~.
\end{equation}
The longest permutation in $S_n$ is $w_0 = (n\,\cdots\,2\,1)$, and its length is given by $\ell(w_0)  = n(n-1)/2$. Observe that, for any $w\in S_n$, we have:\footnote{Here, we are following the convention for permutation decomposition where we act with $w$ first.}
\begin{equation}\label{w0w}
    (w_0w)(i)\,=\, n\, +\, 1\, -\, w(i)~.
\end{equation}

\begin{table}[t]
    \centering
        \renewcommand{\arraystretch}{1.1}
\begin{equation*}
\begin{array}{c||c|c|c|c|c}
w & (2\,1\,3) & (1\,3\,2) & (2\,3\,1) & (3\,1\,2) & (3\,2\,1)\\
\hline
\hline
{\rm r}^{w} &\left(
\begin{array}{ccc}
 0 & 1 & 1 \\
 1 & 2 & 2 \\
 1 & 2 & 3 \\
\end{array}
\right) & \left(
\begin{array}{ccc}
 1 & 1 & 1 \\
 1 & 1 & 2 \\
 1 & 2 & 3 \\
\end{array}
\right)&\left(
\begin{array}{ccc}
 0 & 1 & 1 \\
 0 & 1 & 2 \\
 1 & 2 & 3 \\
\end{array}
\right)&\left(
\begin{array}{ccc}
 0 & 0 & 1 \\
 1 & 1 & 2 \\
 1 & 2 & 3 \\
\end{array}
\right)&\left(
\begin{array}{ccc}
 0 & 0 & 1 \\
 0 & 1 & 2 \\
 1 & 2 & 3 \\
\end{array}
\right)\\
\hline
{\rm k}^{w}&\left(
\begin{array}{ccc}
 1 & 0 & 0 \\
 1 & 0 & 0 \\
 2 & 1 & 0 \\
\end{array}
\right)&\left(
\begin{array}{ccc}
 0 & 0 & 0 \\
 1 & 1 & 0 \\
 2 & 1 & 0 \\
\end{array}
\right)&\left(
\begin{array}{ccc}
 1 & 0 & 0 \\
 2 & 1 & 0 \\
 2 & 1 & 0 \\
\end{array}
\right)&\left(
\begin{array}{ccc}
 1 & 1 & 0 \\
 1 & 1 & 0 \\
 2 & 1 & 0 \\
\end{array}
\right)&\left(
\begin{array}{ccc}
 1 & 1 & 0 \\
 2 & 1 & 0 \\
 2 & 1 & 0 \\
\end{array}
\right)\\
\hline
{\rm a}^{w_0w} & \left(
\begin{array}{ccc}
 0 & 1 & 0 \\
 0 & 1 & 1 \\
 1 & 1 & 1 \\
\end{array}
\right)&\left(
\begin{array}{ccc}
 0 & 0 & 1 \\
 1 & 0 & 1 \\
 1 & 1 & 1 \\
\end{array}
\right)&\left(
\begin{array}{ccc}
 0 & 1 & 0 \\
 1 & 1 & 0 \\
 1 & 1 & 1 \\
\end{array}
\right)&\left(
\begin{array}{ccc}
 1 & 0 & 0 \\
 1 & 0 & 1 \\
 1 & 1 & 1 \\
\end{array}
\right)&\left(
\begin{array}{ccc}
 1 & 0 & 0 \\
 1 & 1 & 0 \\
 1 & 1 & 1 \\
\end{array}
\right)\\
\hline
{\rm r}^{w_0w}&\left(
\begin{array}{ccc}
 0 & 1 & 1 \\
 0 & 1 & 2 \\
 1 & 2 & 3 \\
\end{array}
\right)&\left(
\begin{array}{ccc}
 0 & 0 & 1 \\
 1 & 1 & 2 \\
 1 & 2 & 3 \\
\end{array}
\right)&\left(
\begin{array}{ccc}
 0 & 1 & 1 \\
 1 & 2 & 2 \\
 1 & 2 & 3 \\
\end{array}
\right)&\left(
\begin{array}{ccc}
 1 & 1 & 1 \\
 1 & 1 & 2 \\
 1 & 2 & 3 \\
\end{array}
\right)&\left(
\begin{array}{ccc}
 1 & 1 & 1 \\
 1 & 2 & 2 \\
 1 & 2 & 3 \\
\end{array}
\right)\\
    \end{array}
    \end{equation*}
    \caption{The rank and dimension matrices associated with the nontrivial permutations $w\in S_3$. For each matrix in the third row, reversing the order of its columns and taking the sum with the corresponding matrix of the fourth row gives us the corresponding matrix of the last row.}
    \label{tab:rank and dim matrices for S3}
\end{table}


\noindent
A special type of permutations are the \textit{transpositions} $s_{ij}$ which exchange the elements at positions $i$ and $j$, meanwhile keeping the rest unchanged. For the case where $i$ and $j$ are consecutive, we denote the transposition by $s_i \equiv s_{i, i+1}$. The permutation group $S_{n}$ is generated by these transpositions. A reduced word of the permutation $w\in S_n$ is of the form $w = s_{i_1} \cdots s_{i_{\ell(w)}}$.

Associated with each permutation $w\in S_n$, one can define the \textit{rank} and \textit{dimension} matrices ${\rm r}^w$ and ${\rm k}^w$ as follows \cite[Section 10.1]{Anderson_Fulton_2023}: 
\begin{equation}\label{rw and kw defn}
    \begin{split}
        &{\rm r}^w_{i,j} \,:=\,\#\{\,l\,\leq\,i\,:\,w(l)\,\leq\,j\,\}~,\\
        &{\rm k}^w_{i,j}\, := \,\#\{\,l\,\leq\, i\,:\,w(l) \,> \,j\,\}~.
    \end{split}
\end{equation}
For any permutation $w$, we have:
\begin{equation}\label{rw and kw connection}
\begin{split}
    &{\rm r}^w_{i,j}\,+\,{\rm k}^w_{i,j}\,=\,i~, \qquad \text{and}\qquad {\rm r}^{w_0w}_{i,j}\,=\,{\rm k}^w_{i,n+1-j}\,+\,{\rm a}_{i,j}^{w_0w}~, 
    \end{split}
\end{equation}
for all $i,j = 1, \cdots, n$. Here we introduced the new matrix ${\rm a}^w$ whose elements are given explicitly by:
\begin{equation}
    {\rm a}^w_{i,j} \,:=\,\#\{\,l\leq i\,:\,w(l)\,=\,j\,\}~.
\end{equation}
For example, in table~\ref{tab:rank and dim matrices for S3} we show the explicit form of the rank and dimension matrices for all nontrivial permutations in $S_3$. We also exhibit the corresponding $\text{a}^{w_0w}$ matrices, using which one can check the second identity in \eqref{rw and kw connection}.

\subsection{Schubert varieties in partial flag manifolds}
Let us now consider the partial flag manifold Fl$(\boldsymbol{k};n)$ defined in~\eqref{partial-flag-defn}. Famously, it can also be constructed as the quotient of $GL(n)$ by a parabolic subgroup as in~\eqref{Fl as Un ov P}.  Alternatively, we also have a description as a real manifold (see e.g.~\cite[section 2.1]{Donagi:2007hi}):
\begin{equation}\label{GoverP presentation}
    {\rm Fl}(\boldsymbol{k};n) \cong \frac{U(n)}{U(k_1)\,\times\,U(k_2-k_1)\,\times\,\cdots\,\times\,U(k_s-k_{s-1})\,\times\,U(n-k_s)}~.
\end{equation}
A Schubert variety $X_{[w]}$ is indexed by an equivalence class in the quotient~\cite[page 171]{Anderson_Fulton_2023}:
\bea\label{Wk defn}
   & {\rm W}^{({\boldsymbol{k}};n)}&\,:=\,&\frac{S_n}{\rm {\rm W}_{(\boldsymbol{k};n)}},
   \\
   & &\,\cong\,&\{w^{\rm min}\in [w] \; : \; w^{\rm min} \; \text{has descents \textbf{at most} at  positions}\; k_1\;, \cdots\;, k_s\}~, 
\eea
where ${\rm W}_{(\boldsymbol{k};n)} ={\rm W}_P$ is the Weyl group of the parabolic subgroup defining the  partial flag manifold (or, equivalently, of the group we quotient by in~\eqref{GoverP presentation}), namely: 
\begin{equation}\label{W_k defn}
    {\rm W}_{(\boldsymbol{k};n)} \,:=\, S_{k_1}\,\times\,S_{k_2-k_1}\,\times\,\cdots\,\times\,S_{k_s-k_{s-1}}\,\times\,S_{n-k_s} \,=\, \langle s_{i} \;:\; i\notin \boldsymbol{k} \rangle~.
\end{equation}
The first line of~\eqref{Wk defn} denotes the set of right cosets $[w]$, but we will always represent each $[w]$ by its minimal representative $w^{\rm min}$, which is the unique element of minimal length in $[w]$.  In what follows, we will then denote the minimal representative by $w$, so that $w\in {\rm W}^{(\boldsymbol{k};n)}$ indexes Schubert varieties denoted by $X_w$. The cardinality of the set ${\rm W}^{(\boldsymbol{k};n)}$ is equal to the Euler characteristic of the partial flag manifold $\chi({\rm Fl}(\boldsymbol{k};n))$ given in~\eqref{dim and chi par flag}. Note that the Bruhat order on $S_n$ induces a Bruhat order on the ${\rm W}^{(\boldsymbol{k};n)}$ induced by the standard Bruhat order on the minimal representatives.

Each permutation $w\in S_{n}$ admits a unique decomposition:
\begin{equation}\label{permutation-decomposition}
    w \,=\, w^{{\boldsymbol{k}}} \,\cdot\, w_{{\boldsymbol{k}}}~,
\end{equation}
where $w^{{\boldsymbol{k}}}\in {\rm W}^{({\boldsymbol{k}};n)}$  is the minimal representative, and $w_{{\boldsymbol{k}}}\in {\rm W}_{({\boldsymbol{k}};n)}$ is then a permutation built out of $\{ s_{i} \;:\; i\notin \boldsymbol{k} \}$. For the longest permutation $w_0\in S_{n}$, one can directly see from \eqref{W_k defn} that the corresponding $w_{0\,{\boldsymbol{k}}}$ is given by: 
\begin{equation}\label{w0k explicit}
    w_{0\, {\boldsymbol{k}}}(i)\, =\, k_\ell\, +\, k_{\ell+1} \,+\, 1\,-\,i~, \qquad \forall\, i \,:\, k_\ell \,< \,i\, \leq\, k_{\ell+1}~.
\end{equation}
For example, for the complete flag Fl$(n)$ case, we have $w_{0\,\boldsymbol{k}} \,=\,(1\,2\,\cdots\,n)$ and $w_0^{\boldsymbol{k}}\,=\,w_0$. 
The decomposition \eqref{permutation-decomposition} will be important in the next section when we discuss the parabolic Schubert and Grothendieck polynomials.  For future reference, we list in table~\ref{tab:longpermu for n=4} some useful data for all the possible partial flag manifolds with $n=4$ and the corresponding $w_{0\,\boldsymbol{k}}$. 

\begin{table}[t]
    \centering
            \renewcommand{\arraystretch}{1.1}
    \begin{equation*}
            \begin{array}{|c||c|c|c|c|c|c|c|}
            \hline
         {\rm Fl}(\boldsymbol{k};4)& \mathbb{CP}^3 & {\rm Gr}(2,4) & {\rm Gr}(3,4) & {\rm Fl}(1,2;4) & {\rm Fl}(1,3;4) & {\rm Fl}(2,3;4) & {\rm Fl}(4) \\
         \hline
        \hline
        \chi &4 &6&4&12&12&12&24\\
        \hline
        {\rm dim}& 3&4&3&5&5&5&6\\
        \hline
        w_{0\,\boldsymbol{k}}&(1\,4\,3\,2)&(2\,1\,4\,3)&(3\,2\,1\,4)&(1\,2\,4\,3)&(1\,3\,2\,4)&(2\,1\,3\,4)&(1\,2\,3\,4)\\
        \hline
        w_0^{\boldsymbol{k}} &(4\,1\,2\,3) &(3\,4\,1\,2)&(2\,3\,4\,1)&(4\,3\,1\,2)&(4\,2\,3\,1)&(3\,4\,2\,1)&(4\,3\,2\,1)\\
        \hline
    \end{array}
        \end{equation*}
    \caption{For each possible partial flag with $n=4$, the second row gives the Euler characteristic, the third row gives the complex dimension, and the last two rows give the decomposition of the longest permutation $w_0 = (4\,3\,2\,1)$ according to \eqref{permutation-decomposition}.}
    \label{tab:longpermu for n=4}
\end{table}

We are finally in a position to define the Schubert varieties in the partial flag manifold. Let us first define a standard (reference) complete flag -- see \cite[page 161]{Anderson_Fulton_2023}:
\begin{equation}\label{E flag}
  E_{\bullet} \: = \: \left( E_1\, \subset\, E_2 \,\subset\, \cdots \,\subset\, E_n \,\equiv\, {\mathbb C}^n \right)~.
  \end{equation}
Here $E_j \;:=\; {\rm Span}_\C\{{\rm e}_{n+1-j}, \cdots, {\rm e}_n\}$, where $\{{\rm e}_{1}, \cdots, {\rm e}_n\}$ is the standard basis on $\C^n$. For each permutation $w\in {\rm W}^{(\boldsymbol{k};n)}$, we define the Schubert variety as follows -- see \cite[page 171]{Anderson_Fulton_2023}:\footnote{Note that, in \protect\cite{Anderson_Fulton_2023}, $X_w$ is defined in terms of the reference flag $E^\bullet \equiv E_{n-\bullet}$ and the dimension matrix ${\rm k}^w$. But, using the second relation in \protect\eqref{rw and kw connection} along with the simple observation that ${\rm r}_{i,j+1}^{w_0w} - {\rm a}_{i,j+1}^{w_0w} = {\rm r}^{w_0w}_{i,j}$, one arrives to the form  we present here.}
\begin{equation}\label{X_w defn}
     X_w \,:=\,  \left\{ \,F_{\bullet}\, \in \,{\rm Fl}(\boldsymbol{k};n) \, | \,
    \dim( F_{k_\ell} \cap E_j ) \,\geq \,{\rm r}^{w_0w}_{k_\ell, j}~,\quad \forall\, 1\,\leq \,\ell\,\leq \,s~,\, \forall\,1\,\leq \,j\,\leq\, n
   \, \right\}~.
\end{equation}
The complex codimension of the Schubert variety is given by the length~\eqref{leng-perm} of the defining permutation, ${\rm codim}(X_w)\,=\,\ell(w)$. In particular,
\be\label{ellwok dim}
\ell(w_0^{\boldsymbol{k}})\,=\, {\rm dim}({\rm Fl}({\boldsymbol{k}};n))~,
\ee
and $[X_{w_0^{\boldsymbol{k}}}]$ gives us the point class.
The Bruhat order on ${\rm W}^{({\boldsymbol{k}};n)}$ gives us the Hasse diagram of Schubert varieties just like for the complete flag manifold.

\subsubsection{Special case: the Grassmannian manifold}
As mentioned already in table~\ref{tab:par flag examples}, the Grassmannian Gr$(k,n)$ can be viewed as a one-step flag Fl$(k;n)$. Then, from \eqref{X_w defn}, its Schubert varieties are given by:
\begin{equation}
    X_w \,=\,  \left\{ F_k\, \in\, {\rm Fl}(k;n) \, | \,
    \dim( F_k  \,\cap\, E_j ) \,\geq\, {\rm r}^{w_0w}_{k, j}~,\quad  \forall\,1\,\leq\, j\,\leq\, n
   \, \right\}~,
\end{equation}
where $w \in S_n$ has descent at most at position $k$. Such a permutation with a single descent is called a Grassmannian permutation. 
To simplify further, recall  from \eqref{rw and kw defn} that we have:
\begin{equation}
    {\rm r}_{k,j}^{w_0 w} \, = \,\#\{\,l\,\leq\,k\,:\,n\,+\,1\,-\,w(l)\,\leq\,j\,\}~.
\end{equation}
It is easily seen that:
\begin{equation}\label{r for Gr(k,n)}
    {\rm r}_{k,j}^{w_0 w} = \begin{cases}
        0~,  \qquad &j\, <\, n\,+\,1\,-\,w(k)\\
        1~,  \qquad &n\,+\,1\,-\,w(k) \,\leq \,j\, <\, n\,+\,1\,-\,w(k-1)\\
        2~, \qquad & n\,+\, 1\,-\,w(k-1) \,\leq\, j \,<\, n\,+\, 1\,-\, w(k-2)\\
        ~~\vdots\\
        k~, \qquad& n\,+\,1\,-\, w(1)\, \leq\, j \,\leq\,n~,
    \end{cases}~,
\end{equation}
where we used the fact that $w(1) < w(2) <\cdots < w(k)$. Therefore, we can write the constraints defining the Schubert varieties as follows:
\begin{equation}
\begin{aligned}
    &\dim(F_k \,\cap\, E_j)\, \geq\, 0~, \quad  j \,<\, n\,+\,1\,-\,w(k)~,\\
    &\dim(F_k \,\cap\, E_j) \,\geq\, 1~, \quad n\,+\,1\,-\,w(k)\, \leq\, j \,<\, n\,+\,1\,-\,w(k-1)~,\\
    &\qquad \qquad\vdots\\
    & \dim(F_k \,\cap\, E_j) \,\geq\, k~, \quad n\,+\,1\,-\,w(1)\, \leq\, j\, \leq\, n~.
\end{aligned} 
\end{equation}
Among those constraints, some are trivial (for example, the ones on the first line, because the dimension should always be non-negative), while some are redundant. For example, consider the constraints $\dim(F_k \,\cap\, E_j)\, \geq\, m$ and $\dim(F_k \,\cap\, E_{j'})\, \geq\, m$ for some fixed integer $m$ with $j < j'$; in this case, the first constraint clearly implies the second. In this way, we can reduce our definition of Schubert varieties to:
\begin{equation}\label{Schubert in Grass defn}
    X_w \,=\,  \left\{ F_k \,\in\, {\rm Fl}(k;n) \, | \,
    \dim( F_k  \,\cap\, E_{n\,+\,1\,-\,w(k+1-a)} ) \,\geq\, a~,\quad  \forall\,1\,\leq \,a\,\leq\, k\,
   \, \right\}~,
\end{equation}
Now, if we let:
\begin{equation}\label{defn par k}
   \lambda_{w,\,a} \,\equiv \,\lambda_a \,= \,w(k+1-a)\, -\, (k\,+\,1\,-\,a)~,
\end{equation}
then we see that:
\begin{equation}
    n\,-\,k\,\ge\, \lambda_1\, \geq \,\lambda_2\, \geq\, \cdots\,\geq\, \lambda_k\, \geq\,0~,
\end{equation}
which defines a partition $\lambda$ whose Young tableau fits inside a $k\times (n-k)$ rectangle.   Schubert varieties of Grassmannian manifolds are labelled by such partitions (see {\it e.g.}~\cite[page 128]{Anderson_Fulton_2023}):
\begin{equation}
    X_\lambda \,=\, \left\{ F_k\, \in \,{\rm Fl}(k;n) \, | \,
    \dim( F_k  \,\cap\, E_{n\,-\,k\,+\,a\,-\,\lambda_a} )\, \geq\, a~,\quad  \forall\,1\,\leq\, a\,\leq\, k
   \, \right\}~.
\end{equation}

\section{Parabolic quantum polynomials for partial flag manifolds}\label{sec: quantum polys}

In this section, we discuss the realization of the (equivariant) quantum cohomology and quantum K-theory rings of partial flag manifolds in the Schubert class basis in terms of the Schubert and Grothendieck polynomials, respectively. We also define the parabolic Whitney polynomials and explain how they are related to the parabolic Grothendieck polynomials, as anticipated in the introduction.

\subsection{Generalities on quantum rings for partial flag manifolds}\label{subsec: qkrings generalities}

In the previous section, we reviewed the QH and QK ring relations for the partial flag manifold ${\rm Fl}(k_1, \cdots, k_s; n)$. Let us focus on the QK ring for definiteness. We denote by
\be
\K\,\equiv\, \Z(y, \qk)
\ee
the field we are working on --- that is, the ring coefficients are rational functions of the equivariant parameters $y_i$ ($i=1,\cdots, n$) and of the quantum parameters $\qk_\ell$ ($\ell=1, \cdots, s$). 
We then have the Whitney presentation~\cite{Gu:2023fpw,Huq-Kuruvilla:2024tsg,ahkmox}:
\be\label{QK ring explicit for all x}
{\rm QK}_T({\rm Fl}({\boldsymbol{k}};n))\, \cong \,\K[x^{(\bullet)}, x_1, \cdots, x_n]^{{\rm W}_G\times {\rm W}_P} / ({\rm QI})~,
\ee
where, as we discussed earlier, $x^{(\bullet)}$ denotes the parameters $x_{a_\ell}^{(\ell)}$, corresponding to the K-theoretic Chern roots for the tautological vector bundles $\CS_\ell$, while $x_i$ denote the K-theoretic Chern roots for the quotient bundles. More precisely,  the exponentiated Chern roots for $\CQ_\ell$ are given by~\eqref{Ktheory roots of Qell}. 
The `quantum' ideal ${\rm QI}$ in~\eqref{QK ring explicit for all x} is the one generated by the QK ring relations given in~\eqref{explicit QI rels for QK}.  Finally, note that we need to take the invariants under ${\rm W}_G\times {\rm W}_P$, where ${\rm W}_G$ denotes the Weyl group of~\eqref{par flag gauge group} and ${\rm W}_P$ denotes the Weyl group of the `parabolic' subgroup of $U(n)$ by which we quotient in~\eqref{GoverP presentation}, namely:%
\footnote{To avoid any confusion, let us insist on the fact that $G$ here denotes the GLSM gauge group~\protect\eqref{par flag gauge group} and {\it not} $G=GL(n)$ as often the case in the mathematical literature when writing the presentation~\protect\eqref{Fl as Un ov P} as $G/P$.}
\be
{\rm W}_G\,= \,\bigtimes_{\ell=1}^s\, S_{k_\ell}~,\qquad \qquad {\rm W}_P \,=\,{\rm W}_{({\boldsymbol{k}};n)}\,=\,\bigtimes_{\ell=1}^{s+1}\,S_{k_{\ell}-k_{\ell-1}}~.
\ee
Note that ${\rm W}_G$ acts on the $x^{(\bullet)}$'s only (that is, $S_{k_\ell}$ permutes the $\CS_\ell$ Chern roots $x^{(\ell)}$), while ${\rm W}_P$ acts on the $x_i$'s only (that is, $S_{k_{\ell}-k_{\ell-1}}$ permutes the $\CQ_\ell$ Chern roots~\eqref{Ktheory roots of Qell}).

Following the discussion in the previous section, it is clear that we can entirely eliminate the $\CS_\ell$ variables $x_\bullet^{(\bullet)}$ from the description of the QK ring~\eqref{QK ring explicit for all x}, which then gives us the simpler presentation, sometimes called the Toda presentation~\cite{ahkmox}:
\be\label{QK ring explicit for x for Q}
{\rm QK}_T({\rm Fl}({\boldsymbol{k}};n)) \,\cong\, \K[x_1, \cdots, x_n]^{{\rm W}_P} / (\t {\rm QI})~.
\ee
Here $\t {\rm QI}$ denotes the reduced ideal obtained from ${\rm QI}$ by eliminating the $x^{(\bullet)}$ variables.  

We are interested in an explicit presentation of the (equivariant) quantum K-theory ring in terms of the Schubert classes $[\CO_w]$ indexed by $w\in  {\rm W}^{({\boldsymbol{k}};n)}$ as defined in~\eqref{Wk defn}. These classes should be represented in the Toda-presentation ring~\eqref{QK ring explicit for x for Q} by some ${\rm W}_P$-invariant polynomials, which we call the {\it parabolic quantum (double) Grothendieck polynomials}, generalizing the quantum (double) Grothendieck polynomials of Lenart--Maeno~\cite{lenart2006quantum} that represent (equivariant) Schubert classes in complete flag manifolds~\cite{MNS23}. Note that the Weyl group ${\rm W}_P$ is trivial in the case of the complete flag, hence the ordinary (quantum) Grothendieck polynomials are elements of $\C[x_1, \cdots, x_n]$, without any symmetry properties in general. 

We will proceed in two steps, as we explained in the introduction. We first define the ${\rm W}_G$-invariant {\it parabolic Whitney polynomials} which represent the Schubert classes in the Whitney presentation~\eqref{QK ring explicit for all x} of the quantum ring, and which happen to be independent of the quantum parameters $q$~\cite{ahkmox}. Starting from the Whitney polynomials, variable elimination in the QK-theory ring gives us the parabolic quantum Grothendieck polynomials in the Toda presentation. We will implement this elimination procedure completely explicitly, which also reproduces the recently-derived Toda presentation of the QK ring of partial flag manifolds~\cite{ahkmox}. The situation in quantum cohomology is analogous and can be understood as a limit from K-theory (the 3d to 2d limit in the GLSM); we will introduce all the relevant cohomological polynomials below as well.

\subsection{Parabolic Whitney polynomials in cohomology and K-theory}
New polynomials were recently introduced by Amini {\it et al}~\cite{ahkmox} to describe Schubert classes in the Whitney presentation~\eqref{QK ring explicit for all x} of the (equivariant) QK-theory ring of a partial flag manifold. In this work, we call these the {\it parabolic (double) Whitney polynomials}, to distinguish them from the Toda-presentation polynomials that we will introduce below. A key observation is that the Whitney polynomials are only functions of the $x^{(\bullet)}$ variables: by construction, they do not depend on the parameters $\qk_\ell$ at all. In the next section, we will give a physics interpretation of the Whitney polynomials as 1d Witten indices of Schubert line defects.

Note also that, since the Whitney polynomials are independent of the quantum parameters, we can think of them as representing classes in the classical ring as well. It is only the relations among them in the quantum ring that will depend on the quantum parameters.

\subsubsection{Parabolic Whitney polynomials in cohomology}
Let us first discuss the cohomological case. 
 Here we define the {\it parabolic (double) Whitney polynomials in cohomology}, or sometimes simply the  $\mathfrak{Z}$-polynomials.%
\footnote{We will interpret them as supersymmetric matrix model partition functions in section~\protect\ref{sec:0d2dsystem}, hence the name.} The  $\mathfrak{Z}$-polynomials are function of the Chern roots $\t\sigma^{(\ell)}$ for $\CS_\ell$ only. They should be understood as representing Schubert classes in the Whitney presentation of the quantum cohomology:
\be\label{QH ring explicit Whitney}
{\rm QH}^\bullet_T({\rm Fl}({\boldsymbol{k}};n))\, \cong\, \K[\t\sigma^{(\bullet)}, \sigma_1, \cdots, \sigma_n]^{{\rm W}_G \times {\rm W}_P} / ({\rm qI})~,
\ee
which is written similarly to the K-theoretic ring~\eqref{QK ring explicit for all x}, with $\K\,\equiv\, \Z(m, \qcoh)$ and the ideal $({\rm qI})$ is generated by the Whitney relations~\eqref{QH par flag explicit}.

The $\mathfrak{Z}$-polynomials can be defined recursively. We first define the Whitney polynomial for the longest permutation  $w_0^{\boldsymbol{k}}\in {\rm W}^{(\boldsymbol{k}; n)}$, which represents the point class, and then for any `parabolic' permutation using standard operations~\cite{ahkmox}. 
The parabolic double Whitney polynomial representing the point class in cohomology is given by a very simple formula:
\be\label{pqw hom poly longest}
\mathfrak{Z}^{({\boldsymbol{k}}; n)}_{w_0^{{\boldsymbol{k}}}}(\t\sigma^{(\bullet)}, m) \;=\; \prod_{\ell=1}^{s} \,\prod_{j=k_\ell}^{k_{\ell+1}-1}\, \prod_{a_\ell=1}^{k_\ell}\, \left(\,\t\sigma_{a_\ell}^{(\ell)} \,-\, m_{n-j}\, \right)~.
\ee
We conjecture this identification since it corresponds simply to the cohomological limit of the $K$-theory construction of~\cite{ahkmox} that we will review below. Note that~\eqref{pqw hom poly longest} is a homogeneous (double) polynomial of degree $\ell(w_0^{{\boldsymbol{k}}})$ (and recall~\eqref{ellwok dim}). 

For any  minimal representative $w\in {\rm W}^{({\boldsymbol{k}};n)}$, as defined in~\eqref{Wk defn}, the parabolic double Whitney polynomial in cohomology is then defined as:
\begin{equation}\label{k-poly-defn}
\mathfrak{Z}^{({\boldsymbol{k}}; n)}_{w}(\t\sigma^{(\bullet)}, m)
\,:=\, (-1)^{\ell(w^{-1} w_0^{{\boldsymbol{k}}})} \,\partial^{(m)}_{w^{-1} w_0^{{\boldsymbol{k}}}}\,\mathfrak{Z}^{({\boldsymbol{k}}; n)}_{w_0^{{\boldsymbol{k}}}}(\t\sigma^{(\bullet)}, m)~.
\end{equation}
Here, $\partial^{(m)}$ is the divided difference operator taken with respect to the equivariant parameters $m$; see~\cite[equations (A.2)--(A.4)]{Closset:2025cfm} for its definition.  Note that $\mathfrak{Z}_w^{({\boldsymbol{k}}; n)}$ is a homogeneous double polynomial of degree $\ell(w)$. These (double) polynomials represent the (equivariant) quantum cohomology classes $[X_w]$ in the Whitney presentation~\eqref{QH ring explicit Whitney}.%
\footnote{While these polynomials and their identification with $[X_w]$ were not explicitly discussed in~\protect\cite{ahkmox}, our claims directly follow from their results by taking the cohomological limit of the K-theoretic results.}
  The non-equivariant limit is obtained by setting $m=0$.

\subsubsection{Parabolic Whitney polynomials in K-theory}\label{subsec:parabolic whitney k}
Moving up to the quantum K-theory, the parabolic double Whitney polynomial for the longest permutation $w_0^{{\boldsymbol{k}}}$ is given by:
\be\label{pqw poly longest}
\mathfrak{W}^{({\boldsymbol{k}}; n)}_{w_0^{{\boldsymbol{k}}}}(x^{(\bullet)}, y) \;=\; \prod_{\ell=1}^{s} \,\prod_{j=k_\ell}^{k_{\ell+1}-1}\, \prod_{a_\ell=1}^{k_\ell}\, \left(\,1\, -\, \frac{x_{a_\ell}^{(\ell)}}{y_{n-j}}\,\right)~,
\ee
which is the natural K-theory uplift of~\eqref{pqw hom poly longest}. 
This quantity was introduced in~\cite[theorem 1.2, theorem 5.11]{ahkmox}, where it is proven that it represents the point class in the K-theory of ${\rm Fl}({\boldsymbol{k}}; n)$. 
 Next, one assigns to each minimal permutation $w\in {\rm W}^{({\boldsymbol{k}};n)}$ the parabolic double Whitney polynomials defined as:
\begin{equation}\label{qpw poly def}
\mathfrak{W}_w^{({\boldsymbol{k}}; n)}(x^{(\bullet)},y)\,  := \,\pi^{(y)}_{w^{-1}w_0^{\boldsymbol{k}}}\,\mathfrak{W}^{({\boldsymbol{k}}; n)}_{w_0^{{\boldsymbol{k}}}}(x^{(\bullet)},y)~.
\end{equation}
Here $\pi^{(y)}$ is the Demazure operator acting on the equivariant parameters $y_i$ as defined in~\cite[(A.22)--(A.24)]{Closset:2025cfm}. 
 The Whitney polynomial~\eqref{qpw poly def} represents the Schubert class $[\CO_w]$ in the Whitney presentation of the equivariant QK ring~\cite[theorem 5.9]{ahkmox}. 
The non-equivariant limit corresponds to setting $y=1$.

\medskip
\noindent
{\bf The 2d limit (QK to QH).} Considering the relations $x\equiv e^{-\beta \sigma}$, $y\equiv e^{-\beta m}$, the $\mathfrak{Z}$-polynomials are recovered in the limit: 
\begin{equation}\label{2d-limit-double-whitney}
    \lim_{\beta\rightarrow 0}\,\left( \beta^{-\ell(w)}\, \mathfrak{W}_{w}^{({\boldsymbol{k}}; n)}\left(\,e^{-\beta \t\sigma^{(\bullet)}},e^{-\beta m}\right)\,\right)\, =\, 
    \mathfrak{Z}_{w}^{({\boldsymbol{k}}; n)} (\t\sigma^{(\bullet)}, m)~,
\end{equation}
as expected on general grounds --- in the 3d GLSM interpretation, this is simply a dimensional reduction to 2d, as we already discussed below~\eqref{xy 2d limit}.

\subsection{Parabolic quantum Schubert polynomials for the QH}

Going back down to the quantum cohomology, let us now review the known construction of ${\rm W}_P$-symmetric polynomials representing Schubert classes $[X_w]$ in the (equivariant) quantum cohomology ring. We are now interested in the Toda presentation:
\be\label{QH ring explicit for x for Q}
{\rm QH}^\bullet_T({\rm Fl}({\boldsymbol{k}};n))\, \cong\, \K[\sigma_1, \cdots, \sigma_n]^{{\rm W}_P} / (\widetilde{\rm qI})~,
\ee
in complete analogy to~\eqref{QK ring explicit for x for Q}, with the ideal $(\widetilde{\rm qI})$ obtained from $({\rm qI})$ in~\eqref{QH ring explicit Whitney} after elimination of the variables $\t\sigma^{(\bullet)}$. The relevant polynomials are the {\it parabolic quantum (double) Schubert polynomials} introduced by Lam and Shimozono~\cite{LamShimozono}.

\medskip
\noindent
{\bf Notation.} In the following, we will denote these polynomials by $\mathfrak{S}_{w} {\scriptsize\begin{bmatrix}
    \qcoh\\
    {\boldsymbol{k}}
\end{bmatrix}} (\sigma, m)$, for any minimal representative $w\in {\rm W}^{({\boldsymbol{k}};n)}$. We are using the following notation:
\begin{equation}
    \begin{bmatrix}
        \qcoh\\
        {\boldsymbol{k}}
    \end{bmatrix}
   \, \equiv\, \begin{bmatrix}
        \qcoh_1 \; & \;\qcoh_2 \; & \;\cdots \;& \;\qcoh_{s}\\
        k_1 \; & \; k_2\;& \; \cdots \; & \; k_s 
    \end{bmatrix}~. 
\end{equation}
For example, for the complete flag case ($s = n-1$), we expect that:
\begin{equation}
    \mathfrak{S}_{w}{\scriptsize\begin{bmatrix}
        \qcoh_1 \; & \; \qcoh_2 \; & \; \cdots \; & \; \qcoh_{n-1}\\
        1\;&\; 2 \; & \; \cdots\; &\; n-1
    \end{bmatrix}} (\sigma, m)\, = \,\mathfrak{S}^{(\qcoh)}_{w} (\sigma, m)~.
\end{equation}
The polynomial appearing on the RHS is the quantum double Schubert polynomial --- for the definition and examples, see~\cite[appendix A]{Closset:2025cfm} and references therein.

\subsubsection{Parabolic quantum elementary polynomials}
To define $\mathfrak{S}_{w} {\scriptsize\begin{bmatrix}
    \qcoh\\
    {\boldsymbol{k}}
\end{bmatrix}} (\sigma, m)$ explicitly, let us first define the $n\times n$ matrix:
\begin{equation}\label{q-par-D-matrix}
    \left(D_{n}{\scriptsize\begin{bmatrix}
        \qcoh\\
        {\boldsymbol{k}}
    \end{bmatrix}}\right)_{ij}\, :=\, \sigma_i\, \delta_{i, j}\,-\, \delta_{j, i+1}\, -\, \sum_{\ell=1}^{s}\,(-1)^{k_{\ell}-k_{\ell-1}}\, \qcoh_\ell\, \delta_{i, k_{\ell+1}}\,\delta_{j, k_{\ell-1}+1} ~, 
\end{equation}
with $i, j = 1, \cdots, n$ and the convention that $k_0=0$ and $k_{s+1}=n$. One can easily check that, in the complete-flag limit, we recover the matrix $D^{(\qcoh)}_{n}$ from the definition of the quantum Schubert polynomials~\cite[equation~(A.18)]{Closset:2025cfm}.

Expanding the matrix \eqref{q-par-D-matrix}, we get \textit{the parabolic quantum elementary polynomials}, which we will denote by ${{\rm E}^{({\boldsymbol{k}};n)}}^j_{i}$ \cite{Astashkevich:1993ks, LamShimozono}: 
\begin{equation}\label{par-q-elementary-polys-defn}
    \det\left(\,D_{k_\ell}{\scriptsize\begin{bmatrix}
        \qcoh\\
        {\boldsymbol{k}}
    \end{bmatrix}} -t \;\mathbb{I}_{k_\ell}\,\right)\, =\, \sum_{a_\ell=0}^{k_\ell} (-t)^{k_\ell - a_\ell}\; {{\rm E}^{({\boldsymbol{k}};n)}}_{a_\ell}^{k_\ell}(\sigma; \qcoh)~,  \qquad \ell\,=\,1, \cdots, s~,
\end{equation}
It is understood here that $D_{k_\ell}{\scriptsize\begin{bmatrix}
    \qcoh\\{\boldsymbol{k}}
\end{bmatrix}}$ is the $k_\ell\times k_\ell$ upper-left submatrix of $D_{n}{\scriptsize\begin{bmatrix}
    \qcoh\\{\boldsymbol{k}}
\end{bmatrix}}$. As part of their definition, these elementary polynomials vanish for the cases $a_\ell = k_\ell = 0$, either $k_\ell$ or $a_\ell$ is negative, or $a_\ell > k_\ell$. Moreover, one can observe that the polynomials $ {{\rm E}^{({\boldsymbol{k}};n)}}_{a_\ell}^{k_\ell}(\sigma; \qcoh)$ depend only on the first $k_\ell$  ($\sigma_1, \cdots, \sigma_{k_\ell}$) of the $n$ Toda variables $\sigma_i$.

\medskip
\noindent
\begin{example} {\bf Parabolic quantum elementary polynomials for $n=4$.} For the partial flags with $n=4$, the matrices $D_4{\scriptsize\begin{bmatrix}
    \qcoh\\{\boldsymbol{k}}
\end{bmatrix}}(\sigma)$ are given in table~\ref{tab:flags with nf=4}. Using these matrices, one can find the following list of examples of parabolic quantum elementary polynomials:
 \begin{align}
     \begin{split}
         &{\rm E^{(2;4)}}^2_1 (\sigma; \qcoh)\, =\, e_1^2(\sigma)~,\\
         &{\rm E^{(1,2;4)}}_2^2(\sigma;\qcoh) = e_2^2(\sigma) +\qcoh_1 ~,\\ 
         &{\rm E^{(1,3;4)}}_2^3(\sigma;\qcoh) = e_2^3(\sigma)~,\\
         &{\rm E^{(2,3;4)}}^3_3 (\sigma; \qcoh) =e_3^3(\sigma)\,-\,\qcoh_1~,\\
         &{\rm E^{(1,2,3;4)}}_2^3(\sigma; \qcoh) = e_2^3(\sigma) + \qcoh_1 + \qcoh_2~.\\
     \end{split}
 \end{align}
In the complete-flag case, we reproduce the polynomials given in~\cite[equation (A.16)]{Closset:2025cfm}.  
 \end{example}

\noindent
The parabolic quantum elementary polynomials defined above satisfy the following set of recurrence relations --- see subsection 3.3 of \cite{buch2005quantum}:
\begin{multline}\label{recurrence relations for qpele}
    {{\rm E}^{({\boldsymbol{k}};n)}}_{a_\ell}^{k_\ell} (\sigma;\qcoh)\,=\,  \sum_{r=0}^{k_\ell-k_{\ell-1}} \,e_{r}\left(\sigma_{k_{\ell-1}+1}, \cdots, \sigma_{k_\ell}\right)\,  {{\rm E}^{({\boldsymbol{k}};n)}}_{a_\ell-r}^{k_{\ell-1}} (\sigma; \qcoh) \\-\,(-1)^{k_\ell-k_{\ell-1}}\, \qcoh_{\ell-1} \, {{\rm E}^{({\boldsymbol{k}};n)}}_{a_\ell+k_{\ell-2}-k_\ell}^{k_{\ell-2}}(\sigma; \qcoh)~,
\end{multline}
for $\ell=1, \cdots, s$ and $a_\ell = 1, \cdots, k_\ell$. 
Note the similarity of these recurrence relations with the quantum relations \eqref{QH par flag explicit} --- indeed, the quantum elementary polynomials here satisfy exactly the same quantum-corrected relations as the elementary polynomials $e_a(\t\sigma^{(\ell)})$ in~\eqref{QH par flag explicit}. This means that the quantum polynomials ${{\rm E}^{({\boldsymbol{k}};n)}}_{a_\ell}^{k_\ell}$ allow us to solve explicitly for the $\t\sigma^{(\bullet)}$ variables in terms of the variables $\sigma_i$ in the Whitney presentation of the ring, as:
\be
e_{a_\ell}(\t\sigma^{(\ell)})=   {{\rm E}^{({\boldsymbol{k}};n)}}_{a_\ell}^{k_\ell} (\sigma;\qcoh)~.
\ee
We will elaborate on this important point more explicitly in the K-theoretic context below.

\begin{center}
\renewcommand{\arraystretch}{1.1}
\begin{longtable}[t]{|c|| c | c |}
\hline
   $\text{Fl}({\boldsymbol{k}}; 4)$
   &$D_{4}{\scriptsize \begin{bmatrix}
    \qcoh\\{\boldsymbol{k}}
\end{bmatrix}}(\sigma)$ &${\rm G}_4{\scriptsize\begin{bmatrix}
    \qk\\{\boldsymbol{k}}
\end{bmatrix}}(x)$ \\
   \hline
   \hline
     $\mathbb{CP}^{3}$~
     &~ $\left(
\begin{array}{cccc}
 \sigma _1 & -1 & 0 & 0 \\
 0 & \sigma _2 & -1 & 0 \\
 0 & 0 & \sigma _3 & -1 \\
 \qcoh_1 & 0 & 0 & \sigma _4 \\
\end{array}
\right)$ &$\begin{pmatrix}
       x_1 &-\frac{1}{1-\qk_1} &0 &0 \\
        0&x_2&-1&0\\
      0&0&x_3&-1 \\
      q_1x_1x_2x_3x_4&0&0&x_4\\
\end{pmatrix}$\\
\hline
    $\text{Gr}(2,4)$~
    ~&~$\left(
\begin{array}{cccc}
 \sigma _1 & -1 & 0 & 0 \\
 0 & \sigma _2 & -1 & 0 \\
 0 & 0 & \sigma _3 & -1 \\
 -\qcoh_1 & 0 & 0 & \sigma _4 \\
\end{array}
\right)$&$\begin{pmatrix}
    x_1 &-1 &0 &0 \\
       0 &x_2&-\frac{1}{1-\qk_1}&0\\
     0 &0&x_3&-1\\
      -q_1 x_1^2 x_3 x_4 &q_1(x_1+x_2)x_3 x_4 &0 &x_4\\
\end{pmatrix}$\\
\hline
     $\text{Gr}(3,4)$
     &~$\left(
\begin{array}{cccc}
 \sigma _1 & -1 & 0 & 0 \\
 0 & \sigma _2 & -1 & 0 \\
 0 & 0 & \sigma _3 & -1 \\
 \qcoh_1 & 0 & 0 & \sigma _4 \\
\end{array}
\right)$&$\begin{pmatrix}
    x_1 &-1&0&0\\
    0 &x_2&-1&0\\
    0&0&x_3&-\frac{1}{1-q_1}\\
q_1 x_1^3x_4&-q_1(x_1^2 +x_1x_2+x_2^2)x_4&q_1(x_1+x_2+x_3)x_4&x_4
\end{pmatrix}$\\
    \hline
    $\text{Fl}(1,2;4)$~ 
    &~$\left(
\begin{array}{cccc}
 \sigma _1 & -1 & 0 & 0 \\
 \qcoh_1 & \sigma _2 & -1 & 0 \\
 0 & 0 & \sigma _3 & -1 \\
 0 & \qcoh_2 & 0 & \sigma _4 \\
\end{array}
\right)$&$\begin{pmatrix}
    x_1&-\frac{1}{1-q_1}&0&0\\
    q_1 x_1 x_2 &x_2&-\frac{1}{1-q_2}&0\\
    0&0&x_3&-1\\
    \qk_1 q_2 x_1x_2x_3 x_4 &q_2 x_2 x_3 x_4&0&x_4\\
\end{pmatrix}$\\
\hline
    $\text{Fl}(1,3;4)$~
    &~$\left(
\begin{array}{cccc}
 \sigma _1 & -1 & 0 & 0 \\
 0 & \sigma _2 & -1 & 0 \\
 \qcoh_1 & 0 & \sigma _3 & -1 \\
 0 & -\qcoh_2 & 0 & \sigma _4 \\
\end{array}
\right)$&$\begin{pmatrix}
    x_1&-\frac{1}{1-\qk_1} & 0 & 0\\
    0&x_2 & -1 & 0\\
    q_1x_1x_2x_3& 0 & x_3 & -\frac{1}{1-\qk_2}\\
    q_1q_2x_1x_2x_3x_4&-q_2x_2^2 x_4&q_2(x_2+x_3)x_4&x_4
\end{pmatrix}$\\
    \hline
    $\text{Fl}(2,3;4)$~
    &~$\left(
\begin{array}{cccc}
 \sigma _1 & -1 & 0 & 0 \\
 0 & \sigma _2 & -1 & 0 \\
- \qcoh_1 & 0 & \sigma _3 & -1 \\
 0 & 0 & \qcoh_2 & \sigma _4 \\
\end{array}
\right)$&$\begin{pmatrix}
    x_1 & -1&0&0\\
    0&x_2&-\frac{1}{1-q_1} &0\\
    -q_1x_1^2x_3&q_1(x_1+x_2)x_3&x_3&-\frac{1}{1-q_2}\\
    -q_1q_2x_1^2x_3x_4 & q_1 q_2(x_1+x_2)x_3x_4&q_2x_3x_4&x_4
\end{pmatrix}$\\
    \hline
    $\text{Fl}(4)$~
    &~ $\left(
\begin{array}{cccc}
 \sigma _1 & -1 & 0 & 0 \\
 \qcoh_1 & \sigma _2 & -1 & 0 \\
 0 & \qcoh_2 & \sigma _3 & -1 \\
 0 & 0 & \qcoh_3 & \sigma _4 \\
\end{array}
\right)$&$\begin{pmatrix}
     x_1 & -\frac{1}{1-q_1} & 0 & 0 \\
 q_1 x_1 x_2 & x_2 & -\frac{1}{1-q_2} & 0 \\
 q_1 q_2 x_1 x_2 x_3 & q_2 x_2 x_3 & x_3 & -\frac{1}{1-q_3} \\
 q_1 q_2 q_3 x_1 x_2 x_3 x_4 & q_2 q_3 x_2 x_3 x_4 & q_3 x_3 x_4 & x_4 
\end{pmatrix}$\\
\hline
\caption{The matrices $D_{4}{\scriptsize \begin{bmatrix}
    \qcoh\\{\boldsymbol{k}}
\end{bmatrix}}(\sigma)$ as defined in~\protect\eqref{q-par-D-matrix} and ${\rm G}_4{\scriptsize\begin{bmatrix}
    \qk\\{\boldsymbol{k}}
\end{bmatrix}}(x)$ as defined in~\protect\eqref{Gn explicit} below, for all 7 possible partial flag manifolds with $n=4$.}
    \label{tab:flags with nf=4}
\end{longtable}
\end{center}

\subsubsection{Parabolic quantum double Schubert polynomials}
We first define the parabolic quantum double Schubert polynomial associated with $w_0^{{\boldsymbol{k}}}$, the longest permutation in ${\rm W}^{({\boldsymbol{k}};n)}$, as defined around \eqref{permutation-decomposition}. Note that $[X_{w_0^{{\boldsymbol{k}}}}]$ is the point class. One defines: 
\begin{equation}
    \mathfrak{S}_{w_0^{{\boldsymbol{k}}}} {\scriptsize\begin{bmatrix}
        \qcoh\\{\boldsymbol{k}}
    \end{bmatrix}} (\sigma, m)\, =\, \prod_{\ell=1}^{s}\;\prod_{i=k_\ell}^{k_{\ell+1}-1} \det\left(\,D_{k_\ell}{\scriptsize\begin{bmatrix}
        \qcoh\\{\boldsymbol{k}}
    \end{bmatrix}}\, -\, m_{n-i} \,\mathbb{I}_{k_\ell}\,\right)~.
\end{equation}
Then, for any $w\in {\rm W}^{({\boldsymbol{k}};n)}$, we define the parabolic quantum Schubert polynomials recursively as~\cite[equation (23)]{LamShimozono}:
\begin{equation}\label{par-schu-poly-defn}
    \mathfrak{S}_{w}{\scriptsize\begin{bmatrix}
        \qcoh\\{\boldsymbol{k}}
    \end{bmatrix}}(\sigma, m) \,:=\, (-1)^{\ell(w^{-1} w_0^{{\boldsymbol{k}}})} \,\partial^{(m)}_{w^{-1} w_0^{{\boldsymbol{k}}}}\,\mathfrak{S}_{w_0^{{\boldsymbol{k}}}} {\scriptsize\begin{bmatrix}
        \qcoh\\{\boldsymbol{k}}
    \end{bmatrix}} (\sigma, m) ~,
\end{equation}
exactly as in~\eqref{k-poly-defn}. One can show that one can obtain the parabolic quantum Schubert polynomials~\eqref{par-schu-poly-defn} from the $\mathfrak{Z}$-polynomial~\eqref{k-poly-defn} by elimination of variables in the Whitney-like presentation of the quantum cohomology ring, as alluded to above. 

\begin{example}
\textbf{Fl$(1,3;4)$.} Let us look at the case ${\boldsymbol{k}} = (1,3)$ and $n=4$. According to table~\ref{tab:longpermu for n=4}, we have 12 possible classes in ${\rm W}^{(1,3;4)}$, and applying~(\ref{permutation-decomposition}), (\ref{w0k explicit}), we have  $w_{0\, {\boldsymbol{k}}} = (1 \, 3 \, 2 \, 4)$, $w_0^{{\boldsymbol{k}}} = (4 \, 2 \, 3 \, 1)$. For example, we find the following polynomials: 
\begin{equation}\label{qSchubert in Fl134}
    \begin{split}
        &\mathfrak{S}_{(1\,2\,4\,3)}{\scriptsize\begin{bmatrix}
             \qcoh_1 & \qcoh_2\\
             1&3
         \end{bmatrix}} (\sigma, m)\, =\, \sigma _1+\sigma _2+\sigma _3-m_1-m_2-m_3~,\\
          &\mathfrak{S}_{(2\,1\,4\,3)}{\scriptsize\begin{bmatrix}
             \qcoh_1 & \qcoh_2\\
             1&3
         \end{bmatrix}} (\sigma, m) \,=\,\left(\sigma _1-m_1\right) \left(\sigma _1+\sigma _2+\sigma _3-m_1-m_2-m_3\right) ~,\\
          &\mathfrak{S}_{(4\,2\,3\,1)}{\scriptsize\begin{bmatrix}
             \qcoh_1 & \qcoh_2\\
             1&3
         \end{bmatrix}} (\sigma, m) \,=\, (\sigma_1-m_2)(\sigma_1-m_3)((\sigma_1-m_1)(\sigma_2-m_1)(\sigma_3-m_1)+\qcoh_1)~,\\
         &\mathfrak{S}_{(2\,3\,4\,1)}{\scriptsize\begin{bmatrix}
             \qcoh_1 & \qcoh_2\\
             1&3
         \end{bmatrix}} (\sigma, m)\, =\,(\sigma_1-m_1)(\sigma_2-m_1)(\sigma_3-m_1)+\qcoh_1~,\\
          &\mathfrak{S}_{(4\,1\,3\,2)}{\scriptsize\begin{bmatrix}
             \qcoh_1 & \qcoh_2\\
             1&3
         \end{bmatrix}} (\sigma, m) \,=\, (\sigma_1-m_3)((\sigma_1-m_1)(\sigma_1-m_2)(\sigma_2+\sigma_3-m_1-m_2)-\qcoh_1)~.
    \end{split}
\end{equation}
\end{example}


\begin{example}
\textbf{Fl$(2,3;4)$.} Similarly, for the case ${\boldsymbol{k}} = (2,3)$ and $n=4$, there are 12 classes in the quotient group ${\rm W}^{(2,3;4)}$. For instance, we have the following polynomials:

\begin{equation}\label{qSchubert in Fl234}
    \begin{split}
       & \mathfrak{S}_{(2\,3\,4\,1)}{\scriptsize\begin{bmatrix}
             \qcoh_1 & \qcoh_2\\
             2&3
         \end{bmatrix}} (\sigma, m) \,=\, \left(\sigma _1-m_1\right) \left(\sigma _2-m_1\right) \left(\sigma _3-m_1\right)-\qcoh_1~,\\
        & \mathfrak{S}_{(1\,2\,4\,3)}{\scriptsize\begin{bmatrix}
             \qcoh_1 & \qcoh_2\\
             2&3
         \end{bmatrix}} (\sigma, m)\,=\, \sigma _1+\sigma _2+\sigma _3-m_1-m_2-m_3~,\\
        & \mathfrak{S}_{(2\,3\,1\,4)}{\scriptsize\begin{bmatrix}
             \qcoh_1 & \qcoh_2\\
             2&3
         \end{bmatrix}} (\sigma, m)\,=\, \left(\sigma _1-m_1\right) \left(\sigma _2-m_1\right)~,\\
         &\mathfrak{S}_{(3\,4\,2\,1)}{\scriptsize\begin{bmatrix}
             \qcoh_1 & \qcoh_2\\
             2&3
         \end{bmatrix}} (\sigma, m)\,=\, (\sigma_1-m_2)(\sigma_2-m_2)((\sigma_1-m_1)(\sigma_2-m_1)(\sigma_3-m_1)-\qcoh_1)~,\\
            &\mathfrak{S}_{(2\,4\,3\,1)}{\scriptsize\begin{bmatrix}
             \qcoh_1 & \qcoh_2\\
             2&3
         \end{bmatrix} }(\sigma, m)\,=\, (\sigma_1+\sigma_2-m_2-m_3)((\sigma_1-m_1)(\sigma_2-m_1)(\sigma_3-m_1)-\qcoh_1)~.
    \end{split}
\end{equation}
\end{example}

\begin{example}
\textbf{Fl$(4)$ case.} Moreover, one easily checks that:
\begin{equation}
    \mathfrak{S}_w{\scriptsize\begin{bmatrix}
        \qcoh_1 & \qcoh_2 &  \qcoh_3\\
        1&2&3
    \end{bmatrix}}(\sigma, m) \,=\, \mathfrak{S}^{(\qcoh)}_{w}(\sigma, m)~, \qquad \forall w\in S_4~,
\end{equation}
where on the RHS we have the quantum double Schubert polynomials~\cite[equation (A.18)]{Closset:2025cfm}.
\end{example}

\noindent
{\bf Grassmannian limit.} Taking the special case of Grassmannian manifold Gr$(k,n)$ ($s=1$), one can see that the polynomial \eqref{par-schu-poly-defn} reduces to the double Schubert polynomial -- see subsection A.1.1 of \cite{Closset:2025cfm} for definition and examples. This is because, in this case, we find:
\begin{equation}
\prod_{i = 1}^{n-k} \det\left(D_k{\scriptsize \begin{bmatrix}
    \qcoh_1\\k
\end{bmatrix}} - m_i\, \mathbb{I}_{k}\right) = \prod_{a=1}^{k} \prod_{i=1}^{n-k} (\sigma_a - m_i)~.
\end{equation}
Note that the quantum parameter $\qcoh_1$ does not appear in the final formula. This is expected since, as reviewed in appendix A of \cite{Closset:2025cfm}, in this case the Schubert classes in the quantum cohomology ring are represented by the double Schubert polynomials associated with the Grassmannian permutations, and those polynomials are independent of $\qcoh_1$.

For example, let us look at the case Gr$(2,4)$. For the Grassmannian permutations $(1\,3\,2\,4)$ and $(2\,4\,1\,3)$, the polynomial \eqref{par-schu-poly-defn} takes the following forms:

\begin{equation}
    \begin{split}
        &\mathfrak{S}_{(1\,3\,2\,4)}{\scriptsize\begin{bmatrix} 
            \qcoh_1\\
            2
        \end{bmatrix} }(\sigma, m) \,=\, e_1^2(\sigma) - e_1^2(m)~, \\
        &\mathfrak{S}_{(2\,4\,1\,3)} {\scriptsize\begin{bmatrix} 
            \qcoh_1\\
            2
        \end{bmatrix}} (\sigma, m) \,=\,\left(\sigma _1-m_1\right) \left(\sigma _2-m_1\right) \left(\sigma _1+\sigma _2-m_2-m_3\right)~.
    \end{split}
\end{equation}
These indeed match the expressions in equation (5.15) of \cite{Closset:2023bdr} for the partitions $[1,0]$ and $[2,1]$, respectively.

\subsection{Parabolic quantum Grothendieck polynomials for QK: a proposal}\label{subsec:pqgp def}

We are now in a position to explain our proposal for the K-theoretic polynomials representing $[\CO_w]$ in the Toda presentation of the QK-theory ring of the partial flag manifold, giving us the natural uplift of the parabolic Schubert polynomials we just reviewed. As in the case of the Whitney polynomials, we will first define the parabolic quantum Grothendieck polynomial for the longest element $w_0^{\boldsymbol{k}} \in {\rm W}^{({\boldsymbol{k}};n)}$, corresponding to the point class $[\CO_{w_0^{\boldsymbol{k}}}]\cong [\CO_\text{pt}]$. Then the polynomial for any $w\in {\rm W}^{({\boldsymbol{k}};n)}$ is obtained recursively using the Demazure operators.

\subsubsection{Parabolic quantum elementary polynomials in K-theory}
Let us first define the $n\times n$ matrix:
\begin{equation}\label{Gn explicit}
    \left({\rm G}_n {\scriptsize\begin{bmatrix}
        q\\{\boldsymbol{k}}
    \end{bmatrix}}(x)\right)_{i,j} \,:=\,\begin{cases}
        x_i~, \qquad &i\,=\,j~,\\
        -1~, \qquad &j\,=\,i+1~, \text{and}~i\neq k_{\ell'}~,\\
        -\frac{1}{1\,-\,q_{\ell'}}~, \qquad &(i,j) \,=\,(k_{\ell'}, k_{\ell'}+1)~,\\
        H^{\ell'}_{m,j}(x;q)~, \qquad &i\,=\,k_{\ell'+1}~, \,k_{m-1}<j\leq k_m~,\,1\leq m\leq {\ell'}~,\\
        0~, \qquad &{\rm otherwise}~,
    \end{cases}
\end{equation}
for $\ell'=1,\cdots, s$, where:
\begin{equation}
    H_{m,j}^{\ell'}(x)\,:=\,(-1)^{k_m-j}\,\left(\,\prod_{l=m}^{\ell'}\qk_l\,\right)\,\left(\,\prod_{p=k_m+1}^{k_{\ell'+1}}x_p\,\right)\,h_{k_m-j+1}(x_{k_{m-1}+1}, \cdots, x_j)~,
\end{equation}
where $h_i$ is the $i$-th complete homogeneous symmetric polynomial and $k_{s+1}=n$ in the above convention. In the last column of table~\ref{tab:flags with nf=4}, we exhibit the matrix G$_4{\scriptsize \begin{bmatrix}
    \qk\\{\boldsymbol{k}}
\end{bmatrix}}(x,y)$ for all 7 possible partial flags with $n=4$.

One can then define the {\it K-theoretic parabolic quantum elementary polynomials} ${{\rm F}^{({\boldsymbol{k}};n)}}^\bullet_\bullet$ by expanding the determinants:
\begin{equation}\label{def Fka}
    \det\left(\,{\rm G}_{k_\ell}\scriptsize\begin{bmatrix}
        \qk\\{\boldsymbol{k}}
    \end{bmatrix}(x)\,-\,\lambda\, \mathbb{I}_{k_\ell\times k_\ell}\,\right) \,= \,\sum_{a_\ell=0}^{k_\ell} (-1)^{k_\ell-a_\ell} \,\lambda^{k_\ell-a_\ell} \,{{\rm F}^{({\boldsymbol{k}};n)}}_{a_\ell}^{k_\ell}(x;\qk)~,
\end{equation}
where, for $\ell = 1, \cdots, s$, G$_{k_\ell}\scriptsize\begin{bmatrix}
    \qk\\{\boldsymbol{k}}
\end{bmatrix}$ is the $k_\ell\times k_\ell$ upper-left submatrix of~\eqref{Gn explicit}. It is straightforward to check in examples that these quantum polynomials satisfy the same recurrence relations as the polynomials $e_{a_\ell}(x^{(\ell)})$ in~\eqref{explicit QI rels for QK}, namely:
\begin{multline}\label{recurence for Fak qk}
    {{\rm F}^{({\boldsymbol{k}};n)}}^{k_\ell}_{a_\ell}(x;q)\,=\,\sum_{r=0}^{k_\ell-k_{\ell-1}}\,e_r(\widehat{x}^{(\ell)})\, {{\rm F}^{({\boldsymbol{k}};n)}}^{k_{\ell-1}}_{a_\ell-r}(x;q)\\\, + \,\frac{q_{\ell-1}}{1-q_{\ell-1}}\,x_{k_{\ell-1}+1}\,\cdots\,x_{k_\ell}\,\left({{\rm F}^{({\boldsymbol{k}};n)}}_{a_{\ell}-k_\ell+k_{\ell-1}}^{k_{\ell-1}}(x;q)\,-\,{{\rm F}^{({\boldsymbol{k}};n)}}^{k_{\ell-2}}_{a_\ell-k_\ell+k_{\ell-1}}(x;q)\right)~.
\end{multline}
As a consistency check, the cohomological limit of~\eqref{recurence for Fak qk}  reproduces the recurrence relation~(\ref{recurrence relations for qpele}). 
The parabolic quantum elementary polynomials are therefore a natural parabolic generalization of the quantum Lenart--Maeno polynomials ${\rm F}^\bullet_\bullet(x;q)$~\cite{lenart2006quantum} which we reviewed in~\cite[appendix A]{Closset:2025cfm}. For instance, for the complete flag case, this reduces to \cite[equation (A.36)]{Closset:2025cfm}. Importantly, this implies that these quantum polynomials give the explicit solution for the $\CS_\ell$ variables $x^{(\ell)}$ in terms of the Toda variables $x_i$, as follows:
 \be\label{ex to Fq}
e_{a_\ell}(x^{(\ell)}) = {{\rm F}^{({\boldsymbol{k}};n)}}^{k_\ell}_{a_\ell}(x;q)~, \qquad \ell=1, \cdots, s~, \quad a_\ell =1, \cdots, k_\ell~.
 \ee
One can easily check that this reproduces the explicit examples discussed in subsection~\ref{examples eliminationQK}. Moreover, setting $\ell=s+1$ in this last equation, we also find the explicit equivariant QK ring relations in the Toda formulation:
\be
e_j(y) \,=\, {{\rm F}^{({\boldsymbol{k}};n)}}^{n}_{j}(x;q)~, \qquad j\,=\,1, \cdots,n~.
\ee
Here, the polynomials ${{\rm F}^{({\boldsymbol{k}};n)}}^{n}_{j}$ are obtained by expanding out the determinant of the full ${\rm G}_n$ matrix~\eqref{Gn explicit}, setting $\ell=s+1$ in~\eqref{def Fka}. In other words, the QK ring relations can be written as:
\be
\det(\,Y \,-\, \lambda \, \mathbb{I}_{n\times n}\,)\, =\,\det\left(\,{\rm G}_{n}\scriptsize\begin{bmatrix}
        \qk\\{\boldsymbol{k}}
    \end{bmatrix}(x)\,-\,\lambda\, \mathbb{I}_{n\times n}\,\right)~,
\ee
where $Y=\diag{(y_1, \cdots, y_n)}$ is a diagonal matrix. The QK ring relations therefore state that  $\det(Y - \lambda )= \prod_{j=1}^n(y_i - \lambda)$ is equal to the characteristic polynomial of ${\rm G}_{n}$. This matches exactly the recent result in~\cite[theorem 3.4]{ahkmox}.%
\footnote{Their matrix on the RHS is different from ours but has the same determinant, as checked in examples.}

Furthermore, by extrapolating from many examples, we conjecture the following explicit form for the top quantum polynomials:
\be
{{\rm F}^{({\boldsymbol{k}};n)}}^{n}_{j} \,:=\, \sum_{\substack{I\subseteq [n]\\|I| = j}}\, 
    \left[\,\prod_{i\in I}\,{x_i} \,\prod^s_{\substack{l=1\\(k_{l-1}, k_l] \cap \mathbb{Z} \cap I \neq \emptyset\\ (k_l,  k_{l+1}] \cap \mathbb{Z} \subset I}}\,\frac{1}{1\,-\,q_l}\,\right]~.
\ee
In the complete-flag case, this expression reduces to the known expression~\cite[equation (A.31)]{Closset:2025cfm}.

\subsubsection{Parabolic quantum double Grothendieck polynomials} 

Given the above, let us explain the correct definition for the {\it parabolic quantum (double) Grothendieck polynomials}. Starting with the polynomial~\eqref{pqw poly longest} for the point class in the Whitney presentation, we note that, 
expanding the product over $a_\ell$ in powers of $y_{n-j}$, one obtains the elementary symmetric polynomials 
$e_{a_\ell} (x^{(\ell)})$:
\be
\prod_{a_\ell=1}^{k_\ell}\, \left(\,1\, -\, \frac{x_{a_\ell}^{(\ell)}}{y_{n-j}}\,\right) \,=\, 1\,+\, \sum_{a_\ell = 1}^{k_\ell} \,{(-1)^{a_\ell}\ov y_{n-j}^{a_\ell}} \,e_{a_\ell}(x^{(\ell)})~.
\ee
Using the Whitney ring relations~\eqref{ex to Fq}, we then naturally define: 
\begin{equation}\label{qpdgp longest}
    \mathfrak{G}_{w_0^{\boldsymbol{k}}} {\scriptsize\begin{bmatrix}
        q\\{\boldsymbol{k}}
    \end{bmatrix}} (x,y)\,:=\,\prod_{\ell=1}^s\,\prod_{j = k_\ell}^{k_{\ell+1} - 1} \,\left[\,1\, +\, \sum_{a_\ell=1}^{k_\ell} \frac{(-1)^{a_\ell}}{y_{n-j}^{a_\ell}}\,{{\rm F}^{({\boldsymbol{k}};n)}}^{k_\ell}_{a_\ell}(x;q)\,\right]~,
\end{equation}
which must represent the point class in the Toda presentation. 
Then, for any minimal permutation $w\in {\rm W}^{({\boldsymbol{k}};n)}$, the corresponding parabolic quantum double Grothendieck polynomial is defined recursively by:
\begin{equation}\label{qpdgp}
    \mathfrak{G}_w {\scriptsize\begin{bmatrix}
        \qk\\{\boldsymbol{k}}
    \end{bmatrix}}(x,y)\,  := \,\pi^{(y)}_{w^{-1}w_0^{\boldsymbol{k}}}\,\mathfrak{G}_{w_0^{\boldsymbol{k}}} {\scriptsize\begin{bmatrix}
        q\\{\boldsymbol{k}}
    \end{bmatrix}} (x,y)~,
\end{equation}
exactly as in~\eqref{qpw poly def}. In the case of the complete flag, $w_0^{\boldsymbol{k}}=w_0$ and our definition reduces to the one for the quantum double Grothendieck polynomials~\cite{lenart2006quantum} written in our physics notation~\cite[equation (A.32)]{Closset:2025cfm}.

Since the Demazure operator only acts on the equivariant variables $y$, acting with $\pi^{(y)}$ commutes with using the $y$-independent identities~\eqref{ex to Fq} inside the ring, and therefore the Grothendieck (Toda-like) and Whitney polynomials all agree inside the quantum ring~\eqref{QK ring explicit for all x}:
\be\label{ref Ww to Gw}
\mathfrak{W}_w^{({\boldsymbol{k}}; n)}(x^{(\bullet)},y) \,\cong\, \mathfrak{G}_w {\scriptsize\begin{bmatrix}
        \qk\\{\boldsymbol{k}}
    \end{bmatrix}}(x,y)~, \qquad\forall \, w\in {\rm W}^{({\boldsymbol{k}};n)}~,
\ee
and so they must represent the same Schubert classes~$[\CO_w]$.

\medskip
\noindent
{\bf Non-equivariant limit.} The double polynomials~\eqref{qpdgp} represent {\it equivariant} QK-theory classes. The polynomials representing $[\CO_w]$ in the non-equivariant ring QK$({\rm Fl}({\boldsymbol{k}}; n))$ are simply defined by setting $y_i=1$ in the above definition {\it after} computing the double polynomials:
\be
    \mathfrak{G}_w {\scriptsize\begin{bmatrix}
        \qk\\{\boldsymbol{k}}
    \end{bmatrix}}(x) \,  := \,    \mathfrak{G}_w {\scriptsize\begin{bmatrix}
        \qk\\{\boldsymbol{k}}
    \end{bmatrix}}(x,y)\Big|_{y_i=1\; \forall i}~.
\ee
Note that it is necessary to consider the equivariant deformation in order to apply the definition~\eqref{qpdgp}. Unlike the case of the classical Grothendieck polynomials, there is no direct definition of the non-equivariant quantum polynomials using Demazure operators with respect to $x$, essentially because the latter are now `quantum' variables ({\it i.e.} they are elements of a non-trivial quotient ring defined by a $q$-deformed ideal) and they do not commute with the classical Demazure operation with respect to $x$. On the other hand, the $y$ variables remain fixed parameters irrespective of $q$.

\medskip
\noindent
\textbf{The 2d (cohomological) limit.} One can easily show that, setting $x\equiv e^{-\beta \sigma}$, $y\equiv e^{-\beta m}$ and matching the 3d and 2d quantum parameters as in~\eqref{qk to qcoh}, the parabolic quantum polynomials \eqref{qpdgp} reduce to their quantum Schubert counterparts defined in \eqref{par-schu-poly-defn} in the following 2d limit:
\begin{equation}\label{2d-limit-double-groth}
    \lim_{\beta\rightarrow 0}\,\left( \beta^{-\ell(w)}\, \mathfrak{G}_{w}{\scriptsize\begin{bmatrix}
        \qk\\{\boldsymbol{k}}
    \end{bmatrix}}\left(\,e^{-\beta \sigma},e^{-\beta m}\right)\,\right)\, =\, \mathfrak{S}_{w}{\scriptsize\begin{bmatrix}
        \qcoh\\{\boldsymbol{k}}
    \end{bmatrix}} (\sigma, m)~.
\end{equation}
This gives us another strong consistency check of our proposal.

\section{Schubert line defects in the 3d GLSM}  \label{sect:proposal defect}

After this long mathematical detour, let us go back to our main physical goal. In this section, we construct and study the half-BPS {\it Schubert line defects} $\SL_w$  in the 3d GLSM of the partial flag manifold Fl$({\boldsymbol{k}};n)$, generalizing the construction for the complete flag discussed in part~I~\cite{Closset:2025cfm}. These line defects are realized by coupling a suitably engineered 1d $\mathcal{N}=2$ supersymmetric quantum mechanics (SQM) to the 3d GLSM, such that they flow in the infrared to the Schubert classes in quantum K-theory:
\be\label{SL and CO}
\SL_w \quad \rightsquigarrow \quad [\CO_w]~, \qquad w\in {\rm W}^{({\boldsymbol{k}};n)}~.
\ee
Throughout, we largely follow the convention of part I~\cite{Closset:2025cfm}, in particular for the quiver diagram notation. We will not repeat the analysis of the vacuum equations here, as it is essentially identical to the complete flag case; see~\cite[section 4]{Closset:2025cfm} for details.

\subsection{General proposal for Schubert line defects as 1d quivers}\label{subsec: gen proposal}
\begin{figure}
    \centering
    \includegraphics[width=0.8\linewidth]{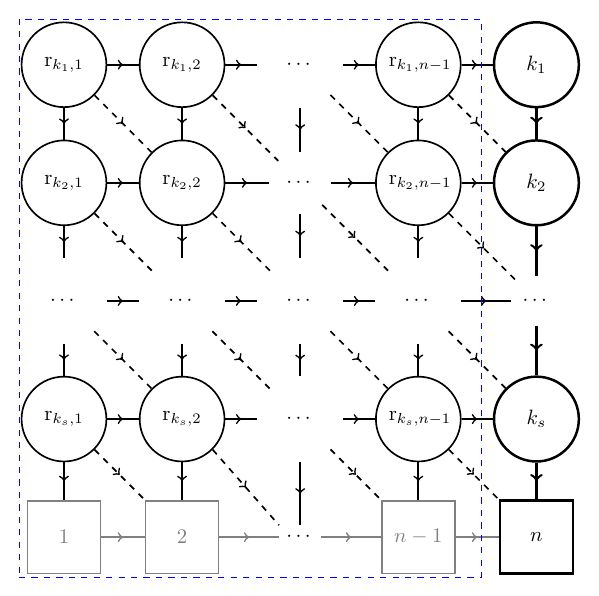}
    \caption{Quiver diagram of the 1d-3d coupled system realizing $X_w\subseteq {\rm Fl}({\boldsymbol{k}};n)$. The last column represents the 3d GLSM to the partial flag manifold Fl$({\boldsymbol{k}}; n)$. The square quiver in the blue box is the 1d quiver gauge theory defining the defect. The circle nodes represent 1d gauge groups of the indicated ranks ${\rm r}_{k_\ell,j}\equiv {\rm r}_{k_{\ell},j}^{w_0w}$.  Horizontal and vertical arrows stand for 1d chiral matter multiplets in the bifundamental representation of the corresponding nodes. Moreover, diagonal arrows correspond to 1d Fermi multiplets in the bifundamental representation of the corresponding nodes. The gray squares in the last row represent fixed background data for the inclusion maps.}
    \label{fig:GenProposal}
\end{figure}

We engineer the Schubert line defects $\SL_w$ in the 3d GLSM for the partial flag manifold Fl$({\boldsymbol{k}};n)$ as the 1d $\CN=2$ quiver gauge theories shown in figure~\ref{fig:GenProposal}. This quiver is essentially the same as in the complete flag case \cite{Closset:2025cfm}, except that we have a rectangular quiver instead of a square quiver. That is, for partial flag manifolds, we have $s+1<n$ rows,  corresponding to the shorter sequence of subspaces in~\eqref{partial-flag-defn}. 
The boldfaced quiver in the right-most column of figure~\ref{fig:GenProposal} summarizes the 3d bulk fields
describing the flag manifold, while the dashed blue box encloses the 1d quiver built out of 1d fields living on the defect. 
For any given Schubert variety~\eqref{X_w defn}, the rank $\text{r}_{k_\ell,j}$ of the unitary gauge group depends on the minimal permutation $w \in {\rm W}^{({\boldsymbol{k}};n)}$ as defined in~\eqref{Wk defn}, as follows:
\begin{equation}
    \text{r}_{k_\ell,j} \,\equiv\, \text{r}_{k_\ell,j}^{w_0 w} \,=\, \# \{\,l \,\leq\, k_\ell \:|\: n\,+\,1\, -\, w(l)\,\leq\, j\,\}~,
\end{equation}
for $\ell=1, \cdots,s$. The rest of our conventions are exactly as in part~I~\cite{Closset:2025cfm}. The solid lines in the 1d $\CN=2$ supersymmetric quiver denote bifundamental chiral multiplets, which correspond to inclusions of vector spaces of fixed dimensions.%
\footnote{For the complete flag case \protect\cite{Closset:2025cfm}, these chiral multiplets correspond to matrices of full rank, which ensures that the linear maps between the vector spaces are injective. This property carries over to the partial flag case, so that the solid lines continue to represent inclusions of subspaces of the appropriate dimensions.}
The dashed diagonal arrows denote Fermi multiplets, which impose commutativity constraints of the crossed square via their $E$-term equations. 
The `bifundamentals' between global symmetry boxes on the last row are merely (non-dynamical) inclusion maps between vector subspaces in a standard reference flag; they serve as an efficient bookkeeping device, as previously explained in~\cite[section 4.2.4]{Closset:2025cfm}.

Upon specializing to ${\boldsymbol{k}} = (1,2,\cdots, n-1)$, the partial flag manifold is simply the complete flag manifold, and our Schubert line defects become the ones discussed in part I~\cite{Closset:2025cfm}.

\begin{figure}[t]
\centering
\begin{tikzcd}[column sep=2em, row sep=3em]
    {V_{1,1}} & {V_{1,2}} & \cdots & {V_{1,n-1}} & {F_{k_1}} \\
    {V_{2,1}} & {V_{2,2}} & \cdots & {V_{2,n-1}} & {F_{k_2}} \\
    \vdots & \vdots & \ddots & \vdots & \vdots \\
    {V_{s,1}} & {V_{s,2}} & \cdots & {V_{s,n-1}} & {F_{k_s}} \\
    {\textcolor{gray}{E_1}} & \textcolor{gray}{E_2} & \cdots & \textcolor{gray}{E_{n-1}} & {\mathbb{C}^n}
    \arrow[from=1-1, to=1-2]
    \arrow[from=1-1, to=2-1]
    \arrow["\circlearrowleft"{description}, draw=none, from=1-1, to=2-2]
    \arrow[from=1-2, to=1-3]
    \arrow[from=1-2, to=2-2]
    \arrow[from=1-3, to=1-4]
    \arrow[from=1-4, to=1-5]
    \arrow[from=1-4, to=2-4]
    \arrow["\circlearrowleft"{description}, draw=none, from=1-4, to=2-5]
    \arrow[from=1-5, to=2-5]
    \arrow[from=2-1, to=2-2]
    \arrow[from=2-1, to=3-1]
    \arrow[from=2-2, to=2-3]
    \arrow[from=2-2, to=3-2]
    \arrow[from=2-3, to=2-4]
    \arrow[from=2-4, to=2-5]
    \arrow[from=2-4, to=3-4]
    \arrow[from=2-5, to=3-5]
    \arrow[from=3-1, to=4-1]
    \arrow[from=3-2, to=4-2]
    \arrow[from=3-4, to=4-4]
    \arrow[from=3-5, to=4-5]
    \arrow[from=4-1, to=4-2]
    \arrow[from=4-1, to=5-1]
    \arrow["\circlearrowleft"{description}, draw=none, from=4-1, to=5-2]
    \arrow[from=4-2, to=4-3]
    \arrow[from=4-2, to=5-2]
    \arrow[from=4-3, to=4-4]
    \arrow[from=4-4, to=4-5]
    \arrow[from=4-4, to=5-4]
    \arrow["\circlearrowleft"{description}, draw=none, from=4-4, to=5-5]
    \arrow[from=4-5, to=5-5]
    \arrow[from=5-1, to=5-2, gray]
    \arrow[from=5-2, to=5-3, gray]
    \arrow[from=5-3, to=5-4, gray]
    \arrow[from=5-4, to=5-5, gray]
\end{tikzcd}

\caption{Diagram of nested vector spaces in $\mathbb{C}^n$, with a fixed reference flag $E_\bullet$ shown in gray. The arrows indicate the inclusions of subspaces, while the circular arrows indicate that the squares commute. Here, $\dim(V_{\ell,j})\, =\, \text{r}_{k_\ell,j}$ and $\dim(F_{k_\ell}) \,=\,\ k_\ell$.}
\label{fig:quiver geometry}
\end{figure}
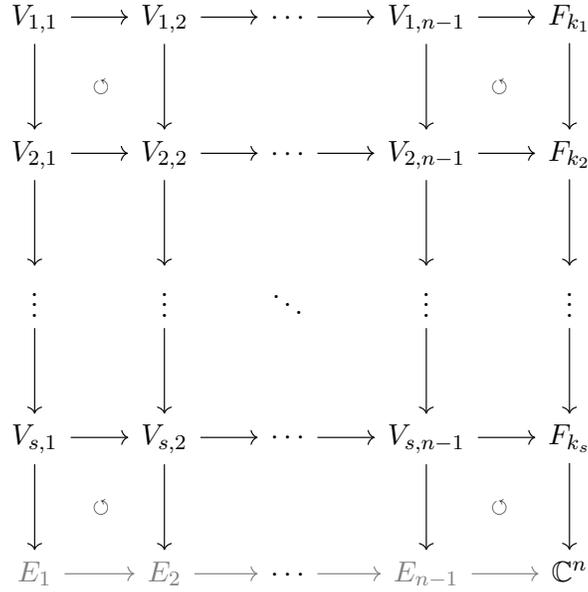

\medskip
\noindent\textbf{The underlying geometry.}
By the same reasoning as in~\cite[section 4]{Closset:2025cfm}, after choosing all 1d (and 3d) real FI parameters to be positive, one can show that the underlying geometry of the 1d-3d coupled system corresponds to the collection of all possible configurations of nested subspaces of the form shown in figure~\ref{fig:quiver geometry}. Projecting this to the 3d bulk direction, that is, to the last column of this quiver, we obtain a collection of partial flags $F_{k_1} \subset F_{k_2} \subset \cdots F_{k_s}\subset \mathbb{C}^n$, subject to the incidence relation:
\begin{equation}
    \dim(F_{k_\ell}\, \cap\, E_j) \,\geq \,\dim(V_{\ell,j} \,\cap\, E_j) \,= \,\dim(V_{\ell,j}) \,=\, \text{r}_{k_\ell,j}~,
\end{equation}
where in the first step, we used the fact that $V_{\ell,j}\,\subset\, F_{k_\ell}$, and in the second step, we used the fact that $V_{\ell,j}\subset E_j$. These are precisely the defining conditions of the Schubert variety $X_w$, confirming that the defect enforces the desired geometric restriction. Physically, this projection is imposed by integrating out the 1d degree of freedom on the defect, leaving us with 1d vacuum constraints on the 3d fields that restrict the 3d GLSM target (at the point of insertion of the defect line on $\Sigma$) to be the Schubert variety $X_w$. Moreover, as we explained in part I~\cite{Closset:2025cfm}, the full 1d-3d coupled systems actually constructs some $\t X_w\rightarrow X_w$, which at least for Grassmannians and complete flag manifolds is known to be a resolution of the Schubert variety (which is generally singular). We discuss some explicit resolutions of Schubert varieties in examples in appendix~\ref{app:geometry}. More generally, we expect that our rectangular quiver construction can be reinterpreted as a Bott--Samelson resolution $\t X_w$ constructed as a quiver Grassmannian~\cite{iezzi2025quivergrassmanniansbottsamelsonresolution}, as discussed in~\cite[appendix B.2]{Closset:2025cfm} for the complete flag manifold.%
\footnote{
We conjecture that the defect construction is giving a smooth resolution of the Schubert variety, based on the fact that we do not see noncompact branches in the GLSM. This was the case in all examples in part I~\protect\cite{Closset:2025cfm} because of the relation between the defect construction and {\it e.g.}~quiver Grassmannians, which are known in mathematics to be smooth.  We do not know of an analogous mathematics result here that guarantees smoothness.}

\begin{figure}
    \centering
    \includegraphics[width=0.8\linewidth]{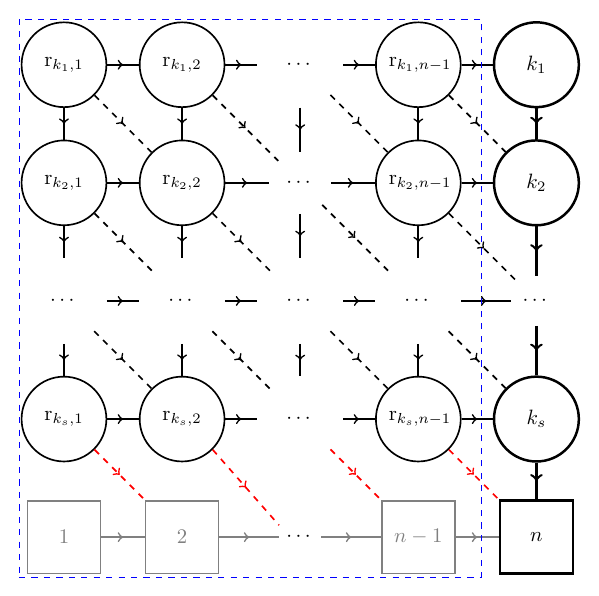}
    \caption{The 1d-3d coupled system after integrating out the massive fermionic and bosonic degrees of freedom of the system given in figure~\ref{fig:GenProposal}. In the last row, the red diagonal arrows now represent Fermi multiplets in the fundamental representation of the gauge group represented by the circle node and have a charge $-1$ under the $U(1)$ subgroup of the bulk flavor symmetry $SU(n)$ indexed by the square node.}
    \label{fig:last-row-reduced}
\end{figure}

\medskip
\noindent\textbf{An integrating out process.} As in the complete flag case \cite{Closset:2025cfm}, certain pairs of 1d chiral and Fermi multiplets acquire mass due to the special form of the $E$-term induced by the fixed chiral background field (bottom row) --- see \cite[subsection 4.2.4]{Closset:2025cfm}. Integrating out these massive pairs leads to a reduced quiver as shown in figure~\ref{fig:last-row-reduced}.

\medskip
\noindent
\textbf{Flavor fugacity assignment.} We adopt the same convention as in \cite{Closset:2025cfm}. For the last row of the quiver, the flavor fugacities associated with $U(j)$ are $y_n\,,\, y_{n-1}\,, \,\cdots\,,\, y_{n-j+1}$, and the surviving Fermi multiplet in figure~\ref{fig:last-row-reduced} charged under $U(1) \subset U(1)^{j+1} \subset U(j+1)$ is assigned fugacity $y_{n-j}$. Note that $GL(j)=U(j)_\C$ in the bottom row is the flavor symmetry corresponding to $E_j$ in the reference flag~\eqref{E flag}, as pictured in figure~\ref{fig:quiver geometry}.

The bulk gauge groups are treated as flavor symmetries from the 1d perspective. For each $U(k_\ell)$ in the last column of the quiver, we denote the flavor fugacities by $x^{(\ell)}_1, x^{(\ell)}_2, \cdots, x^{(\ell)}_{k_\ell}$ for $\ell=1,2,\cdots, s$, which, as discussed in subsection \ref{subsec: 3d GLSM and QK}, correspond to the K-theoretic Chern roots of the tautological bundle $\mathcal{S}_\ell$ on the partial flag manifold.

\medskip
\noindent
\textbf{Reducing the quivers via 1d Seiberg-like dualities.} As we discussed in section 2 of the accompanying paper \cite{Closset:2025cfm}, one can use known dualities between 1d $\CN=2$ unitary gauge theories~\cite{Closset:2025zyl} to simplify the form of the square quiver \ref{fig:GenProposal} defining the Schubert line defects. In the next subsection, we apply these duality moves for the line defects of Gr$(k,n)$ to show that, in this case, our proposal reduces to the  Grothendieck lines introduced in~\cite{Closset:2023bdr}. We also exhibit the   duality-reduced 1d quivers for Schubert defects in Fl$(1,2;4)$, Fl$(1,3;4)$, and Fl$(2,3;4)$ in appendix \ref{subsec:defects124}, \ref{subsec:defects134}, and \ref{subsec:defects234}, respectively.

\subsection{1d flavored Witten indices and parabolic quantum polynomials} 

Given the Schubert line defect $\SL_w$ defined as a 1d $\CN=2$ quiver, one can compute its Witten index as a function of the flavor fugacities $y_i$ and of the 3d gauge parameters $x^{(\ell)}_{a_\ell}$ (which also appear as flavor fugacities from the 1d perspective). Given the correspondence~\eqref{SL and CO} between Schubert line defect and Schubert classes in (quantum, equivariant) K-theory, one should expect that the Witten index computes the Chern character of the Schubert class:
\be\label{Iw eq chT again}
    \mathcal{I}^{({\rm 1d})}[\SL_w] \,=\, {\rm ch}_T(\CO_w)~,
\ee
as we already found in previous works on 3d GLSMs with Grassmannian manifold target spaces~\cite{Closset:2023bdr, Gu:2025tda}. For flag manifolds, the key question is whether we use the variables corresponding to Chern roots of the tautological bundles $\CS_\ell$ or of the quotient bundles $\CQ_\ell$, as explained in section~\ref{subsec: qkrings generalities}. The flavored Witten $ \mathcal{I}^{({\rm 1d})}_w \equiv \mathcal{I}^{({\rm 1d})}[\SL_w]$ of our 1d quiver is a function of the $\CS_\ell$ variables $x^{(\ell)}$ only, and it does not depend on $\qk_\ell$ (as there is no natural coupling of the 1d gauge theory to the 3d topological symmetries). Hence, we expect that it should coincide with the parabolic Whitney polynomials introduced in subsection~\ref{subsec:parabolic whitney k} above.

\medskip
\noindent
\textbf{Conjecture: Witten equals Whitney.} For any minimal representative  $w \in {\rm W}^{({\boldsymbol{k}};n)}$, the flavored Witten index of the 1d Schubert defect $\SL_w$ is equal to the parabolic double Whitney polynomial~\eqref{qpw poly def}, namely:
\begin{equation}\label{conjecture Whitney equal index}
\mathcal{I}^{({\rm 1d})}_w(x^{(\bullet)}, y) \,=\, \mathfrak{W}_w^{({\boldsymbol{k}}; n)}(x^{(\bullet)},y)~.
\end{equation}
It then follows from~\cite[theorem 5.9]{ahkmox} that the Witten index $\mathcal{I}^{({\rm 1d})}_w$ indeed represents the quantum K-theory class $[\CO_w]$ in QK$_T({\rm Fl}(\boldsymbol{k}; n))$, and therefore~\eqref{Iw eq chT again} is true when suitably interpreted. (In particular, it is true in the classical K ring, since the index is independent of $q$ anyway.)  In the following and in the appendices, we will give much evidence for the conjecture~\eqref{conjecture Whitney equal index}, mostly by explicit computations in examples. 

Finally, as explained in the previous section, one can perform an elimination of variables inside the QK ring in the Whitney presentation. Then, given~\eqref{ref Ww to Gw} and assuming the conjecture~\eqref{conjecture Whitney equal index} holds, the Witten index $\mathcal{I}^{({\rm 1d})}_w$ becomes the parabolic quantum double Grothendieck polynomial~\eqref{qpdgp}.

\subsubsection{Computing the 1d flavored Witten indices}
The flavored Witten indices of the 1d $\CN=2$  SQM quiver of figure \ref{fig:last-row-reduced} can be computed using supersymmetric localization~\cite{Hori:2014tda}. Following the conventions from~\cite{Closset:2023bdr,Closset:2025zyl,Closset:2025cfm}, we have:
\begin{equation}\label{1d index JK}
    \mathcal{I}^{({\rm 1d})}_{w}(x^{(\bullet)},y)\, =\, \oint_{\rm JK}\, ({\rm dM})\,{\rm Z}_{\rm chiral}^{\rm ver}\,{\rm Z}_{\rm chiral}^{\rm hor}\,{\rm Z}_{\rm Fermi}^{\rm black}\,{\rm Z}_{\rm Fermi}^{\rm red}~.
\end{equation}
The measure $({\rm dM})$ is given by:
\begin{equation}
    ({\rm dM})\,:=\,\prod_{\ell = 1}^s\,\prod_{i=1}^{n-1} \,\left[\,\Delta^{(k_\ell,i)}(z)\, \frac{1}{{\rm r}_{k_\ell,j}\,!}\,\prod_{\alpha=1}^{{\rm r}_{k_\ell,i}}\,\frac{-\,dz_\alpha^{(k_\ell, i)}}{2\pi i \,z_\alpha^{(k_\ell,i)}}\,\right]~,
\end{equation}
with $\Delta^{(k_\ell,i)}$ being the Vandermonde determinant factor coming from the 1d gauge W-bosons:
\begin{equation}
    \Delta^{(k_\ell,i)}(z)\,:=\,\prod_{1\leq \alpha\neq\beta\leq{\rm r}_{k_\ell, i}}\left(1-\frac{z^{(k_\ell,i)}_\alpha}{z^{(k_\ell,i)}_\beta}\right)~.
\end{equation}
Meanwhile, the different components in the integrand of the JK residue have the following explicit forms:
\begin{align}\label{matter contr 1d index}
    \begin{split}
        &{\rm Z}_{\rm chiral}^{\rm ver} \,:=\,\prod_{\ell=1}^{s-1}\,\prod_{i=1}^{n-1}\,\prod_{\alpha=1}^{{\rm r}_{k_\ell,i}}\,\prod_{\beta=1}^{{\rm r}_{k_{\ell+1},i}}\,\left(\,1\,-\,\frac{z^{(k_\ell,i)}_{\alpha}}{z^{(k_{\ell+1},i)}_{\beta}}\,\right)^{-1} ~,\\
        &{\rm Z}_{\rm chiral}^{\rm hor} \,:=\,\prod_{\ell=1}^{s}\,\left[\,\prod_{\alpha=1}^{{\rm r}_{k_\ell,n-1}}\prod_{a = 1}^{k_\ell}\left(\,1\,-\,\frac{z_{\alpha}^{(k_\ell,n-1)}}{x_{a}^{(\ell)}}\,\right)^{-1}\,\prod_{i=1}^{n-2}\,\prod_{\beta=1}^{{\rm r}_{k_\ell,i}}\,\prod_{\gamma=1}^{{\rm r}_{k_{\ell},i+1}}\,\left(\,1\,-\,\frac{z^{(k_\ell,i)}_{\beta}}{z^{(k_{\ell},i+1)}_{\gamma}}\,\right)^{-1}\,\right] ~,\\ 
        &{\rm Z}^{\rm black}_{\rm Fermi} \,:=\, \prod_{\ell=1}^{s-1}\,\left[\,\prod_{\gamma=1}^{{\rm r}_{k_\ell,n-1}}\,\prod_{a=1}^{k_{\ell+1}}\,\left(\,1\,-\,\frac{z^{(k_\ell,n-1)}_\gamma}{x_{a}^{(\ell+1)}}\,\right)\,\prod_{i = 1}^{n-1}\,\prod_{\alpha=1}^{{\rm r}_{k_\ell,i}}\,\prod_{\beta = 1}^{{\rm r}_{k_{\ell+1},i+1}}\,\left(\,1\,-\,\frac{z^{(k_\ell,i)}_\alpha}{z_\beta^{(k_{\ell+1},i+1)}}\,\right)\,\right]\,~,\\
        &{\rm Z}^{\rm red}_{\rm Fermi} \,:=\prod_{i=1}^{n-1}\,\prod_{\alpha=1}^{{\rm r}_{k_s,i}}\,\left(\,1\,-\,\frac{z^{(k_s,i)}_\alpha}{y_{n-i}}\,\right) ~,\\
    \end{split}
\end{align}
where, in the last contribution, and as shown in the last row in figure \ref{fig:last-row-reduced}, we used the observations made earlier about the integrating out of the massive chiral and Fermi modes at each node in the last line of the quiver defect. At the level of the index, this is exhibited by the cancellation between their corresponding contributions.

When computing the contour integrals in \eqref{1d index JK}, the JK residue prescription instructs us to pick up only the poles coming from the contributions ${\rm Z}^{\rm ver}_{\rm chiral}$ and ${\rm Z}^{\rm hor}_{\rm chiral}$. This is related to the assumption that we made earlier concerning the 1d FI parameters being positive.

\medskip
\noindent
\textbf{Explicit examples.} As an example, let us look at the partial flags with $n=4$. As shown in table~\ref{tab:longpermu for n=4}, there are 7 possible partial flags. We studied the complete flag manifold Fl$(4)$ in appendix C of part I~\cite{Closset:2025cfm}. For the three Grassmannian cases, $\mathbb{CP}^3, {\rm Gr}(3,4)$ and Gr$(2,4)$, we refer to subsection \ref{subsec:Grass case} below. For the three remaining cases, Fl$(1,2;4)$, Fl$(1,3;4)$, and Fl$(2,3;4)$, we present the corresponding 1d quiver defects and the end result of the Witten index computation (and of the explicit variable elimination as well) in appendices \ref{app:124}, \ref{app:134}, and \ref{app:234}, respectively. In all examples, one can check that the conjecture~\eqref{conjecture Whitney equal index} indeed holds true.

\subsection{A special case: the Grassmannian manifold}\label{subsec:Grass case}

Having understood the general case, it is interesting to reconsider the simpler case of the Grassmannian ${\rm Gr}(k,n)= {\rm Fl}(k;n)$, the one-step flag manifold. This case is rather degenerate because $\CQ_1=\CS_1$ (the actual quotient bundle of the Grassmannian would be $\CQ=\CQ_2$, in our notation). This implies that the elimination procedure that underlies~\eqref{ref Ww to Gw} is trivial, because we only need to set $x_i = x^{(1)}_a$ for $i=a= 1,\cdots, k$. Hence, the parabolic Whitney polynomial and the parabolic quantum Grothendieck polynomial are literally identical as polynomials of $k$ variables $x_a$. Since the Whitney polynomials are independent of $q$, we see that the quantum Grothendieck polynomials are equal to the classical Grothendieck polynomials for any Grassmannian permutations $w$.%
\footnote{The previously-known mathematical coincidence~\cite{buch-lmi} is thus nicely explained.} Our wider perspective thus clarifies previous constructions of these latter polynomials from Grothendieck line defects~\cite{Closset:2023bdr}.

\begin{figure}[t!]
    \centering
    \includegraphics[width=1\linewidth]{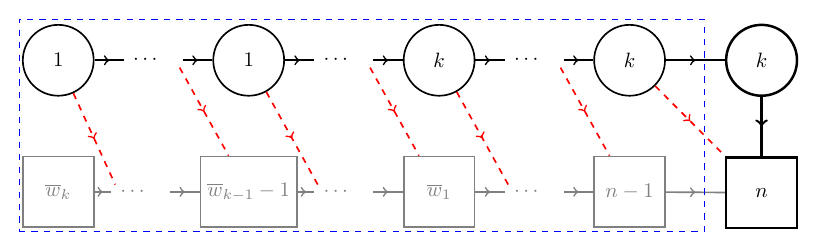}
    \caption{The 1d-3d coupled system defining the Schubert line defects in the Grassmannian manifold Gr$(k,n)$ case. Here we are integrating out the massive 1d fermionic and bosonic degrees of freedom as done in figure \ref{fig:last-row-reduced}.}
    \label{fig:defectGrassGeneral}
\end{figure}

Let us now discuss why our Schubert line defects are equivalent to the Grothendieck lines~\cite{Closset:2023bdr}. The rectangular quiver of figure~\eqref{fig:last-row-reduced} reduces to two rows as shown in figure~\ref{fig:defectGrassGeneral}, where the rank matrix was computed in~\eqref{r for Gr(k,n)}. In the following, we will also use the shorthand notation $\overline{w}_i = \overline{w}(i) \equiv (w_0w)(i)$, where $w_0w$ is defined in \eqref{w0w}. 
One can simplify this quiver further using the duality moves discussed in~\cite[section 2.2]{Closset:2025cfm}. For instance, locally at the nodes of rank $a$, with $a<k$, one can apply type I duality consecutively on these nodes to end up with the following local form of the quiver defect:
\begin{equation}
    \raisebox{-0.5\height}{\includegraphics[width=0.84\linewidth]{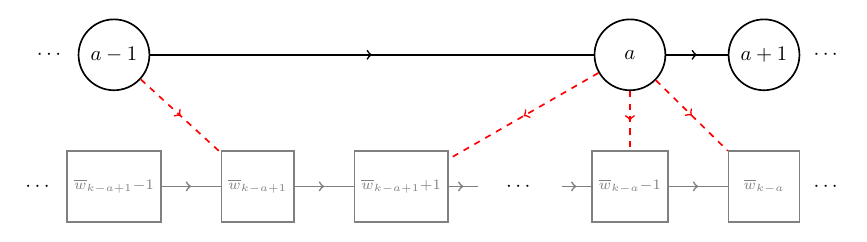}}
    \label{fig:reduce at a}
\end{equation}
In this way, we end up with a single node of rank $a$ attached with $M_a$ Fermi multiplets with:
\begin{equation}\label{Ma}
    M_a\,\equiv\,  \overline{w}_{k-a} \,- \,\overline{w}_{k-a+1} \,=\, \lambda_a\,-\,\lambda_{a+1}\,+1~,
\end{equation}
where, in the second equality, we used the definition of the $k$-partition \eqref{defn par k}. 
One can follow the same logic for the nodes of rank $k$. In this case, we find that we end up with one single node of rank $k$ coupled to $M_k$ Fermi multiplets with:
\begin{equation}\label{Mk}
    M_k\,:=\, n\,-\,\overline{w}_1\,=\, \lambda_k~.
\end{equation}
Putting these observations together, we find that, upon applying type I duality -- see section 2 of \cite{Closset:2025cfm} -- to the quiver defect given in figure \ref{fig:defectGrassGeneral}, we end with the following quiver:
\begin{equation}
    \raisebox{-0.5\height}{\includegraphics[width=0.64\linewidth]{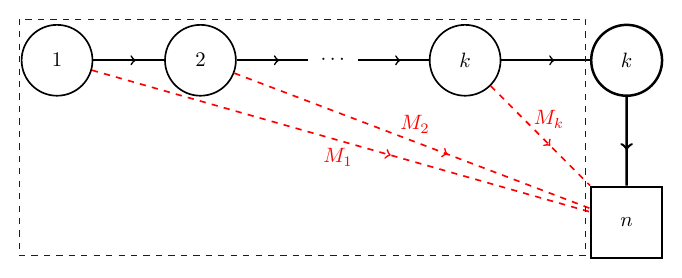}}
    \label{fig:intermediatereduction}
\end{equation}
Here, we mean to imply that each one of the Fermi fields is charged under a particular $U(1)$ subgroup of the global $SU(n)$, not in the antifundamental of the whole $SU(n)$.  We adopt this new convention, matching \cite{Closset:2023bdr}, for the remainder of this section.

\medskip
\noindent
\textbf{Further simplifications.} For the connection between the construction here and that of \cite{Closset:2023bdr} to be more apparent, let us consider the generic Grassmannian permutation $w$ such that the corresponding $k$-partition is of the form $\lambda = [\lambda_1, \cdots, \lambda_p, 0,\cdots, 0]$ for some $p\leq k$. Note that, for $\tilde{a} = p+1, \cdots, k-1$, $M_{\tilde{a}}\,=\,1$, and $M_{k} = 0$. 
In this case, one can apply type II duality~\cite[section 2.2]{Closset:2025cfm} for these $\tilde{a}$ gauge nodes. Indeed, starting from the right-hand part of the 1d quiver diagram, and applying this duality consecutively moving to the left, we end up with the following final form of the defect:
\begin{equation}
    \raisebox{-0.5\height}{\includegraphics[width=0.64\linewidth]{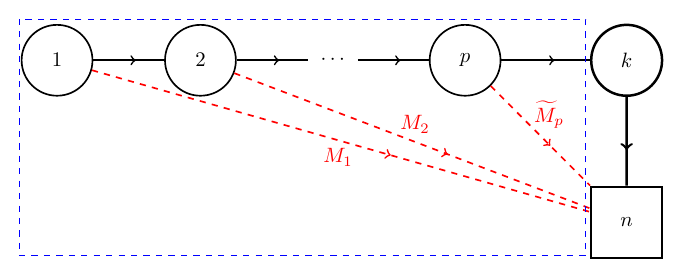}}
    \label{fig:finalreduction}
\end{equation}
In this quiver diagram, for $a =1, \cdots, p-1$, the number of Fermi multiplets $M_a$ is given in \eqref{Ma}. Meanwhile, for the last node, we have:
\begin{equation}\label{tildeMp}
    \widetilde{M}_p \,= \,M_p\, + \,k\,-\,p\,-\,1\,=\,\lambda_p\,+\,k\,-\,p~.
\end{equation}
The final form of this quiver defect matches that in figure 3 of \cite{Closset:2023bdr}, where the explicit expressions for $M_a$ match those given in \cite[equation (3.63)]{Closset:2023bdr}.

\begin{figure}[h]
    \centering
    \includegraphics[width=1\linewidth]{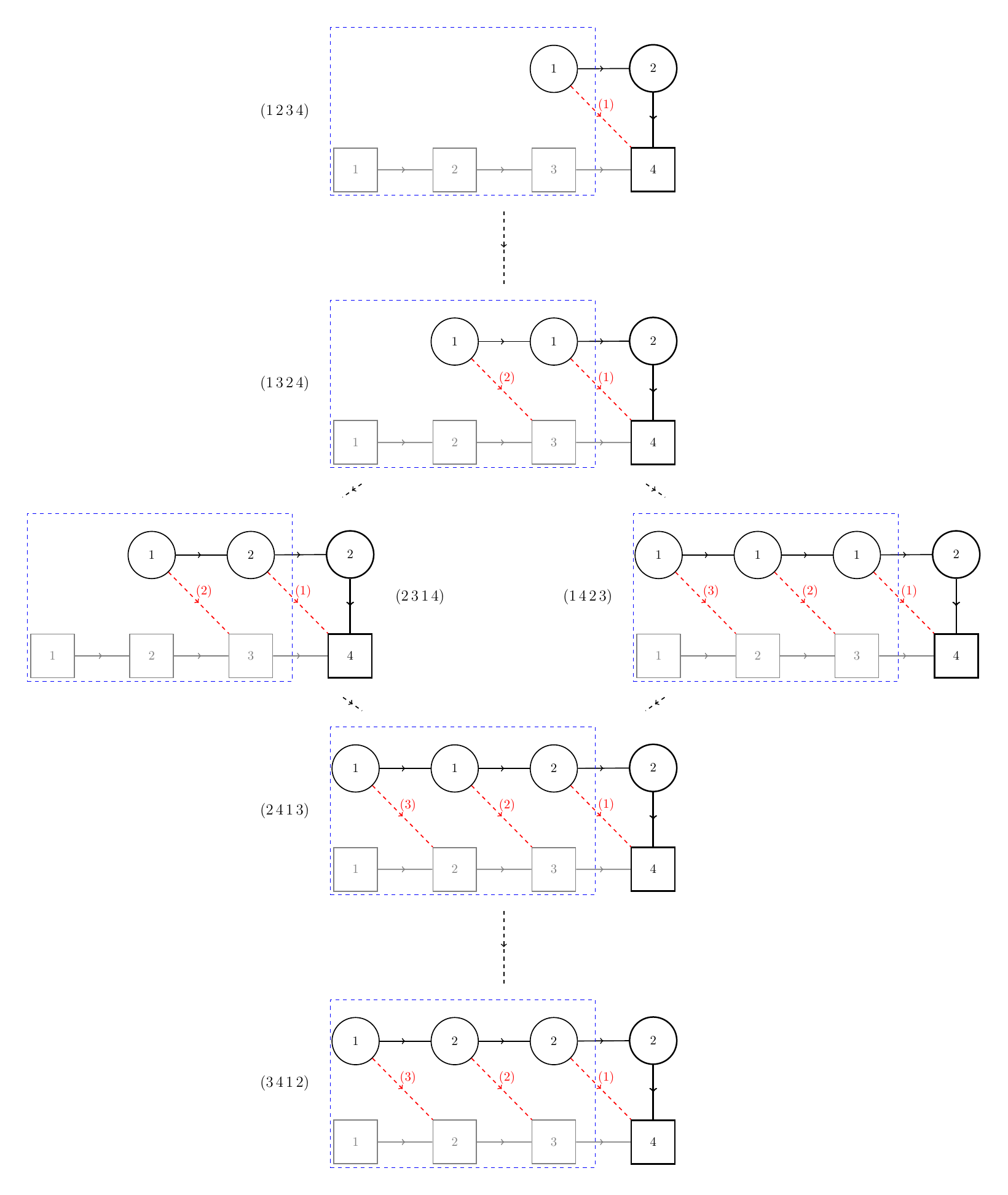}
    \caption{Hasse diagram for all possible Schubert defects in the Grassmannian manifold Gr$(2,4)$. The index next to each red arrow is the flavor index of the leftover Fermi multiplet after integrating out the massive chiral and Fermi multiplets.}
    \label{fig:HasseGr24}
\end{figure}

\medskip
\noindent
\textbf{1d indices and Grothendieck polynomials.} The above observations simplify the calculation of the 1d indices in the Grassmannian case, because the 1d Witten index~\eqref{1d index JK} for the Grothendieck lines indexed by the $k$-partition $\lambda$ can be written as~\cite{Closset:2023bdr}:
\begin{equation}
    \mathfrak{G}_\lambda(x,y) \,=\,\frac{\det_{1\leq a,b\leq k}\,\left(\,x_a^{b-1}\,\prod_{i=1}^{\lambda_b+k-b}\,(\,1\,-\,x_a\,y^{-1}_i\,)\,\right)}{\prod_{1\leq \,a\,<\,b\,\leq k}\,(\,x_b\,-\,x_a\,)}~,
\end{equation}
which is the double Grothendieck polynomial for a Grassmannian permutation, see~\eqref{defn par k}. 

\medskip
\noindent
\textbf{Example: Gr$(2,4)$.} As an explicit example of how the above reduction works, let us take the case Gr$(2,4)$. For this partial flag, there are six possible Schubert varieties indexed by the permutations ${\rm W}^{(2;4)}\cong\{(1\,2\,3\,4),(1\,3\,2\,4),(2\,3\,1\,4),(1\,4\,2\,3),(2\,4\,1\,3),(3\,4\,1\,2)\}$. Using the general proposal given in figure~\ref{fig:defectGrassGeneral}, we get the quiver defects shown in figure~\ref{fig:HasseGr24}.

After following the algorithm mentioned above of applying dualities for each one of these defects, one ends up with the Hasse diagram shown in~\cite[figure 4]{Closset:2023bdr}. Computing the 1d Witten indices for the defects in figure~\ref{fig:HasseGr24}, we indeed reproduce the corresponding Grothendieck polynomials listed in~\cite[equation (3.72)]{Closset:2023bdr}.

\section{Schubert point defects in the 2d GLSM}\label{sec:0d2dsystem}

In this last section, we briefly discuss the reduction of the quantum K-theory construction to quantum cohomology. That is, we consider the dimensional reduction of the 3d GLSM to 2d, which maps the Schubert line defect $\SL_w$ wrapping the circle on $\Sigma \times S^1$ to a {\it Schubert point defect} $O_w$ in the 2d GLSM. Similar to~\eqref{SL and CO}, these point defects engineer Schubert classes in the quantum cohomology of the partial flag manifold: 
\be\label{qcoh and omegaw}
O_w \quad \rightsquigarrow \quad [\Omega_w]~, \qquad w\,\in\, {\rm W}^{({\boldsymbol{k}};n)}~.
\ee
 Moreover, by the same logic, the Witten index of $\SL_w$ reduces to an $\CN=2$ supersymmetric matrix model partition function $\mathcal{I}_{w}^{({\rm 0d})}$ for the 0d degrees of freedom on the defect. By dimensional reduction, the `Whitney = Witten' conjecture~\eqref{conjecture Whitney equal index} identifies these partition functions to the parabolic double Whitney polynomials in cohomology~\eqref{k-poly-defn}, namely:
\begin{equation}\label{conjecture Whitney equal Z qcoh}
\mathcal{I}^{({\rm 0d})}_w(\t\sigma^{(\bullet)}, m) \,=\, \mathfrak{Z}^{({\boldsymbol{k}}; n)}_{w}(\t\sigma^{(\bullet)}, m)~.
\end{equation}
 As for the Witten indices, these polynomials do not depend on the quantum parameters $\qcoh_\ell$, but one can use the quantum cohomology ring relations~\eqref{QH par flag explicit} to eliminate the variables $\t\sigma^{(\bullet)}$ in favor of the $\CQ_\ell$ Chern roots $\{\sigma_i\}\equiv \{\widehat{\sigma}^{(\ell)}\}$, giving us the parabolic quantum Schubert polynomials \eqref{par-schu-poly-defn}.

\subsection{Schubert point defects and 0d matrix models}
The point Schubert defect is defined in the UV of the 2d GLSM in terms of an $\mathcal{N}=2$ supersymmetric matrix model with gauge group $G = \times_{\ell, i}\, U({\rm r}_{k_\ell,i})$ and 0d bifundamental chiral and Fermi multiplets connecting them as shown in the figure~\ref{fig:GenProposal}. The partition function of this matrix model can be written as the JK residue~\cite{Franco:2016tcm,Closset:2017yte}:
\begin{equation}\label{0d JK par func}
    \mathcal{I}_{w}^{({\rm 0d})}(\sigma, m)\,=\,\oint_{\rm JK} \;(dM) \;Z_{\rm chiral}^{\rm vec}\;Z_{\rm chiral}^{\rm hor}\;Z_{\rm Fermi}^{\rm black}\;Z_{\rm Fermi}^{\rm red}~.
\end{equation}
The integrand receives contributions from the bifundamental matter multiplets:
\begin{align}
    \begin{split}
        &{Z}_{\rm chiral}^{\rm ver} \,:=\,\prod_{\ell=1}^{s-1}\,\prod_{i=1}^{n-1}\,\prod_{\alpha=1}^{{\rm r}_{k_\ell,i}}\,\prod_{\beta=1}^{{\rm r}_{k_{\ell+1},i}}\,\left(\,{u^{(k_\ell,i)}_{\alpha}}-{u^{(k_{\ell+1},i)}_{\beta}}\,\right)^{-1} ~,\\
        &{Z}_{\rm chiral}^{\rm hor} \,:=\,\prod_{\ell=1}^{s}\,\left[\,\prod_{\alpha=1}^{{\rm r}_{k_\ell,n-1}}\,\prod_{a = 1}^{k_\ell}\,\left(\,{u_{\alpha}^{(k_\ell,n-1)}}-{\widetilde{\sigma}_{a}^{(\ell)}}\,\right)^{-1}\,\prod_{i=1}^{n-2}\,\prod_{\beta=1}^{{\rm r}_{k_\ell,i}}\,\prod_{\gamma=1}^{{\rm r}_{k_\ell,i+1}}\,\left(\,{u^{(k_\ell,i)}_{\beta}}-{u^{(k_\ell,i+1)}_{\gamma}}\,\right)^{-1}\,\right] ~,\\ 
        &{Z}^{\rm black}_{\rm Fermi} \,:=\, \prod_{\ell=1}^{s-1}\,\left[\,\prod_{\gamma=1}^{{\rm r}_{k_\ell,n-1}}\prod_{a=1}^{k_{\ell+1}}\left(\,{u^{(k_\ell,n-1)}_\gamma}-{\widetilde{\sigma}_{a}^{(\ell+1)}}\,\right)\prod_{i = 1}^{n-1}\,\prod_{\alpha=1}^{{\rm r}_{k_\ell,i}}\,\prod_{\beta = 1}^{{\rm r}_{k_{\ell+1},i+1}}\,\left(\,{u^{(k_\ell,i)}_\alpha}-{u_\beta^{(k_{\ell+1},i+1)}}\,\right)\,\right]\,~,\\
        &{Z}^{\rm red}_{\rm Fermi} \,:=\,\prod_{i=1}^{n-1}\,\prod_{\alpha=1}^{{\rm r}_{k_s,i}}\,\left(\,{u^{(k_s,i)}_\alpha}-{m_{n-i}}\,\right) ~.
    \end{split}
\end{align}
As for the measure $(dM)$ in \eqref{0d JK par func}, it is given more explicitly by:
\begin{equation}
    (dM) \,:=\, \prod_{\ell=1}^{s}\,\prod_{i=1}^{n-1}\,\left[\,\Delta^{(k_\ell,i)}_{\rm 0d}(u)\,\frac{1}{{\rm r}_{k_\ell,i}\,!}\,\prod_{\alpha=1}^{{\rm r}_{k_\ell,i}}\,\frac{du_\alpha^{(k_\ell,i)}}{2\pi i }\,\right]~,
\end{equation}
with $\Delta_{\rm 0d}^{(k_\ell,i)}$ being the Vandermonde determinant factor:
\begin{equation}
    \Delta_{\rm 0d}^{(k_\ell,i)}(u)\,:=\,\prod_{1\leq \alpha\neq \beta\leq {\rm r}_{k_\ell,i}}\,\left(u^{(k_\ell,i)}_\alpha - u_\beta^{(k_\ell,i)}\right)~.
\end{equation}
When computing the residues, we pick all the poles coming from $Z_{\rm chiral}^{\rm ver}$ and $Z_{\rm chiral}^{\rm hor}$.

\medskip
\noindent
\textbf{Examples and explicit computations.} Similarly to the 1d indices in the previous section, our evidence for the conjecture~\eqref{conjecture Whitney equal Z qcoh} comes from looking at various limits, checking consistency, and by computing both sides of the equality in many explicit examples. In particular, we can study all partial flag manifolds with $n=4$, like in the 1d-3d case. For the complete flag Fl$(4)$, see appendix C of the accompanying work \cite{Closset:2025cfm}. The partial flag manifolds Fl$(1,2;4)$, Fl$(1,3;4)$, and Fl$(2,3;4)$ are discussed in appendices~\ref{app:124}, \ref{app:134} and \ref{app:234},  respectively.


\subsubsection{The special case of  Gr\texorpdfstring{$(k,n)$}{k,n}}
In the Grassmannian case, the Schubert defect of~figure \ref{fig:GenProposal} reduces to the 1d quiver in figure~\ref{fig:defectGrassGeneral}. In the K-theoretic context, where we interpreted this structure as a 1d-3d coupled system, we were able to use 1d dualities to reduce the form of the quiver to~\eqref{fig:finalreduction}. 
In the case at hand, it is likely that analogous 0d dualities exist, but a 0d duality is simply an identity between integrals. Here, we will directly argue that the partition function of the matrix model in figure \ref{fig:defectGrassGeneral} matches that in \eqref{fig:finalreduction} when both are viewed as 0d-2d coupled systems.

With that aim, let us focus on the subquiver that has the $U(a)$ 0d gauge nodes. At the level of the partition function, we have the following contribution:
\begin{equation}\label{ath residue}
    \prod_{j=\overline{w}_{k-a+1}}^{\overline{w}_{k-a}-1} \,\frac{1}{a!}\,\oint_{\rm JK} \,\frac{d^au^{(j)}}{(2\pi i)^a}\,\Delta^{(j)}_{\rm 0d}(u)\,\mathcal{Z}_j(u,y)~,
\end{equation}
with, 
\begin{equation}
    \Delta^{(j)}_{\rm 0d}(u) \,=\, \prod_{1\leq \alpha\neq \beta\leq a}(u^{(j)}_\alpha - u_\beta^{(j)})~,
\end{equation}
and, 
\begin{equation}\label{mathcalZj}
  \mathcal{Z}_j(u,y)\, =\,   \begin{cases}
      \frac{ \prod_{\alpha=1}^a(u_{\alpha}^{(j)}-m_{n-j})}{\prod_{\alpha,\beta=1}^a(u_\alpha^{(j)}-u_\beta^{(j+1)})}  ~, \qquad & j\,=\,\overline{w}_{k-a+1}, \cdots, \overline{w}_{k-a}-2~,\\
       \frac{ \prod_{\alpha=1}^a(u_{\alpha}^{(\overline{w}_{k-a}-1)}-m_{n-\overline{w}_{k-a}+1})}{\prod_{\alpha=1}^a\prod_{\beta=1}^{a+1}(u_\alpha^{(\overline{w}_{k-a}-1)}-u_\beta^{(\overline{w}_{k-a})})} ~,\qquad &j \,=\,\overline{w}_{k-a}-1~.
    \end{cases}
\end{equation}
Here we are using the shorthand notation $u^{(j)} \equiv u^{(k,j)}$.

Let us point out at this stage that, to be able to treat the above residue integral independently from the rest, we are assuming that we are performing the full nested residue computation recursively, starting with the $U(1)$ 0d gauge node associated with $j=1$. For instance, from the first glance at \eqref{ath residue}, one might worry about the poles of $u^{(\overline{w}_{k-a}-1)}_\bullet$ coming from the chiral bifundamental between $U(a-1)$ at position $j=\overline{w}_{k-a+1}-1$ and $U(a)$ at position $j=\overline{w}_{k-a+1}$. But these poles will have already been taken care of when performing the residues associated with the $U(a-1)$ gauge nodes.

Now, let us perform the residue integrals \eqref{ath residue} recursively. Note that, when we get to the node $j = \overline{w}_{k-a+1}, \cdots,\overline{w}_{k-a}-2$, we will have to perform the residue to the following \eqref{mathcalZj}:
\begin{equation}
        \Delta_{\rm 0d}^{(j)}(u)\,  \frac{ \prod_{\alpha=1}^a\,(\,u_{\alpha}^{(j)}\,-\,m_{n-j}\,)}{\prod_{\alpha,\beta=1}^a\,(\,u_\alpha^{(j)}\,-\,u_\beta^{(j+1)}\,)}~.
\end{equation}
The presence of the Vandermonde determinant in the numerator instructs us to compute the residue at:
\begin{equation}\label{residue a}
    u_\alpha^{(j)} \,=\, u_\alpha^{(j+1)}~, \qquad \alpha = 1, \cdots, a~,
\end{equation}
or any $S_a$ permutation of these. These $a!$ residues will have the same value, and this overcounting is taken care of by the $\frac{1}{a!}$ factor in \eqref{ath residue}. The residue at \eqref{residue a} reads:
\begin{equation}
    \Delta_{\rm 0d}^{(j+1)}(u)\,\frac{\prod_{\alpha=1}^a\,(\,u_\alpha^{(j+1)}\,-\,m_{n-j}\,)}{\prod_{1\,\leq\, \alpha\,\neq\, \beta\,\leq\, a}(u_\alpha^{(j+1)}\,-\,u_\beta^{(j+1)})}~,
\end{equation}
where the Vandermonde determinant in the numerator is now expressed in terms of the $u^{(j+1)}$ parameters. This is exactly the factor that we get in the denominator. 
 In summary, the gauge nodes indexed by $j= \overline{w}_{k-a+1}, \cdots,\overline{w}_{k-a}-2$ can be integrated out recursively, leaving behind the following term:
\begin{equation}
    \prod_{j=\overline{w}_{k-a+1}}^{\overline{w}_{k-a}-2}\,\prod_{\alpha=1}^{a} \,(u_\alpha^{(\overline{w}_{k-a}\,-\,1)} \,-\, m_{n-j})~.
\end{equation}
In fact, putting this contribution in the residue integral for $j = \overline{w}_{k-a}-1$, then, the integral \eqref{ath residue} becomes that of the 0d $U(a)$ gauge node in \eqref{fig:reduce at a}. Performing the above steps recursively for each $a = 1, \cdots, k$ in figure \ref{fig:defectGrassGeneral}, one can see that the final form of the residue integral is that of the quiver \eqref{fig:finalreduction}.

Our result here agrees with the proposal given in section 5 of \cite{Closset:2023bdr}. In particular, one can now perform the `reduced' residue integral of the 0d matrix model similarly, as was done in that work, to get the double Schubert polynomials:
\begin{equation}
    \mathfrak{S}_\lambda(\sigma,m)\,=\,\frac{\det_{1\leq a,b\leq k}\left(\prod_{i=1}^{\lambda_b+k-b}\,(\,\sigma_a\,-\,m_i\,)\,\right)}{\prod_{1\,\leq\, a\,<\,b\,\leq\, k}\,(\,\sigma_a\,-\,\sigma_b\,)}~,
\end{equation}
where we used the relations \eqref{sigma11=sigma1}, and the  partition $\lambda=\lambda_w$ was defined in~\eqref{defn par k}.

\section{Conclusions}\label{sec:conclusion}
We constructed a family of Schubert line defects $\SL_w$ in 3d GLSMs with target $X = {\rm Fl}(\boldsymbol{k};n)$ a partial flag manifold, allowing us to physically realize the equivariant quantum K-theory ring QK$_T(X)$ in the basis spanned by the Schubert classes $[\CO_w]$. This construction subsumes previous constructions for Grassmannian manifolds~\cite{Closset:2023bdr} and for the complete flag manifold~\cite{Closset:2025cfm}. We similarly constructed Schubert point defects in the 2d GLSM.

Since the line defects $\SL_w$ are constructed as 1d $\CN=2$ supersymmetric gauged quantum mechanics, the natural observables associated with them are their flavored Witten indices, which are non-trivial polynomials in the 3d gauge and flavor Coulomb-branch parameters. Mathematically, they are polynomials in the K-theoretic Chern roots of the tautological vector bundles of $X$. We have conjectured that these polynomials are exactly the parabolic Whitney polynomials recently introduced in the mathematical literature~\cite{ahkmox} (and here renamed by us). We provided a large amount of evidence for this conjecture, from exact computations and by showing that these polynomials naturally lead us to a number of other `quantum' polynomials that represent the classes $[\CO_w]$ in various presentations of the quantum K-theory ring, including the quantum (double) Grothendieck polynomials. As a mathematical side-quest, we derived a new set of such polynomials, the parabolic quantum (double) Grothendieck polynomials that naturally represent $[\CO_w]$ in a Toda presentation of QK$_T(X)$ (in terms of only $n$ variables $x_i$).

In the future, it would be very interesting to better understand whether the full algebraic structure underpinning these polynomials, as reviewed in section~\ref{sec: quantum polys}, can be understood purely from physical arguments. In particular, the Demazure difference operator seemingly relates different Schubert line defects through certain difference equations, but a physical understanding of this fact is lacking.  It would also be worthwhile to further develop computational algorithms similar to those discussed in~\cite{Closset:2023vos, Closset:2023bdr} to efficiently work out the structure coefficients of the quantum K-theory ring. 

\section*{Acknowledgments}
We would like to thank C.~Lenart, L.~Mihalcea, M.~Shimozono, and W.~Xu for useful conversations.
C.~Closset is a Royal Society
University Research Fellow.  The research of W.~Gu is funded by the National Natural Science Foundation of China (NSFC) with Grant No.12575077. O.~Khlaif is a Junior Research Fellow at the Philippe Meyer Institute, and, in the early stages of this project, he was supported by the School of Mathematics at the University of Birmingham.
E.~Sharpe and H.~Zhang were partially supported by NSF grant PHY-2310588. H.~Zou was partially supported by the National Natural Science Foundation of China (NSFC) with Grant No.~12405083 and~12475005 and the Shanghai Pujiang Program with Grant No.~24PJA119.

 \appendix


\newpage
\section{Schubert defects in Fl\texorpdfstring{$(1,2;4)$}{124}} \label{app:124}

\subsection*{All possible Schubert varieties}
In this partial flag, we have 12 possible Schubert varieties indexed by the equivalence classes in ${\rm W}^{(1,2;4)}$. These classes and their corresponding dimensions are summarized in the following table:
\begin{table}[!ht]
\centering
\begin{subtable}{0.45\textwidth}
\centering
\begin{equation*}
        \begin{array}{|c||c|c|c|}
    \hline
        w&{\ell(w)}  & \dim(X_w) \\
        \hline
        \hline
        {(1\,2\,3\,4)} & 0  & 5\\
        \hline
        {(1\,3\,2\,4)} &  1 & 4 \\
        \hline
        {(2\,1\,3\,4)} &  1 & 4\\
        \hline
         {(1\,4\,2\,3)} &  2 & 3 \\
 
        \hline
         {(2\,3\,1\,4)} & 2 & 3 \\
         \hline
           {(3\,1\,2\,4)} & 2& 3\\
        \hline
        \end{array}
    \end{equation*}
\end{subtable}
\hfill
\begin{subtable}{0.45\textwidth}
\centering
\begin{equation*}
        \begin{array}{|c||c|c|c|}
    \hline
         w&{\ell(w)} & {\dim(X_w)} \\
        \hline
     \hline
        {(2\,4\,1\,3)} &  3 &2\\
               \hline
        {(4\,1\,2\,3)} &  3&2 \\
        \hline
        {(3\,2\,1\,4)} &  3&2\\
        \hline
        {(3\,4\,1\,2)} &  4&1\\
        \hline
        {(4\,2\,1\,3)} &  4&1\\
        \hline
        {(4\,3\,1\,2)} & 5 &0\\
        \hline
        \end{array}
    \end{equation*}
\end{subtable}
\end{table}
These Schubert varieties are connected to each other via the following Hasse diagram:
\begin{figure}[h!]
    \centering
    \includegraphics[width=0.75\linewidth]{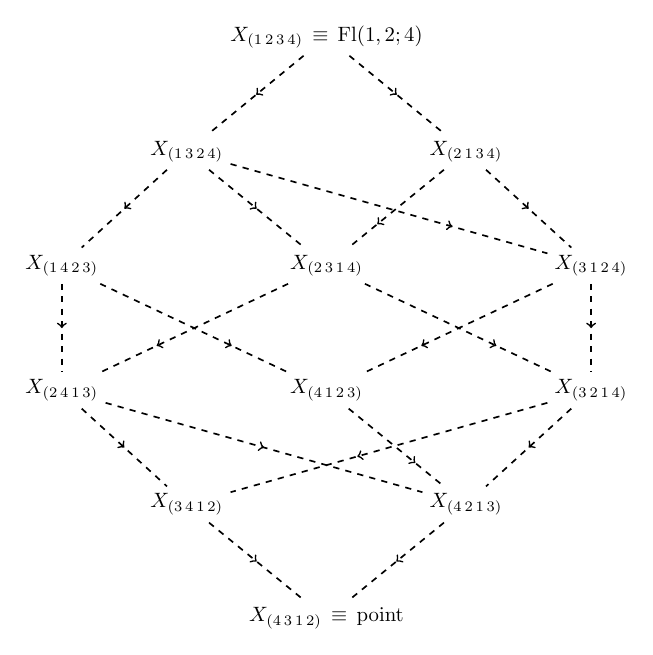}
\end{figure}

\newpage
\subsection*{Coupled systems defining the Schubert defects}\label{subsec:defects124}
For each one of the 12 permutation classes $[w]\in {\rm W}^{(1,2;4)}$, let us now write down the 1d-3d (0d-2d) coupled system defining the Schubert line (point) defect. Here we follow the conventions of figure \ref{fig:GenProposal} for the general proposal. Viewed as Schubert line defects, in the last column, we also include the reduced form of the quiver after applying the duality moves reviewed in section 2 of \cite{Closset:2025cfm}.
\begin{table}[!ht]
\centering
\begin{subtable}{0.45\textwidth}
\centering
\begin{equation*}
        \begin{array}{|c||c|c|c|}
    \hline
        w&{\rm General\,\,proposal}&  {\rm Reduced \,\,quiver}\\
        \hline
        \hline
        \raisebox{6ex}{(1\,2\,3\,4)} & \includegraphics[scale = 0.3]{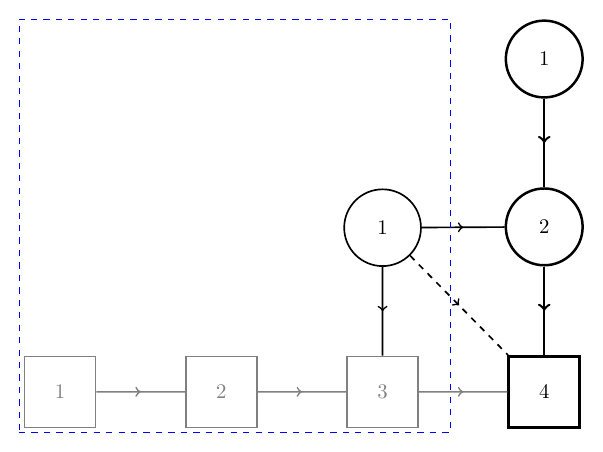} &\includegraphics[scale = 0.3]{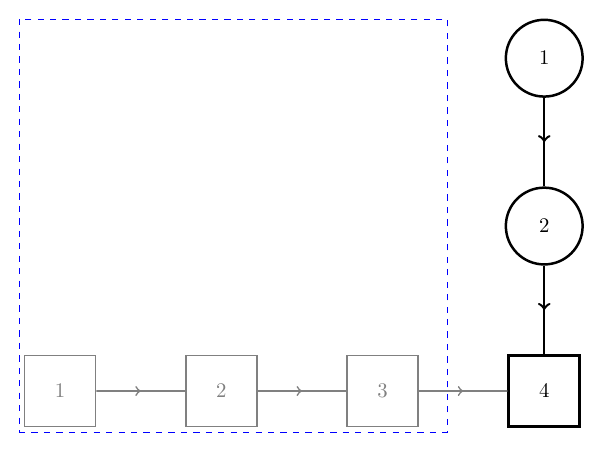}\\
        \hline
        \raisebox{6ex}{(1\,3\,2\,4)} & \includegraphics[scale = 0.3]{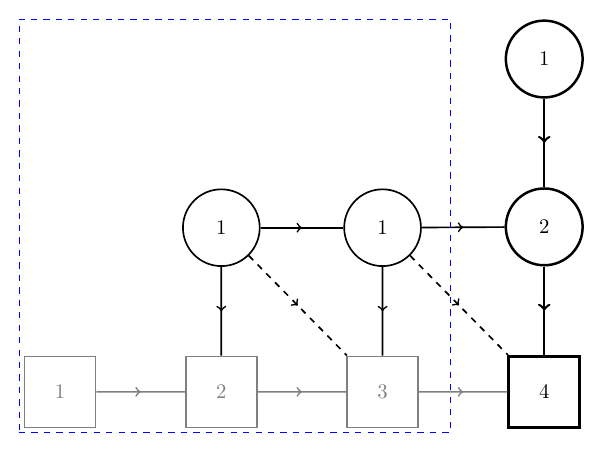} & \includegraphics[scale = 0.3]{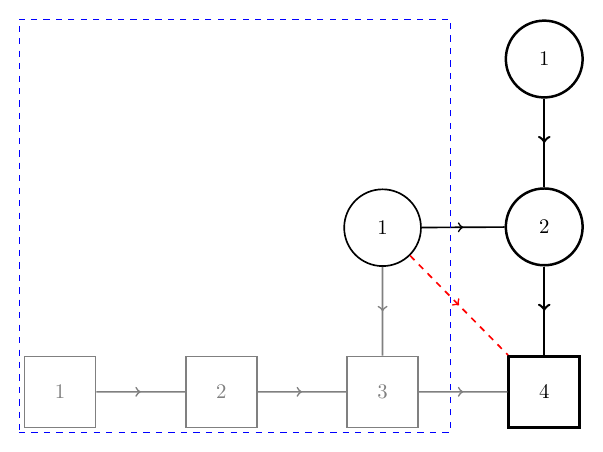}\\
         \hline
        \raisebox{6ex}{(2\,1\,3\,4)} & \includegraphics[scale = 0.3]{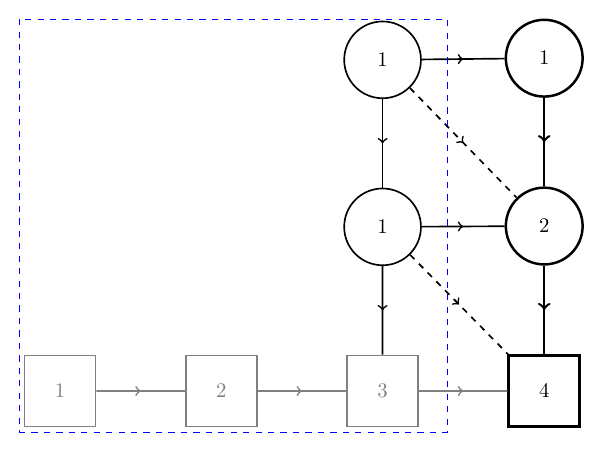} & \includegraphics[scale = 0.3]{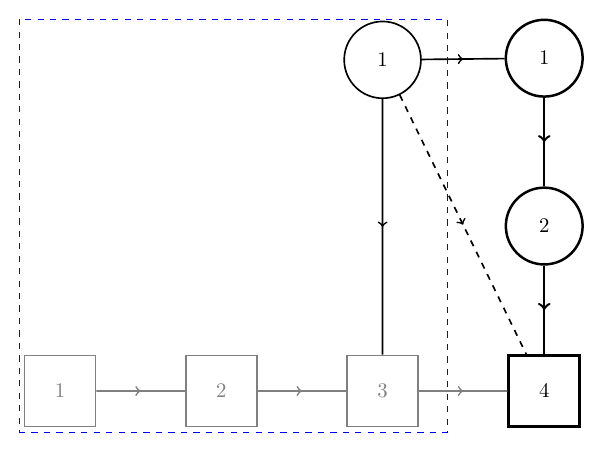}\\
        \hline
         \raisebox{6ex}{(1\,4\,2\,3)} &  \includegraphics[scale = 0.3]{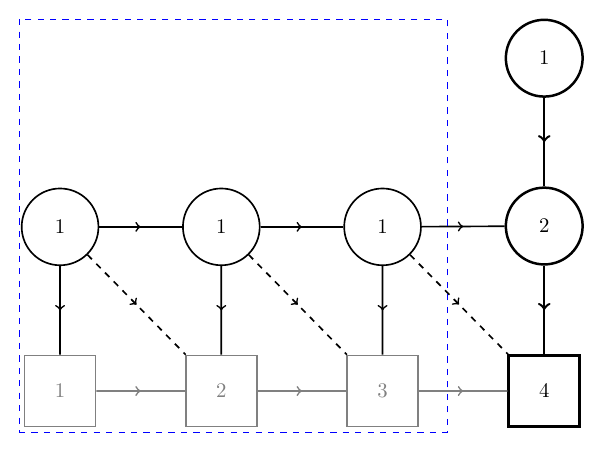} & \includegraphics[scale = 0.3]{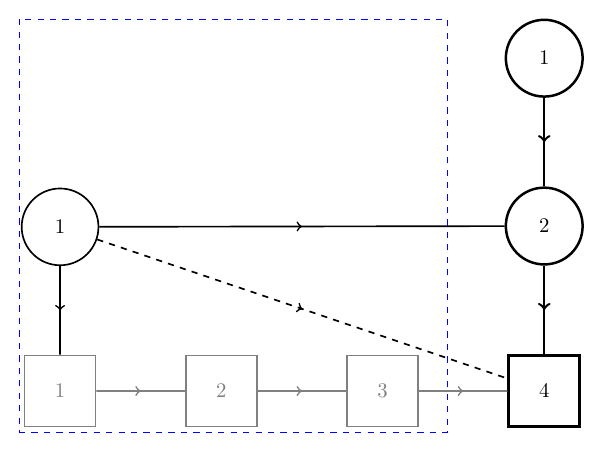}\\
       
        \hline
         \raisebox{6ex}{(2\,3\,1\,4)} & \includegraphics[scale = 0.3]{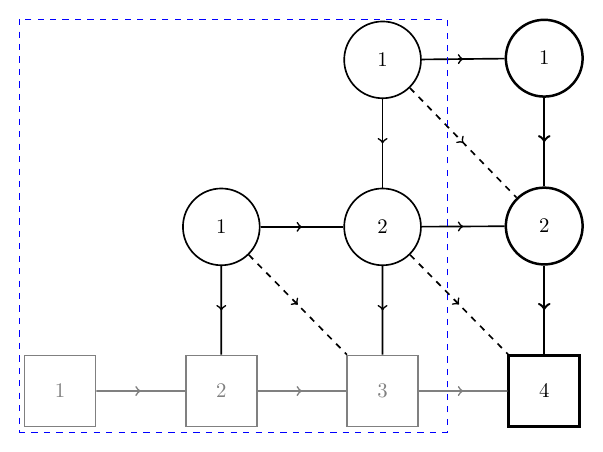} & \includegraphics[scale = 0.3]{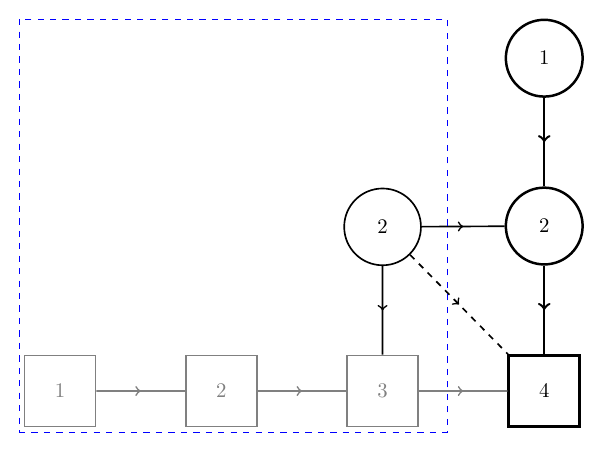}\\
          \hline
           \raisebox{6ex}{(3\,1\,2\,4)} &\includegraphics[scale = 0.3]{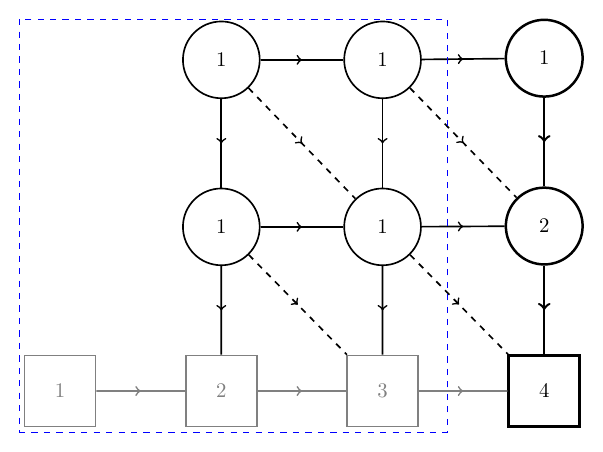} & \includegraphics[scale = 0.3]{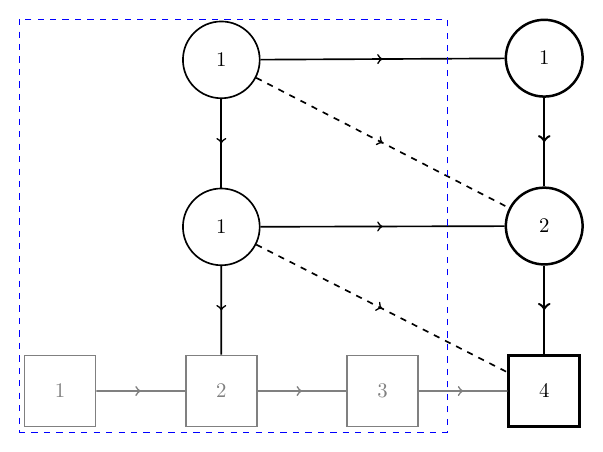}\\
        \hline
        \end{array}
    \end{equation*}
\end{subtable}
\hfill
\begin{subtable}{0.45\textwidth}
\centering
\begin{equation*}
        \begin{array}{|c||c|c|c|}
    \hline
        w&{\rm General\,\,proposal}&  {\rm Reduced \,\,quiver}\\
        \hline
       \hline
        \raisebox{6ex}{(2\,4\,1\,3)} & \includegraphics[scale = 0.3]{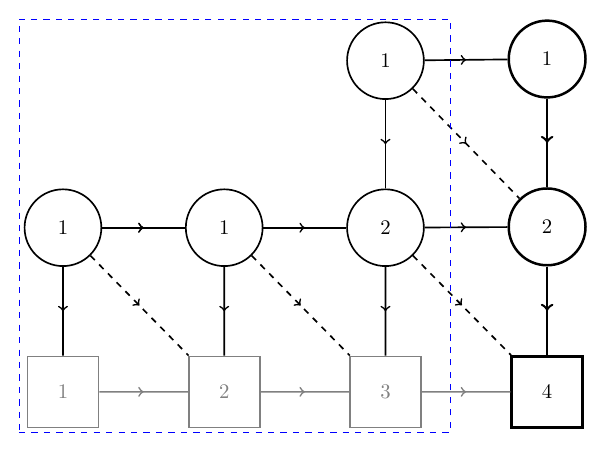} & \includegraphics[scale = 0.3]{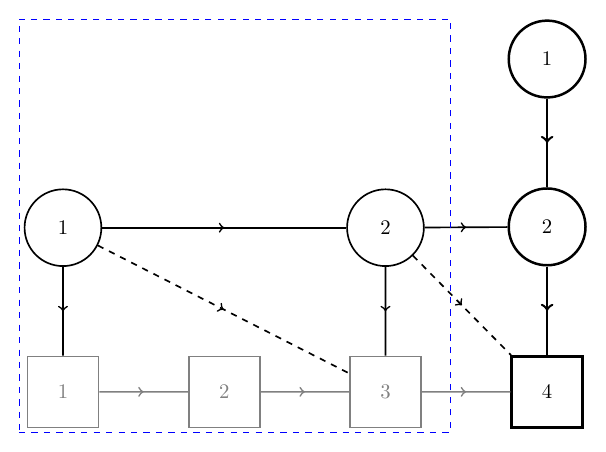}\\
     \hline
        \raisebox{6ex}{(4\,1\,2\,3)} & \includegraphics[scale = 0.3]{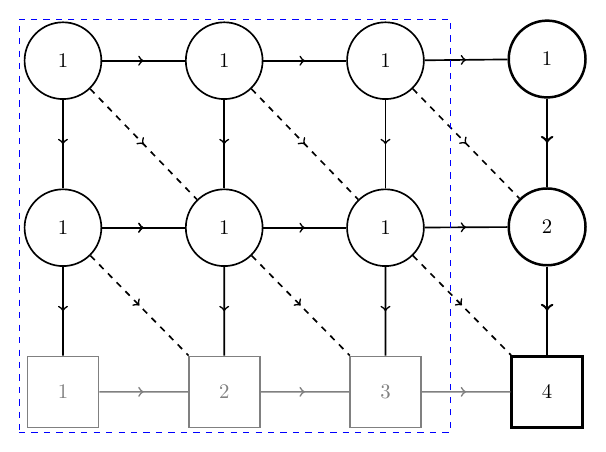} & \includegraphics[scale = 0.3]{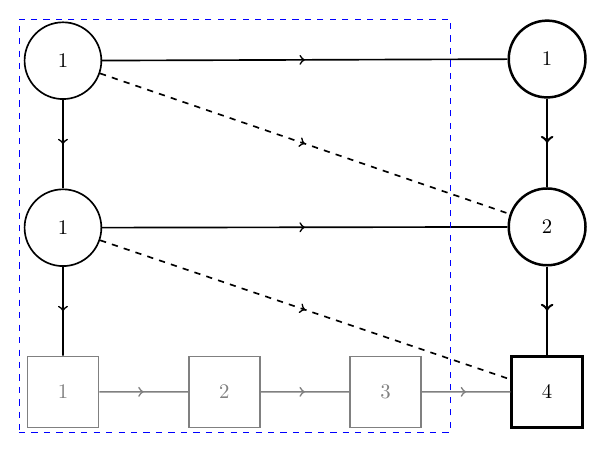}\\
        \hline
        \raisebox{6ex}{(3\,2\,1\,4)} & \includegraphics[scale = 0.3]{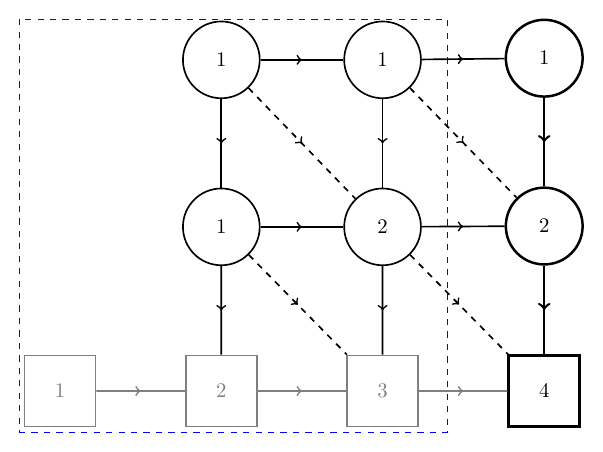} & \includegraphics[scale = 0.3]{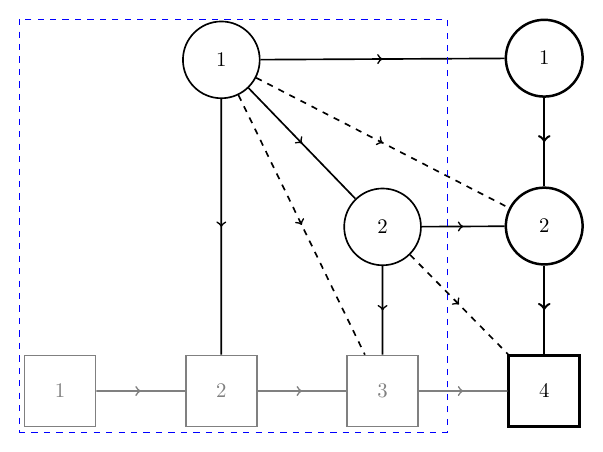}\\
        \hline
        \raisebox{6ex}{(3\,4\,1\,2)} & \includegraphics[scale = 0.3]{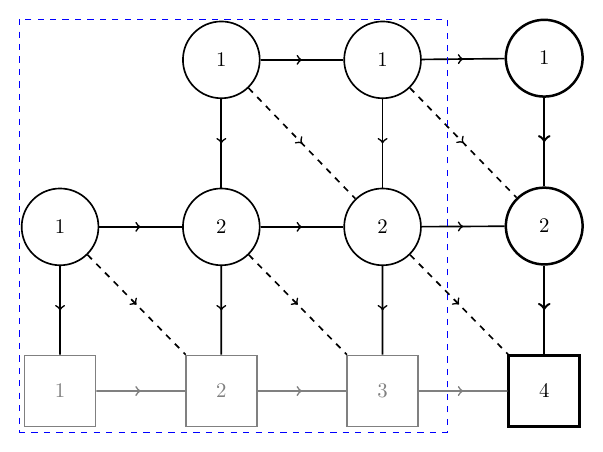} & \includegraphics[scale = 0.3]{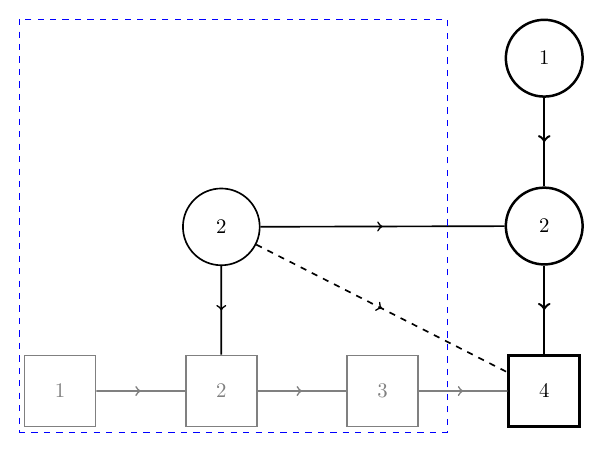}\\
        \hline
        \raisebox{6ex}{(4\,2\,1\,3)} & \includegraphics[scale = 0.3]{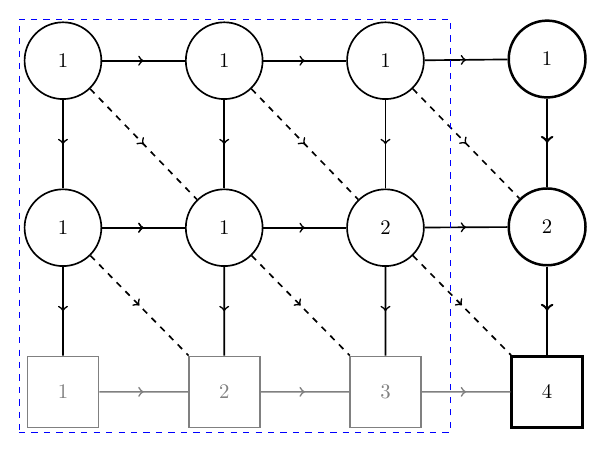} & \includegraphics[scale = 0.3]{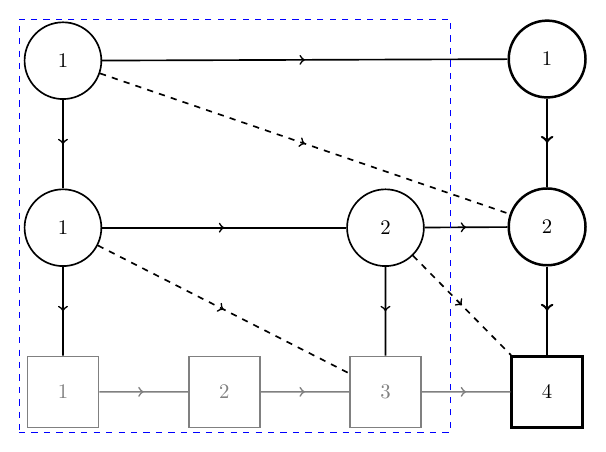}\\
        \hline
        \raisebox{6ex}{(4\,3\,1\,2)} & \includegraphics[scale = 0.3]{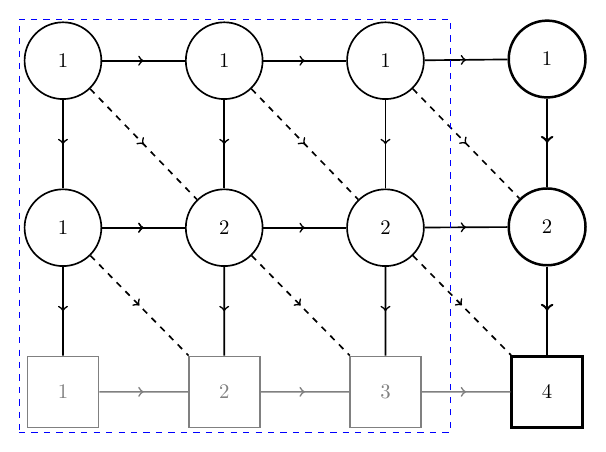} & \includegraphics[scale = 0.3]{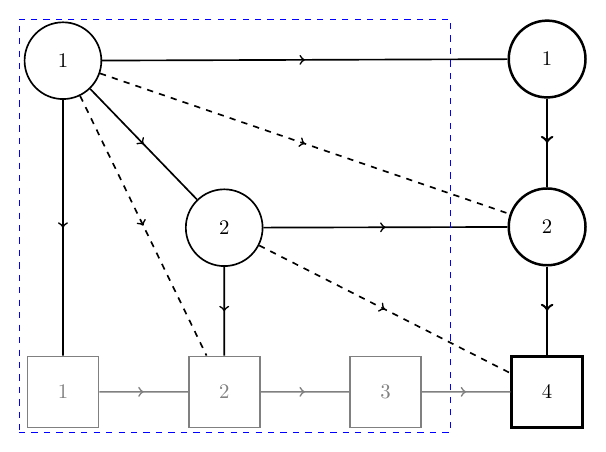}\\
        \hline
        \end{array}
    \end{equation*}
\end{subtable}
\end{table}
\newpage
\subsection*{1d indices and parabolic quantum Grothendieck polynomials }
\begin{center}
    \renewcommand{\arraystretch}{1}
    \begin{longtable}[!h]{|c||c|c|}
    \hline
    $w$&$\mathcal{I}_w^{(\rm 1d)}(x,y)$&$\mathcal{I}_w^{(\rm 1d)}{\scriptsize\begin{bmatrix}
        \qk_1&\qk_2\\
        1&2
    \end{bmatrix}}(x,y)$\\
    \hline
    \hline
          $(1\,3\,2\,4)$   &\parbox{7cm}{\centering $1-\frac{x_1^{(2)}x_2^{(2)}}{y_1y_2}$}&\parbox{7cm}{\centering $1-\frac{x_1x_2}{(1-q_1)\,y_1y_2}$}\\
       \hline
      $(2\,1\,3\,4)$   & \parbox{7cm}{\centering $1-\frac{x_1^{(1)}}{y_1}$}& \parbox{7cm}{\centering $1-\frac{x_1}{y_1}$}\\
    \hline 
    $(1\,4\,2\,3)$   &\parbox{7cm}{\centering$1-\frac{x_2^{(2)} x_1^{(2)}}{y_1 y_2}-\frac{x_1^{(2)}x_2^{(2)} }{y_1 y_3}-\frac{x_1^{(2)}x_2^{(2)}}{y_2 y_3}+\frac{ x_1^{(2)}x_2^{(2)\,2}}{y_1 y_2 y_3}+\frac{x_1^{(2)\,2}x_2^{(2)} }{y_1 y_2 y_3}$} &\parbox{7cm}{\centering$1-\frac{x_1 x_2}{(1-q_1)\,y_1 y_2}-\frac{x_1x_2}{(1-q_1)\,y_1 y_3}-\frac{x_1x_2}{(1-q_1)\,y_2 y_3}+\frac{ x_1x_2^{2}}{(1-q_1)\,y_1 y_2 y_3}+\frac{x_1^{2}x_2 }{(1-q_1)\,y_1 y_2 y_3}$}\\
    \hline
      $(2\,3\,1\,4)$   &\parbox{7cm}{\centering$1-\frac{x_1^{(2)}}{y_1}-\frac{x_2^{(2)}}{y_1}+\frac{x_1^{(2)}x_2^{(2)} }{y_1^2}$} &\parbox{7cm}{\centering$1-\frac{x_1}{y_1}-\frac{x_2}{y_1}+\frac{x_1x_2}{(1-q_1)\,y_1^2}$} \\
          \hline
      $(3\,1\,2\,4)$   & \parbox{7cm}{\centering $1-\frac{x_1^{(1)}}{y_1}-\frac{x_1^{(1)}}{y_2}+\frac{x_1^{(1)}x_1^{(2)} }{y_1 y_2}+\frac{x_1^{(1)}x_2^{(2)} }{y_1 y_2}-\frac{x_1^{(2)} x_2^{(2)}}{y_1 y_2}$}&\parbox{7cm}{\centering $1-\frac{x_1}{y_1}-\frac{x_1}{y_2}+\frac{x_1^2 }{y_1 y_2}-\frac{q_1\,x_1 x_2}{(1-q_1)\,y_1 y_2}$}\\
    \hline
      $(2\,4\,1\,3)$   &\parbox{7cm}{\centering $1-\frac{x_1^{(2)}}{y_1}-\frac{x_2^{(2)}}{y_1}+\frac{x_1^{(2)}x_2^{(2)} }{y_1^2}-\frac{x_1^{(2)}x_2^{(2)} }{y_2 y_3}+\frac{x_1^{(2)\,2}x_2^{(2)} }{y_1 y_2 y_3}+\frac{x_1^{(2)}x_2^{(2)\,2} }{y_1 y_2 y_3}-\frac{x_1^{(2)\,2}x_2^{(2)\,2} }{y_1^2 y_2 y_3}$} &\parbox{7cm}{\centering $1-\frac{x_1}{y_1}-\frac{x_2}{y_1}+\frac{x_1x_2}{(1-q_1)\,y_1^2}-\frac{x_1x_2}{(1-q_1)\,y_2 y_3}+\frac{x_1^{2}x_2 }{(1-q_1)\,y_1 y_2 y_3}+\frac{x_1x_2^{2} }{(1-q_1)\,y_1 y_2 y_3}-\frac{x_1^{2}x_2^{2} }{(1-q_1)^2\,y_1^2 y_2 y_3}$}\\
          \hline
      $(4\,1\,2\,3) $  &\parbox{7cm}{\centering $1-\frac{x_1^{(1)}}{y_1}-\frac{x_1^{(1)}}{y_2}-\frac{x_1^{(1)}}{y_3}+\frac{x_1^{(1)} x_2^{(2)}}{y_1 y_2}+\frac{x_1^{(1)} x_2^{(2)}}{y_1 y_3}+\frac{x_1^{(1)} x_2^{(2)}}{y_2y_3}+\frac{x_1^{(1)} x_1^{(2)}}{y_1 y_2}-\frac{x_1^{(2)}x_2^{(2)} }{y_1 y_2}+\frac{x_1^{(1)} x_1^{(2)}}{y_1 y_3}-\frac{x_1^{(2)}x_2^{(2)} }{y_1 y_3}+\frac{x_1^{(1)} x_1^{(2)}}{y_2 y_3}-\frac{ x_1^{(2)}x_2^{(2)}}{y_2
   y_3}-\frac{x_1^{(1)} x_1^{(2)\,2}}{y_1 y_2 y_3}+\frac{x_2^{(2)} x_1^{(2)\,2}}{y_1 y_2 y_3}+\frac{x_2^{(2)\,2} x_1^{(2)}}{y_1 y_2 y_3}-\frac{x_1^{(1)} x_2^{(2)\,2}}{y_1
   y_2 y_3}-\frac{x_1^{(1)}x_1^{(2)} x_2^{(2)} }{y_1 y_2 y_3}$} &\parbox{7cm}{\centering $1-\frac{x_1}{y_1}-\frac{x_1}{y_2}-\frac{x_1}{y_3}+\frac{x_1 x_2}{y_1 y_2}+\frac{x_1 x_2}{y_1 y_3}+\frac{x_1 x_2}{y_2y_3}+\frac{x_1^2}{y_1 y_2}-\frac{x_1x_2 }{(1-q_1)\,y_1 y_2}+\frac{x_1^2}{y_1 y_3}-\frac{x_1x_2}{(1-q_1)\,y_1 y_3}+\frac{x_1^2}{y_2 y_3}-\frac{ x_1x_2}{(1-q_1)\,y_2
 y_3}-\frac{x_1^3}{y_1 y_2 y_3} + \frac{\qk_1 x_1 x_2^2}{(1-\qk_1) y_1 y_2 y_3} + \frac{2\qk_1 x_1^2 x_2}{(1-\qk_1)y_1 y_2 y_3}$}\\
    \hline
      $(3\,2\,1\,4)$   & \parbox{7cm}{\centering $1-\frac{x_1^{(1)}}{y_2}-\frac{x_1^{(2)}}{y_1}-\frac{x_2^{(2)}}{y_1}+\frac{x_1^{(1)}x_1^{(2)} }{y_1 y_2}+\frac{x_1^{(1)}x_2^{(2)} }{y_1 y_2}+\frac{x_1^{(2)} x_2^{(2)}}{y_1^2}-\frac{x_1^{(1)}x_1^{(2)} x_2^{(2)} }{y_1^2 y_2}$}&\parbox{7cm}{\centering $1-\frac{x_1}{y_2}-\frac{x_1}{y_1}-\frac{x_2}{y_1}+\frac{x_1^2 }{y_1 y_2}+\frac{x_1x_2 }{y_1 y_2}+\frac{x_1 x_2}{(1-q_1)\,y_1^2}-\frac{x_1^2 x_2 }{(1-q_1)\,y_1^2 y_2}$}\\
    \hline 
    $(3\,4\,1\,2) $  & \parbox{7cm}{\centering$1-\frac{x_1^{(2)}}{y_1}-\frac{x_1^{(2)}}{y_2}-\frac{x_2^{(2)}}{y_1}-\frac{x_2^{(2)}}{y_2}+\frac{x_1^{(2)\,2}}{y_1 y_2}+\frac{2 x_1^{(2)} x_2^{(2)} }{y_1 y_2}+\frac{x_2^{(2)\,2}}{y_1 y_2}+\frac{x_1^{(2)}x_2^{(2)}
   }{y_1^2}+\frac{x_1^{(2)}x_2^{(2)} }{y_2^2}-\frac{x_1^{(2)\,2}x_2^{(2)} }{y_1^2 y_2}-\frac{x_1^{(2)\,2}x_2^{(2)}}{y_1 y_2^2}-\frac{x_1^{(2)}x_2^{(2)\,2} }{y_1^2 y_2}-\frac{x_1^{(2)}x_2^{(2)\,2} }{y_1 y_2^2}+\frac{x_1^{(2)\,2}x_2^{(2)\,2} }{y_1^2 y_2^2}$}&\parbox{7cm}{\centering$1-\frac{x_1}{y_1}-\frac{x_1}{y_2}-\frac{x_2}{y_1}-\frac{x_2}{y_2}+\frac{x_1^{2}}{y_1 y_2}+\frac{2 x_1 x_2 }{y_1 y_2}+\frac{x_2^{2}}{y_1 y_2}+\frac{x_1x_2
   }{(1-q_1)\,y_1^2}+\frac{x_1x_2 }{(1-q_1)\,y_2^2}-\frac{x_1^{2}x_2 }{(1-q_1)\,y_1^2 y_2}-\frac{x_1^{2}x_2}{(1-q_1)\,y_1 y_2^2}-\frac{x_1x_2^{2} }{(1-q_1)\,y_1^2 y_2}-\frac{x_1x_2^{2} }{(1-q_1)\,y_1 y_2^2}+\frac{x_1^{2}x_2^{2} }{(1-q_1)^2\,y_1^2 y_2^2}$}\\
    \hline
      $(4\,2\,1\,3)$   & \parbox{7cm}{\centering$\frac{1}{y_1^2 y_2 y_3}(x^{(2)}_1-y_1)(x^{(2)}_2-y_1)(x^{(1)}_1  x^{(2)}_1 + x^{(1)}_1  x^{(2)}_2 -x^{(1)}_1  y_2 - x^{(1)}_1 y_3  - x^{(2)}_1 x^{(2)}_2 + y_2 y_3 )$ } &\parbox{7cm}{\centering $1-\frac{x_1}{y_1}-\frac{x_1}{y_2}-\frac{x_1}{y_3}-\frac{x_2}{y_1}+\frac{x_1^2}{y_1 y_2}+\frac{x_1^2}{y_1 y_3}+\frac{x_1^2}{y_2 y_3}+\frac{x_1 x_2}{y_1 y_2}+\frac{x_1 x_2}{y_1 y_3}+\frac{x_1 x_2}{\left(1-q_1\right) y_1^2}-\frac{q_1 x_1 x_2}{\left(1-q_1\right) y_2 y_3} -\frac{x_2 x_1^2}{\left(1-q_1\right) y_1^2 y_2}-\frac{x_2 x_1^2}{\left(1-q_1\right) y_1^2 y_3}-\frac{\left(1-2 q_1\right) x_2 x_1^2}{\left(1-q_1\right) y_1 y_2 y_3}+\frac{q_1 x_2^2 x_1}{\left(1-q_1\right) y_1 y_2 y_3} -\frac{x_1^3}{y_1 y_2 y_3}+ \frac{x_1^3x_2 }{\left(1-q_1\right) y_1^2 y_2 y_3}-\frac{q_1 x_1^2x_2^2 }{(1-q_1)^2 y_1^2 y_2 y_3}$}\\
    \hline
      $(4\,3\,1\,2)$   & \parbox{7cm}{\centering $- \frac{(x^{(1)}_1 - y_3)}{y_1^2 y_2^2 y_3} (x^{(2)}_1 - y_1)(x^{(2)}_2 - y_1) (x^{(2)}_1 - y_2)(x^{(2)}_2 - y_2)$ } & \parbox{7cm}{\centering $1-\frac{x_1}{y_1}-\frac{x_1}{y_2}-\frac{x_1}{y_3}-\frac{x_2}{y_1}-\frac{x_2}{y_2} + \frac{x_2 x_1}{\left(1-q_1\right) y_1^2}+\frac{x_2 x_1}{\left(1-q_1\right) y_2^2}+\frac{x_1^2}{y_1 y_2}+\frac{x_1^2}{y_1 y_3}+\frac{x_1^2}{y_2 y_3}+\frac{2 x_2 x_1}{y_1 y_2}+\frac{x_2 x_1}{y_1 y_3}+\frac{x_2 x_1}{y_2 y_3}+\frac{x_2^2}{y_1 y_2} -\frac{x_2 x_1^2}{(1-q_1) y_1^2 y_2}-\frac{x_2 x_1^2}{(1-q_1) y_1 y_2^2}-\frac{x_2 x_1^2}{(1-q_1) y_1^2 y_3}-\frac{x_2 x_1^2}{(1-q_1) y_2^2 y_3}-\frac{x_2^2 x_1}{(1-q_1) y_1^2 y_2}-\frac{x_2^2 x_1}{(1-q_1) y_1 y_2^2}-\frac{x_1^3}{y_1 y_2 y_3}-\frac{2 x_2 x_1^2}{y_1 y_2 y_3}-\frac{x_2^2 x_1}{y_1 y_2 y_3}+\frac{x_1^3 x_2}{(1-q_1) y_1^2 y_2 y_3}+\frac{x_1^3 x_2}{(1-q_1) y_1 y_2^2 y_3}+\frac{x_1^2 x_2^2}{(1-q_1) y_1^2 y_2 y_3}+\frac{x_1^2 x_2^2 }{(1-q_1) y_1 y_2^2 y_3}+\frac{x_1^2 x_2^2}{(1-q_1)^2 y_1^2 y_2^2} - \frac{x_1^3 x_2^2}{(1-q_1)^2 y_1^2 y_2^2 y_3}$} \\
    \hline
    \end{longtable}
\end{center}

\subsection*{0d partition functions and parabolic quantum Schubert polynomials}
\begin{center}
    \renewcommand{\arraystretch}{1.1}
    \begin{longtable}[!h]{|c||c|c|}
    \hline
    $w$&$\mathcal{I}_w^{(\rm 0d)}(\widetilde{\sigma},m)$&$\mathcal{I}_w^{(\rm 0d)}{\scriptsize\begin{bmatrix}
        \qcoh_1&\qcoh_2\\
        1&2
    \end{bmatrix}}(\sigma,m)$\\
    \hline
    \hline
    $(1\,3\,2\,4)$   & $\widetilde{\sigma}_1^{(2)}+\widetilde{\sigma}_2^{(2)}-m_1-m_2$ & $\sigma_1 + \sigma_2 -m_1-m_2$\\
        \hline
      $(2\,1\,3\,4)$   & \parbox{6cm}{\centering $\widetilde{\sigma}_1^{(1)}-m_1$} & \parbox{6cm}{\centering $\sigma_1-m_1$}\\
    \hline 
    $(1\,4\,2\,3)$   & \parbox{6cm}{\centering $\widetilde{\sigma}_2^{(2)} (\widetilde{\sigma}_1^{(2)}-m_2-m_3)+(\widetilde{\sigma}_1^{(2)}-m_2) (\widetilde{\sigma}_1^{(2)}-m_3)-m_1 (\widetilde{\sigma}_1^{(2)}+\widetilde{\sigma}_2^{(2)}-m_2-m_3)+\widetilde{\sigma}_2^{(2)\,2}$}&\parbox{6cm}{\centering$\sigma _2 \left(\sigma _1-m_2-m_3\right)+\left(\sigma _1-m_2\right) \left(\sigma _1-m_3\right)-m_1 \left(\sigma _1+\sigma _2-m_2-m_3\right)+\sigma _2^2-\qcoh_1$}\\
      \hline
      $(2\,3\,1\,4)$   &\parbox{6cm}{\centering $(\widetilde{\sigma}_1^{(2)}-m_1) (\widetilde{\sigma}_2^{(2)}-m_1)$} &\parbox{6cm}{\centering $(\sigma_1-m_1) (\sigma_2-m_1)+\qcoh_1$} \\
          \hline
      $(3\,1\,2\,4)$   & \parbox{6cm}{\centering $\widetilde{\sigma}_1^{(1)} (\widetilde{\sigma}_1^{(2)}-m_2)+(\widetilde{\sigma}_1^{(1)}-\widetilde{\sigma}_1^{(2)}) \widetilde{\sigma}_2^{(2)}-m_1(\widetilde{\sigma}_1^{(1)}-m_2)$}&\parbox{6cm}{\centering $\left(\sigma _1-m_1\right) \left(\sigma _1-m_2\right)-\qcoh_1$}\\
    \hline
      $(2\,4\,1\,3)$   &\parbox{6cm}{\centering $(\widetilde{\sigma}_1^{(2)}-m_1) (\widetilde{\sigma}_2^{(2)}-m_1) (\widetilde{\sigma}_1^{(2)}+\widetilde{\sigma}_2^{(2)}-m_2-m_3)$} &\parbox{6cm}{\centering $((\sigma_1-m_1) (\sigma_2-m_1)+\qcoh_1) (\sigma_1+\sigma_2-m_2-m_3)$}\\
          \hline
      $(4\,1\,2\,3)$   & \parbox{6cm}{\centering $(\widetilde{\sigma}_1^{(2)}-\widetilde{\sigma}_1^{(1)}) \widetilde{\sigma}_2^{(2)} (m_2+m_3-\widetilde{\sigma}_1^{(2)})+\widetilde{\sigma}_1^{(1)} (m_2-\widetilde{\sigma}_1^{(2)}) (m_3-\widetilde{\sigma}_1^{(2)})+m_1 (m_2 (\widetilde{\sigma}_1^{(1)}-m_3)+\widetilde{\sigma}_1^{(1)} (m_3-\widetilde{\sigma}_1^{(2)})+(\widetilde{\sigma}_1^{(2)}-\widetilde{\sigma}_1^{(1)}) \widetilde{\sigma}_2^{(2)})+(\widetilde{\sigma}_1^{(1)}-\widetilde{\sigma}_1^{(2)}) \widetilde{\sigma}_2^{(2)\,2}$}&\parbox{6cm}{\centering $m_2 (\sigma _1 (m_3-\sigma _1)+\qcoh_1)+m_1 ((m_2-\sigma _1) (\sigma _1-m_3)+\qcoh_1)+m_3 \qcoh_1-m_3 \sigma _1^2-2 \qcoh_1 \sigma _1-\qcoh_1 \sigma _2+\sigma _1^3$}\\
    \hline
      $(3\,2\,1\,4)$   & \parbox{6cm}{\centering $(\widetilde{\sigma}^{(1)}_1-m_2) (\widetilde{\sigma}_1^{(2)}-m_1) (\widetilde{\sigma}_2^{(2)}-m_1)$} &\parbox{6cm}{\centering $(\sigma _1-m_2) ((\sigma _1-m_1) (\sigma _2-m_1)+\qcoh_1)$}\\
    \hline 
    $(3\,4\,1\,2)$   &\parbox{6cm}{\centering $(\widetilde{\sigma}_1^{(2)}-m_1) (\widetilde{\sigma}_1^{(2)}-m_2) (\widetilde{\sigma}_2^{(2)}-m_1) (\widetilde{\sigma}_2^{(2)}-m_2)$} &\parbox{6cm}{\centering $((\sigma _1-m_1) (\sigma _2-m_1)+\qcoh_1) ((\sigma _1-m_2) (\sigma _2-m_2)+\qcoh_1)$}\\
    \hline
      $(4\,2\,1\,3)$   & \parbox{6cm}{\centering $(\widetilde{\sigma}^{(2)}_1 - m_1)(\widetilde{\sigma}^{(2)}_2 - m_1) (\widetilde{\sigma}^{(1)}_1 \widetilde{\sigma}^{(2)}_1 + \widetilde{\sigma}^{(1)}_1 \widetilde{\sigma}^{(2)}_2 - \widetilde{\sigma}^{(2)}_1 \widetilde{\sigma}^{(2)}_2 - \widetilde{\sigma}^{(1)}_1 m_2 - \widetilde{\sigma}^{(1)}_1 m_3 + m_2 m_3) $ } & \parbox{6cm}{\centering $((\sigma_1 -m_1)(\sigma_2 -m_1)+\qcoh_1)((\sigma_1-m_2)(\sigma_1-m_3)-\qcoh_1)$}\\
    \hline
      $(4\,3\,1\,2)$   & \parbox{6cm}{\centering $(\widetilde{\sigma}^{(1)}_1 - m_3)(\widetilde{\sigma}^{(2)}_1 - m_1)(\widetilde{\sigma}^{(2)}_1 - m_2)(\widetilde{\sigma}^{(2)}_2 - m_1)(\widetilde{\sigma}^{(2)}_2 - m_2)$} & \parbox{6cm}{\centering $(\sigma_1-m_1)(\sigma_1-m_2)(\sigma_1-m_3)(\sigma_2-m_1) (\sigma_2-m_2) - \qcoh_1 (\sigma_1 - m_3)(2\sigma_1\sigma_2 - (\sigma_1+\sigma_2)(m_1+m_2) + m_1^2 +m_2^2 + \qcoh_1)$ }\\
    \hline
    \end{longtable}
\end{center}


\section{Schubert defects in Fl\texorpdfstring{$(1,3;4)$}{134}}\label{app:134}
\subsection*{All possible Schubert varieties}
In this partial flag, we have 12 possible Schubert varieties indexed by the equivalence classes in ${\rm W}^{(1,3;4)}$. These classes and their corresponding dimensions are summarized in the following table:
\begin{table}[!ht]
\centering
\begin{subtable}{0.45\textwidth}
\centering
\begin{equation*}
        \begin{array}{|c||c|c|c|}
    \hline
        w&{\ell(w)} & {\dim(X_w)}\\
        \hline
        \hline
        {(1\,2\,3\,4)} & 0 &5\\
        \hline
        {(1\,2\,4\,3)} &  1 &4\\
        \hline
         {(2\,1\,3\,4)} &  1 &4 \\
        \hline
        {(1\,3\,4\,2)} &  2&3 \\
        \hline
        {(2\,1\,4\,3)} &  2&3\\
        \hline
        {(3\,1\,2\,4)} & 2&3\\

        \hline
        \end{array}
    \end{equation*}
\end{subtable}
\hfill
\begin{subtable}{0.45\textwidth}
\centering
\begin{equation*}
        \begin{array}{|c||c|c|c|}
    \hline
         w&{\ell(w)}&{\dim(X_w)}  \\
        \hline
        \hline
         {(2\,3\,4\,1)} & 3 &2 \\
            \hline
        {(3\,1\,4\,2)} &  3&2\\
        \hline
        {(4\,1\,2\,3)} &  3&2\\
        \hline
        {(3\,2\,4\,1)} &  4&1\\
        \hline
        {(4\,1\,3\,2)} &  4&1 \\
        \hline
        {(4\,2\,3\,1)} & 5 &0\\
        \hline
        \end{array}
    \end{equation*}
\end{subtable}
\end{table}

These Schubert varieties are connected via the following Hasse diagram:
\begin{figure}[h!]
    \centering
    \includegraphics[width=0.75\linewidth]{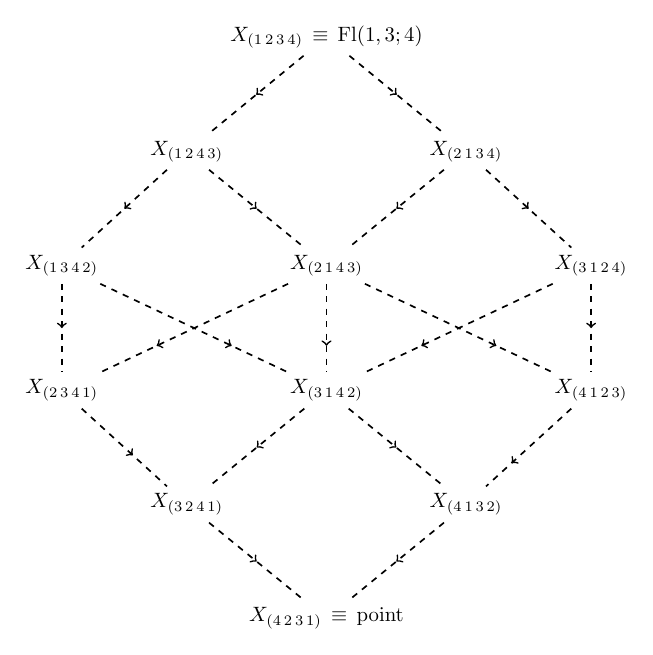}
\end{figure}

\newpage
\subsection*{Coupled systems defining the Schubert defects}\label{subsec:defects134}
For each one of the 12 permutation classes $[w]\in {\rm W}^{(1,3;4)}$, let us now write down the 1d-3d (0d-2d) coupled system defining the Schubert line (point) defect. Here we follow the conventions of figure \ref{fig:GenProposal} for the general proposal. Viewed as Schubert line defects, in the last column, we also include the reduced form of the quiver after applying the duality moves reviewed in section 2 of \cite{Closset:2025cfm}.
    \begin{table}[!h]
\centering
\begin{subtable}{0.45\textwidth}
\centering
\begin{equation*}
        \begin{array}{|c||c|c|c|}
    \hline
        w&{\rm General\,\,proposal} & {\rm Reduced \,\,quiver}\\
        \hline
        \hline
        \raisebox{6ex}{(1\,2\,3\,4)}   & \includegraphics[scale=0.3]{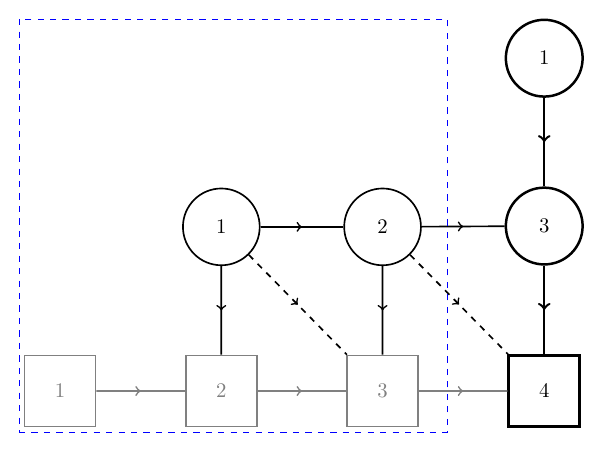}&\includegraphics[scale=0.3]{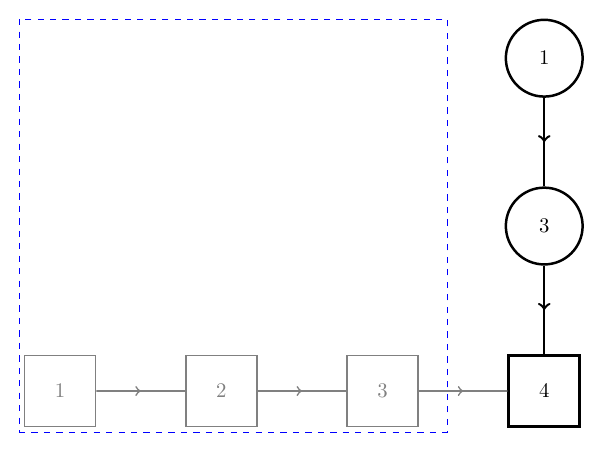}\\
           \hline
           \raisebox{6ex}{(1\,2\,4\,3)}   & \includegraphics[scale=0.3]{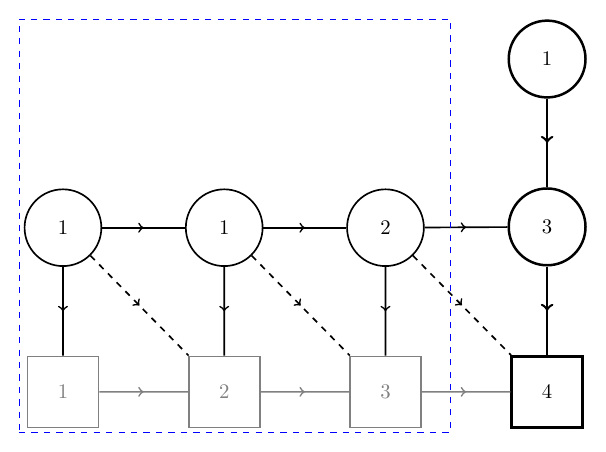}& \includegraphics[scale=0.3]{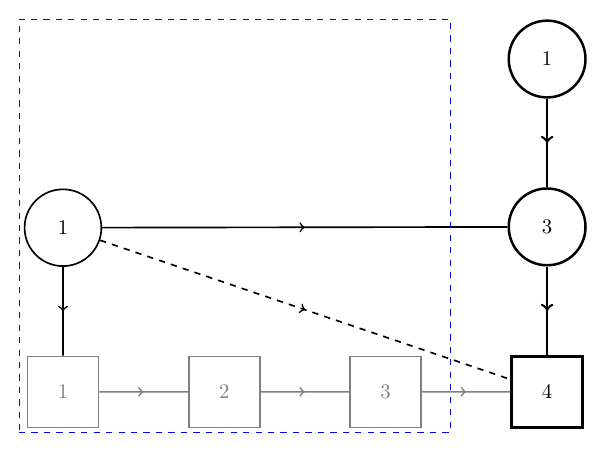}\\
           \hline
        \raisebox{6ex}{(2\,1\,3\,4)}   & \includegraphics[scale=0.3]{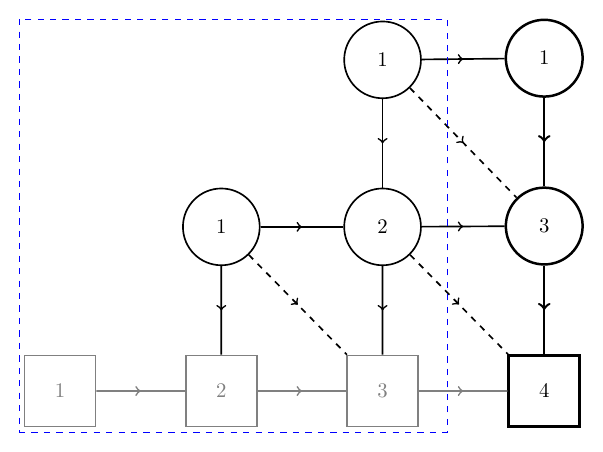}& \includegraphics[scale=0.3]{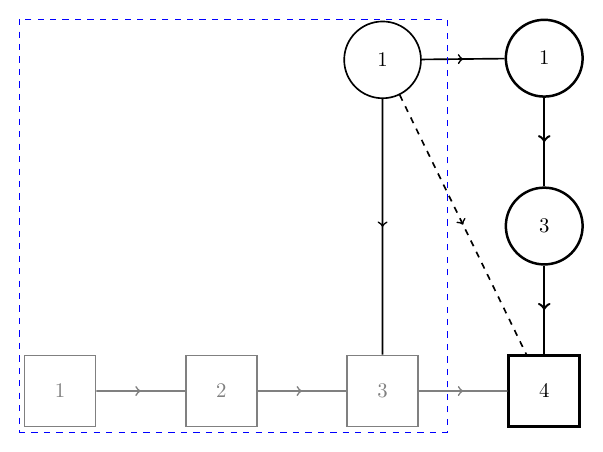}\\
            \hline
           \raisebox{6ex}{(1\,3\,4\,2)} &\includegraphics[scale = 0.3]{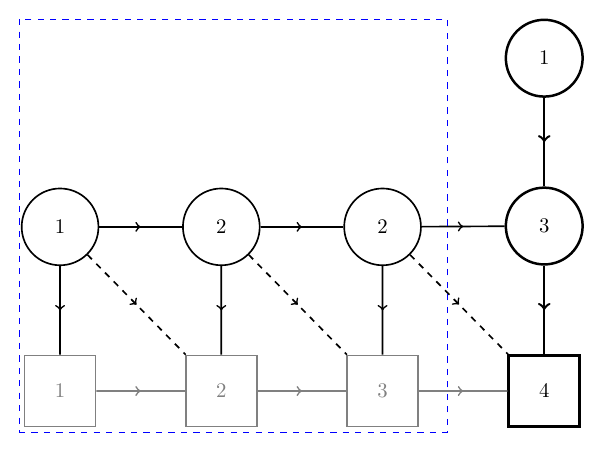}&\includegraphics[scale = 0.3]{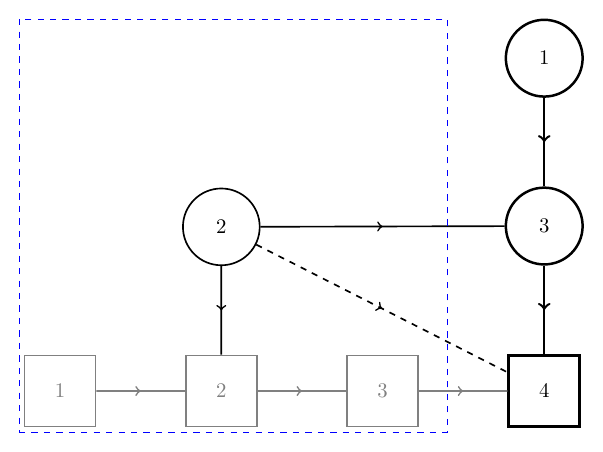}\\
           \hline
           \raisebox{6ex}{(2\,1\,4\,3)} &\includegraphics[scale = 0.3]{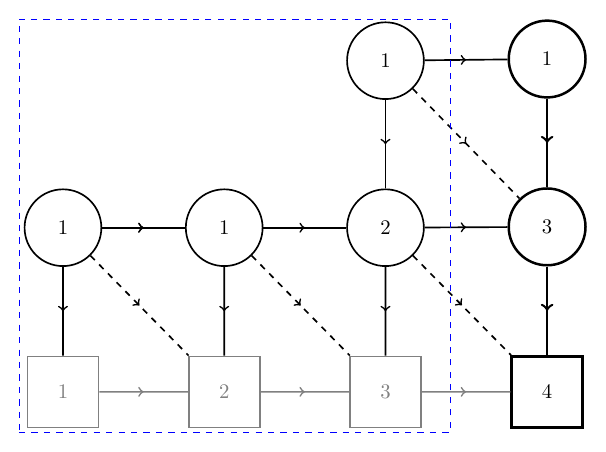} & \includegraphics[scale=0.3]{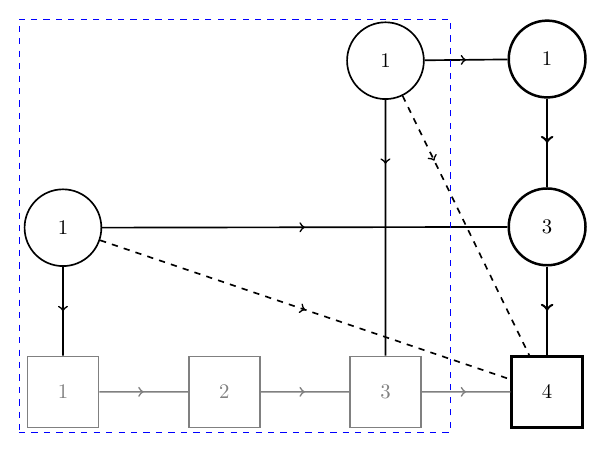}\\
                   \hline
             \raisebox{6ex}{(3\,1\,2\,4)} &\includegraphics[scale = 0.3]{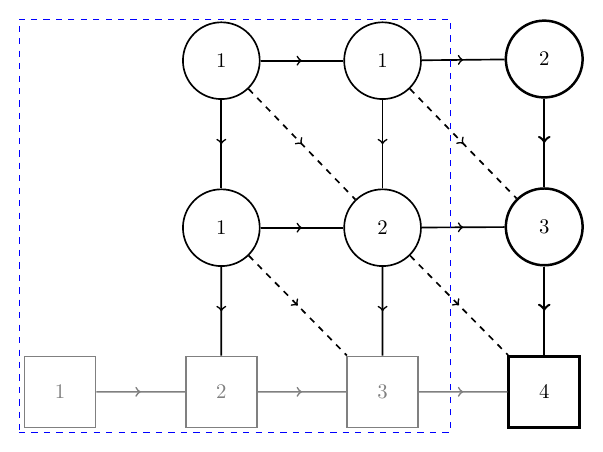}&\includegraphics[scale = 0.3]{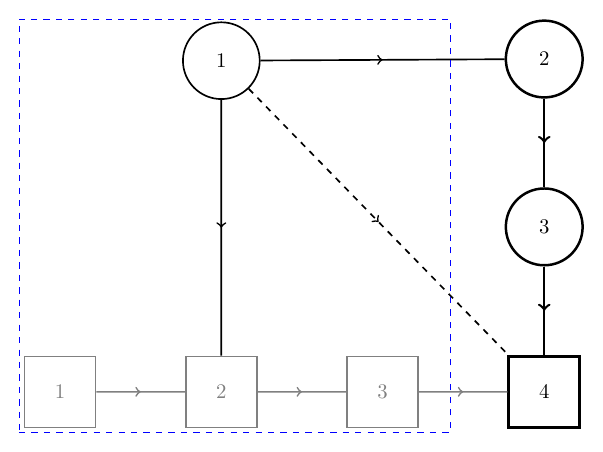}\\
           \hline
           
        \end{array}
    \end{equation*}
\end{subtable}
\hfill
\begin{subtable}{0.45\textwidth}
\centering
\begin{equation*}
        \begin{array}{|c||c|c|c|}
    \hline
        w&{\rm General\,\,proposal}&  {\rm Reduced \,\,quiver}\\
        \hline
        \hline
           \raisebox{6ex}{(2\,3\,4\,1)} &\includegraphics[scale = 0.3]{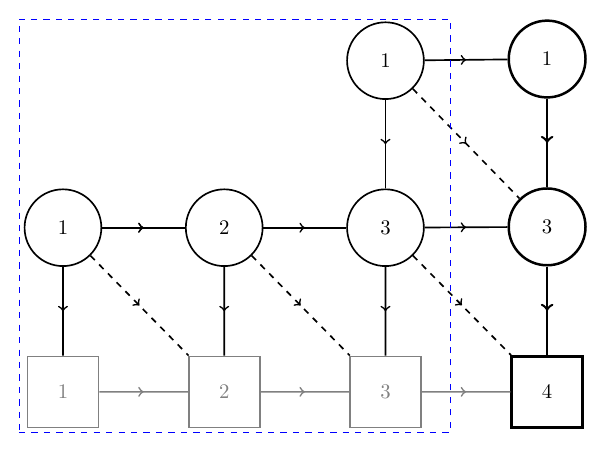} &\includegraphics[scale=0.3]{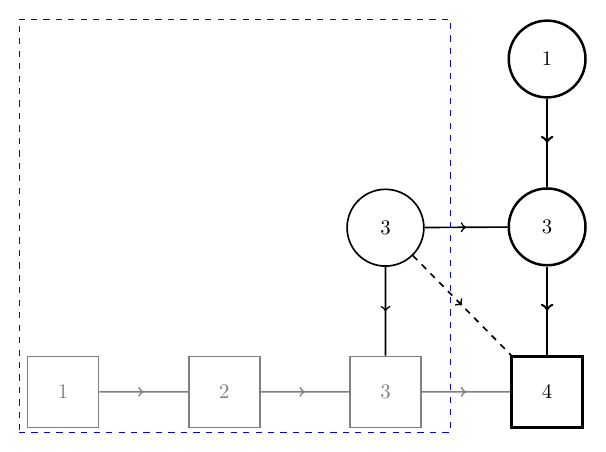}\\
                      \hline
           \raisebox{6ex}{(3\,1\,4\,2)}   & \includegraphics[scale=0.3]{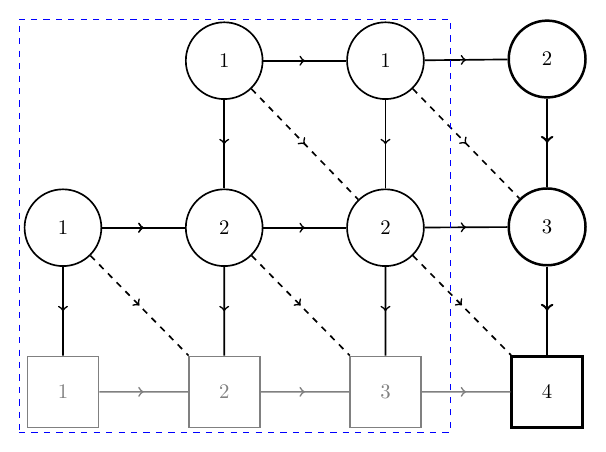}&\includegraphics[scale=0.3]{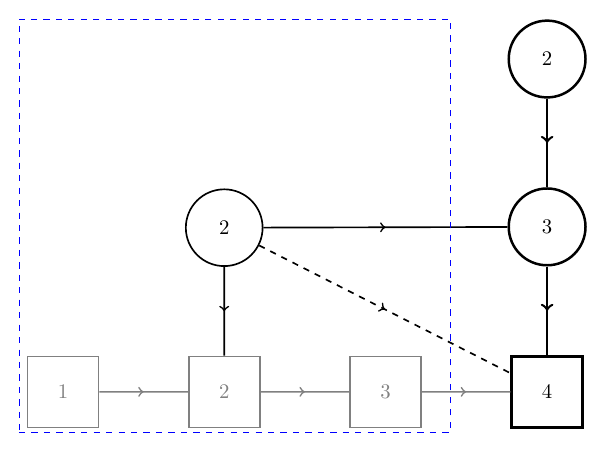}\\
               \hline
           \raisebox{6ex}{(4\,1\,2\,3)}&\includegraphics[scale=0.3]{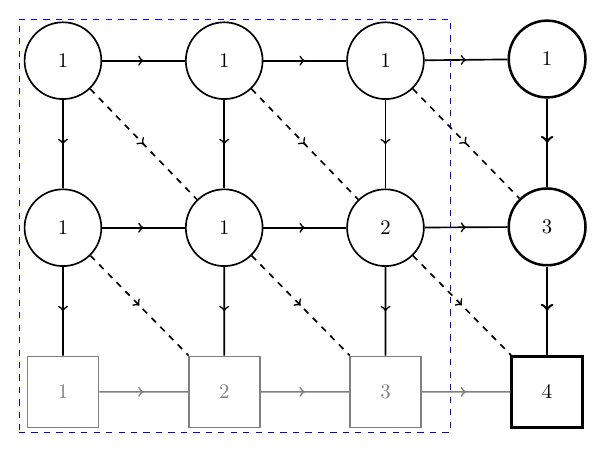} & \includegraphics[scale=0.3]{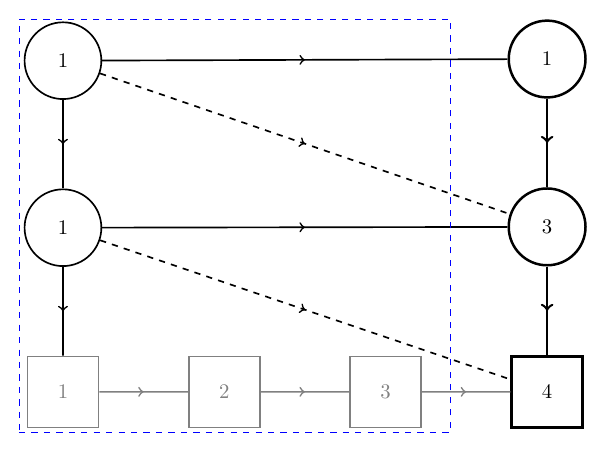} \\
                      \hline
          \raisebox{6ex}{(3\,2\,4\,1)}   &\includegraphics[scale=0.3]{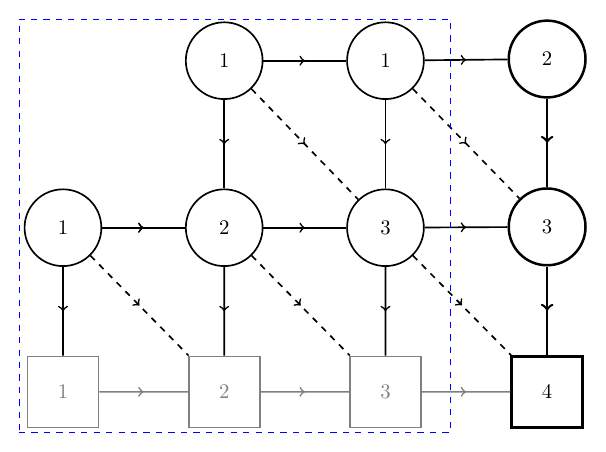} &\includegraphics[scale=0.3]{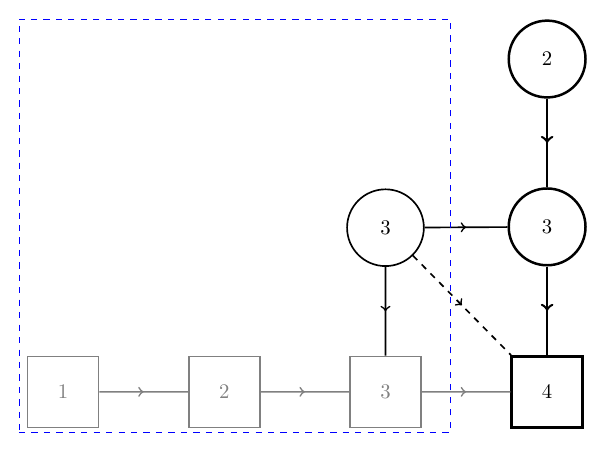}\\
           \hline
                \raisebox{6ex}{(4\,1\,3\,2)}   &\includegraphics[scale=0.3]{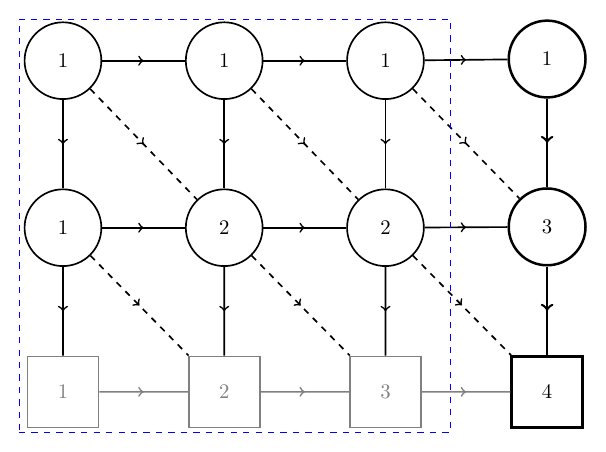} &  \includegraphics[scale=0.3]{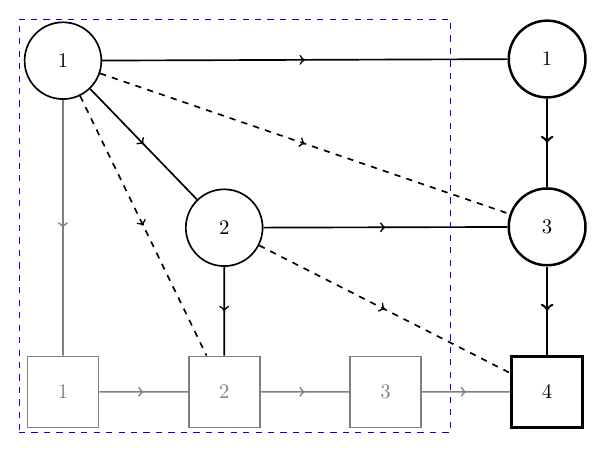}\\
           \hline
  \raisebox{6ex}{(4\,2\,3\,1)}  &\includegraphics[scale=0.3]{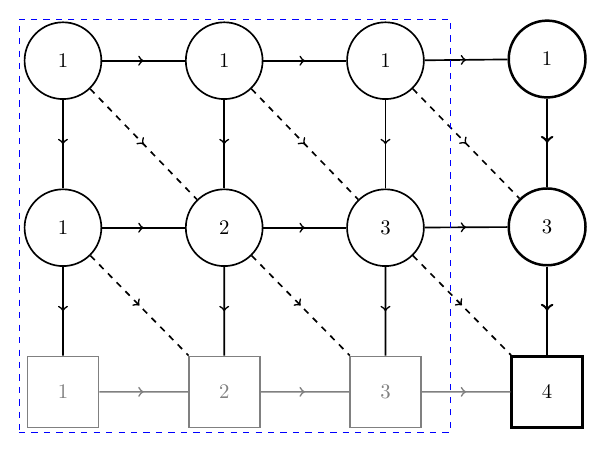} & \includegraphics[scale=0.3]{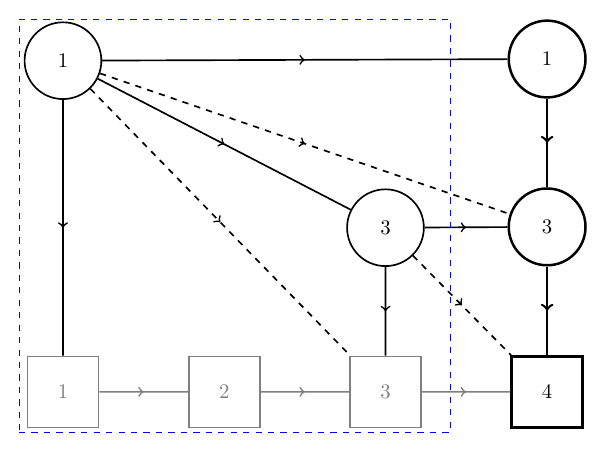}\\
           \hline
        \end{array}
    \end{equation*}
\end{subtable}
\end{table}

\newpage
\subsection*{1d indices and parabolic quantum Grothendieck polynomials}
\begin{center}
\begin{longtable}[!h]{|c||c|c|}
    \hline
    $w$&$\mathcal{I}_w^{(\rm 1d)}(x,y)$&$\mathcal{I}_w^{(\rm 1d)}{\scriptsize\begin{bmatrix}
        \qk_1&\qk_2\\
        1&3
    \end{bmatrix}}(x,y)$\\
    \hline
    \hline
    $(1\,2\,4\,3)$ & \raisebox{0ex}{\parbox{7cm}{\centering $1-\frac{x_1^{(2)} x_2^{(2)} x_3^{(2)}}{y_1 y_2 y_3}$}} &\raisebox{0ex}{\parbox{6.5cm}{\centering $1-\frac{x_1 x_2 x_3}{(1-q_1)\,y_1 y_2 y_3}$}} \\
        \hline
        $(2\,1\,3\,4)$ & \raisebox{0ex}{\parbox{7cm}{\centering $1-\frac{x_1^{(1)}}{y_1}$}} &\raisebox{0ex}{\parbox{6.5cm}{\centering $1-\frac{x_1 }{y_1}$}} \\
                \hline
        $(1\,3\,4\,2)$ &\parbox{7cm}{\centering $1 - \frac{x_1^{(2)} x_2^{(2)}}{
 y_1 y_2} - \frac{x_1^{(2)} x_3^{(2)}}{y_1 y_2} - \frac{
 x_2^{(2)} x_3^{(2)}}{y_1 y_2}+ \frac{x_1^{(2)} x_2^{(2)} x_3^{(2)}}{y_1 y_2^2} + \frac{x_1^{(2)} x_2^{(2)} x_3^{(2)}}{y_1^2 y_2} $} & \parbox{6.5cm}{\centering $1 - \frac{x_1 x_2}{
 y_1 y_2} - \frac{x_1 x_3}{y_1 y_2} - \frac{
 x_2 x_3}{y_1 y_2}+ \frac{x_1 x_2 x_3}{(1-\qk_1)\, y_1 y_2^2} + \frac{x_1 x_2 x_3}{(1-\qk_1)\,y_1^2 y_2} $}\\
        \hline
        $(2\,1\,4\,3)$& \raisebox{0ex}{\parbox{7cm}{\centering $1-\frac{x_1^{(1)}}{y_1}-\frac{x_1^{(2)} x_2^{(2)} x_3^{(2)}}{y_1 y_2 y_3}+\frac{x_1^{(1)} x^{(2)}_1 x_2^{(2)} x_3^{(2)} }{y_1^2 y_2 y_3}$}} & \raisebox{0ex}{\parbox{6.5cm}{\centering $1-\frac{x_1}{y_1}-\frac{x_1 x_2 x_3}{(1-\qk_1) \,y_1 y_2 y_3}+\frac{x_1^2 x_2 x_3 }{(1-\qk_1)\,y_1^2 y_2 y_3}$}}\\
 \hline
 $(3\,1\,2\,4)$& $\left(1-\frac{x_1^{(1)}}{y_1}\right)\left(1-\frac{x_1^{(1)}}{y_2}\right)$&$\left(1-\frac{x_1}{y_1}\right)\left(1-\frac{x_1}{y_2}\right) $\\
  \hline
        $(2\,3\,4\,1)$&\raisebox{0ex}{\parbox{7cm}{\centering $1-\frac{x_1^{(2)}}{y_1}-\frac{x_2^{(2)}}{y_1}-\frac{x_3^{(2)}}{y_1}+\frac{x_1^{(2)} x_2^{(2)} }{y_1^2}+\frac{x^{(2)}_1 x_3^{(2)} }{y_1^2}+\frac{x_2^{(2)} x^{(2)}_3}{y_1^2}-\frac{x_1^{(2)} x_2^{(2)} x_3^{(2)} }{y_1^3}$}}&\raisebox{0ex}{\parbox{6.5cm}{\centering $1-\frac{x_1}{y_1}-\frac{x_2}{y_1}-\frac{x_3}{y_1}+\frac{x_1x_2 }{y_1^2}+\frac{x_1 x_3 }{y_1^2}+\frac{x_2 x_3}{y_1^2}-\frac{x_1 x_2 x_3 }{(1-\qk_1)\,y_1^3}$}}\\
                \hline
        $(3\,1\,4\,2)$ &\parbox{7cm}{\centering $1-\frac{x_1^{(1)}}{y_1}-\frac{x_1^{(1)}}{y_2}+\frac{x_1^{(1)}x_1^{(2)} }{y_1y_2}+\frac{x_1^{(1)} x_2^{(2)}}{y_1y_2}+\frac{x_1^{(1)}x_3^{(2)}}{y_1 y_2}-\frac{x_1^{(2)} x_2^{(2)}}{y_1 y_2}-\frac{x_1^{(2)} x_3^{(2)}}{y_1 y_2}-\frac{x_2^{(2)} x_3^{(2)}}{y_1 y_2}+\frac{x_1^{(2)}
   x_2^{(2)} x_3^{(2)}}{y_1^2 y_2}+\frac{x_1^{(2)} x_2^{(2)} x_3^{(2)}}{y_1 y_2^2}-\frac{x_1^{(1)}x_1^{(2)}
   x_2^{(2)} x_3^{(2)} }{y_1^2 y_2^2}$} &\parbox{6.5cm}{\centering $1-\frac{x_1}{y_1}-\frac{x_1}{y_2}+\frac{x_1^2 }{y_1y_2}+\frac{x_1 x_2}{y_1y_2}+\frac{x_1x_3}{y_1 y_2}-\frac{x_1 x_2}{y_1 y_2}-\frac{x_1 x_3}{y_1 y_2}-\frac{x_2 x_3}{y_1 y_2}+\frac{x_1
   x_2 x_3}{(1-\qk_1)\,y_1^2 y_2}+\frac{x_1 x_2 x_3}{(1-\qk_1)\,y_1 y_2^2}-\frac{x_1^2
   x_2 x_3 }{(1-\qk_1)\,y_1^2 y_2^2}$}\\
              \hline
        $(4\,1\,2\,3)$&\parbox{7cm}{\centering $1-\frac{x_1^{(1)}}{y_1}-\frac{x_1^{(1)}}{y_2}-\frac{x_1^{(1)}}{y_3}+\frac{x_1^{(1)\,2}}{y_1 y_2}+\frac{x_1^{(1)\,2}}{y_1 y_3}+\frac{x_1^{(1)\,2}}{y_2 y_3}-\frac{x_1^{(1)\,2}x_1^{(2)} }{y_1 y_2 y_3}-\frac{x_1^{(1)\,2}x_2^{(2)} }{y_1 y_2 y_3}-\frac{x_1^{(1)\,2}x_3^{(2)}
   }{y_1 y_2 y_3}+\frac{x_1^{(1)}x_1^{(2)} x_2^{(2)} }{y_1 y_2
   y_3}+\frac{x_1^{(1)}x_1^{(2)} x_3^{(2)} }{y_1 y_2 y_3}+\frac{x_1^{(1)}x_2^{(2)} x_3^{(2)} }{y_1 y_2 y_3}-\frac{x_1^{(2)} x_2^{(2)} x_3^{(2)}}{y_1
   y_2 y_3}$}&\parbox{6.5cm}{\centering $1-\frac{x_1}{y_1}-\frac{x_1}{y_2}-\frac{x_1}{y_3}+\frac{x_1^{2}}{y_1 y_2}+\frac{x_1^{2}}{y_1 y_3}+\frac{x_1^{2}}{y_2 y_3}-\frac{x_1^{3} }{y_1 y_2 y_3}-\frac{\qk_1 x_1 x_2 x_3}{(1-\qk_1)\,y_1
   y_2 y_3}$}\\
 \hline
        $(3\,2\,4\,1)$ &\parbox{7cm}{\centering $1-\frac{x_1^{(1)}}{y_2}-\frac{x_1^{(2)}}{y_1}-\frac{x_2^{(2)}}{y_1}-\frac{x_3^{(2)}}{y_1}+\frac{x_1^{(1)} x_1^{(2)} }{y_1 y_2}+\frac{x_1^{(1)}x_2^{(2)} }{y_1 y_2}+\frac{x_1^{(1)}x_3^{(2)} }{y_1 y_2}+\frac{x_1^{(2)} x_2^{(2)}}{y_1^2}+\frac{x_1^{(2)}
   x_3^{(2)}}{y_1^2}+\frac{x_2^{(2)} x_3^{(2)}}{y_1^2}-\frac{x_1^{(1)}x_1^{(2)} x_2^{(2)} }{y_1^2 y_2}-\frac{x_1^{(1)}x_1^{(2)}
   x_3^{(2)} }{y_1^2 y_2}-\frac{x_1^{(1)}x_2^{(2)} x_3^{(2)} }{y_1^2 y_2}-\frac{x_1^{(2)} x_2^{(2)} x_3^{(2)}}{y_1^3}+\frac{ x_1^{(1)}x_1^{(2)} x_2^{(2)} x_3^{(2)}}{y_1^3
   y_2}$} & \parbox{6.5cm}{\centering $1-\frac{x_1}{y_2}-\frac{x_1}{y_1}-\frac{x_2}{y_1}-\frac{x_3}{y_1}+\frac{x_1^2 }{y_1 y_2}+\frac{x_1x_2 }{y_1 y_2}+\frac{x_1x_3 }{y_1 y_2}+\frac{x_1 x_2}{y_1^2}+\frac{x_1
   x_3}{y_1^2}+\frac{x_2 x_3}{y_1^2}-\frac{x_1^2 x_2 }{y_1^2 y_2}-\frac{x_1^2
   x_3 }{y_1^2 y_2}-\frac{x_1x_2 x_3 }{y_1^2 y_2}-\frac{x_1 x_2 x_3}{(1-\qk_1)\,y_1^3}+\frac{ x_1^2 x_2 x_3}{(1-\qk_1)\,y_1^3
   y_2}$}\\
           \hline
        $(4\,1\,3\,2)$& \parbox{7cm}{\centering $1-\frac{x_1^{(1)}}{y_1}-\frac{x_1^{(1)}}{y_2}-\frac{x_1^{(1)}}{y_3}+\frac{x_1^{(1)\,2}}{y_1 y_3}+\frac{x_1^{(1)\,2}}{y_2 y_3}+\frac{x_1^{(1)}x_1^{(2)} }{y_1
   y_2}+\frac{x_1^{(1)}x_2^{(2)} }{y_1 y_2}+\frac{ x_1^{(1)}x_3^{(2)}}{y_1 y_2}-\frac{x_1^{(2)} x_{2}^{(2)}}{y_1 y_2}-\frac{x_1^{(2)}
   x_3^{(2)}}{y_1 y_2}-\frac{x_2^{(2)} x_3^{(2)}}{y_1 y_2}-\frac{x_1^{(1)\,2}x_1^{(2)} }{y_1 y_2 y_3}-\frac{x_1^{(1)\,2}x_2^{(2)} }{y_1 y_2 y_3}-\frac{x_1^{(1)\,2}x_3^{(2)} }{y_1 y_2
   y_3}+\frac{x_1^{(1)}x_1^{(2)} x_2^{(2)} }{y_1 y_2 y_3}+\frac{x_1^{(1)}x_1^{(2)}
   x_3^{(2)} }{y_1 y_2 y_3}+\frac{ x_1^{(1)}x_2^{(2)} x_3^{(2)}}{y_1 y_2 y_3}+\frac{x_1^{(2)} x_2^{(2)} x_3^{(2)}}{y_1^2 y_2}+\frac{x_1^{(2)} x_2^{(2)} x_3^{(2)}}{y_1 y_2^2}-\frac{x_1^{(1)}x_1^{(2)} x_2^{(2)} x_3^{(2)} }{y_1^2 y_2 y_3}-\frac{x_1^{(1)}x_1^{(2)} x_2^{(2)}
   x_3^{(2)} }{y_1 y_2^2 y_3}-\frac{x_1^{(1)}x_1^{(2)} x_2^{(2)} x_3^{(2)} }{y_1^2 y_2^2}+\frac{x_1^{(1)\,2}x_1^{(2)} x_2^{(2)} x_3^{(2)} }{y_1^2 y_2^2 y_3}$} &\parbox{6.5cm}{\centering $1-\frac{x_1}{y_1}-\frac{x_1}{y_2}-\frac{x_1}{y_3}+\frac{x_1^{2}}{y_1 y_3}+\frac{x_1^{2}}{y_2 y_3}+\frac{x_1^2 }{y_1
   y_2}+\frac{x_1x_2 }{y_1 y_2}+\frac{ x_1x_3}{y_1 y_2}-\frac{x_1 x_{2}}{y_1 y_2}-\frac{x_1
   x_3}{y_1 y_2}-\frac{x_2 x_3}{y_1 y_2}-\frac{x_1^{3} }{y_1 y_2 y_3}-\frac{x_1^{2}x_2 }{y_1 y_2 y_3}-\frac{x_1^{2}x_3 }{y_1 y_2
   y_3}+\frac{x_1^{2} x_2 }{y_1 y_2 y_3}+\frac{x_1^2
   x_3 }{y_1 y_2 y_3}+\frac{ x_1x_2x_3}{y_1 y_2 y_3}+\frac{x_1 x_2 x_3}{(1-\qk_1)\,y_1^2 y_2}+\frac{x_1 x_2 x_3}{(1-\qk_1)\,y_1 y_2^2}-\frac{x_1^{2} x_2 x_3 }{(1-\qk_1)\,y_1^2 y_2 y_3}-\frac{x_1^2 x_2
   x_3}{(1-\qk_1)\,y_1 y_2^2 y_3}-\frac{x_1^2 x_2 x_3}{(1-\qk_1)\,y_1^2 y_2^2}+\frac{x_1^{3} x_2x_3}{(1-\qk_1)\,y_1^2 y_2^2 y_3}$}\\
        \hline
        $(4\,2\,3\,1)$ &\parbox{7cm}{\centering $1-\frac{x_1^{(1)}}{y_2}-\frac{x_1^{(1)}}{y_3}-\frac{x_1^{(2)}}{y_1}-\frac{x_2^{(2)}}{y_1}-\frac{x_3^{(2)}}{y_1}+\frac{x_1^{(1)\,2}}{y_2 y_3}+\frac{x_1^{(1)}x_1^{(2)} }{y_1 y_2}+\frac{x_1^{(1)}x_2^{(2)} }{y_1 y_2}+\frac{x_1^{(1)}x_3^{(2)}
   }{y_1 y_2}+\frac{x_1^{(2)}
   x_2^{(2)}}{y_1^2}+\frac{x_1^{(2)} x_3^{(2)}}{y_1^2}+\frac{x_2^{(2)} x_3^{(2)}}{y_1^2} +\frac{x_1^{(1)}x_1^{(2)}}{y_1 y_3}+\frac{x_1^{(1)}x_2^{(2)}}{y_1 y_3}+\frac{x_1^{(1)}x^{(2)}_3
   }{y_1 y_3}-\frac{x_1^{(1)\,2}x_1^{(2)} }{y_1 y_2 y_3}-\frac{x_1^{(1)\,2}x_2^{(2)}}{y_1 y_2 y_3}-\frac{x_1^{(1)\,2}x_3^{(2)}}{y_1 y_2 y_3}-\frac{x_1^{(1)} x_1^{(2)} x_2^{(2)}}{y_1^2 y_2}-\frac{x_1^{(1)}x_1^{(2)} x_3^{(2)}}{y_1^2 y_2}-\frac{x_1^{(1)}x_2^{(2)} x_3^{(2)} }{y_1^2
   y_2}-\frac{x_1^{(1)}x_1^{(2)} x_2^{(2)} }{y_1^2 y_3}-\frac{x_1^{(1)}x_1^{(2)} x_3^{(2)} }{y_1^2 y_3}-\frac{x_1^{(1)}x_2^{(2)} x_3^{(2)} }{y_1^2
   y_3}-\frac{x_1^{(2)} x_2^{(2)} x_3^{(2)}}{y_1^3}+\frac{x_1^{(1)\,2}x_1^{(2)} x_2^{(2)}
   }{y_1^2 y_2 y_3}+\frac{x_1^{(1)\,2} x^{(2)}_1 x^{(2)}_3 }{y_1^2 y_2 y_3}+\frac{x_1^{(1)\,2}x_2^{(2)} x_3^{(2)}}{y_1^2 y_2 y_3}+\frac{x_1^{(1)}x_1^{(2)} x_2^{(2)} x_3^{(2)}}{y_1^3 y_2}+\frac{x_1^{(1)}x_1^{(2)} x_2^{(2)}x_3^{(2)}}{y_1^3 y_3}-\frac{x_1^{(1)\,2}x_1^{(2)} x_2^{(2)}
   x_3^{(2)}}{y_1^3 y_2 y_3}$}&\parbox{6.5cm}{\centering $1-\frac{x_1}{y_2}-\frac{x_1}{y_3}-\frac{x_1}{y_1}-\frac{x_2}{y_1}-\frac{x_3}{y_1}+\frac{x_1^{2}}{y_2 y_3}+\frac{x_1^2}{y_1 y_2}+\frac{x_1x_2}{y_1 y_2}+\frac{x_1x_3
   }{y_1 y_2}+\frac{x_1^2
   x_2}{y_1^2}+\frac{x_1x_3}{y_1^2}+\frac{x_2x_3}{y_1^2} +\frac{x_1^2}{y_1 y_3}+\frac{x_1x_2}{y_1 y_3}+\frac{x_1x_3
   }{y_1 y_3}-\frac{x_1^{3}}{y_1 y_2 y_3}-\frac{x_1^{2}x_2}{y_1 y_2 y_3}-\frac{x_1^{2}x_3}{y_1 y_2 y_3}-\frac{x_1^2  x_2}{y_1^2 y_2}-\frac{x_1^2x_3}{y_1^2 y_2}-\frac{x_1x_2 x_3 }{y_1^2
   y_2}-\frac{x_1^2 x_2}{y_1^2 y_3}-\frac{x_1^2 x_3 }{y_1^2 y_3}-\frac{x_1x_2x_3 }{y_1^2
   y_3}-\frac{x_1 x_2 x_3}{(1-\qk_1)\,y_1^3}+\frac{x_1^{3} x_2
   }{y_1^2 y_2 y_3}+\frac{x_1^{3}x_3 }{y_1^2 y_2 y_3}+\frac{x_1^{2}x_2 x_3}{y_1^2 y_2 y_3}+\frac{x_1^{2} x_2 x_3}{(1-\qk_1)\,y_1^3 y_2}+\frac{x_1^{2} x_2x_3}{(1-\qk_1)\,y_1^3 y_3}-\frac{x_1^{3}x_2
   x_3}{(1-\qk_1)\,y_1^3 y_2 y_3}$}\\
 \hline
\end{longtable}
\end{center}


\subsection*{0d partition functions and parabolic quantum Schubert polynomials}
\begin{center}
    \renewcommand{\arraystretch}{1.1}
    \begin{longtable}[!h]{|c||c|c|}
    \hline
        $w$ & $\mathcal{I}_w^{(\rm 0d)}(\widetilde{\sigma},m)$ &$ \mathcal{I}_w^{(\rm 0d)}{\scriptsize\begin{bmatrix}
            \qcoh_1&\qcoh_2\\1&3
        \end{bmatrix}}(\sigma,m)$\\
        \hline
               \hline
       $(1\,2\,4\,3)$   &$e_1(\widetilde{\sigma}^{(2)}) - e_1^3(m)$
&$e_1^3(\sigma) - e_1^3(m)$\\
       \hline
       $(2\,1\,3\,4)$ &$\widetilde{\sigma}_1^{(1)}-m_1$&$\sigma_1-m_1$\\
       \hline
       $(1\,3\,4\,2)$&\parbox{7cm}{\centering $m_1 (m_2-\widetilde{\sigma}_1^{(2)}-\widetilde{\sigma} _2^{(2)}-\widetilde{\sigma}_3^{(2)})+(m_2-\widetilde{\sigma}_1^{(2)}) (m_2-\widetilde{\sigma}_2^{(2)})+\widetilde{\sigma}_3^{(2)} (-m_2+\widetilde{\sigma}_1^{(2)}+\widetilde{\sigma}_2^{(2)})+m_1^2$}&\parbox{6.6cm}{\centering $m_1 \left(m_2-\sigma _1-\sigma _2-\sigma _3\right)+\left(m_2-\sigma _1\right) \left(m_2-\sigma _2\right)+\sigma _3 \left(-m_2+\sigma _1+\sigma _2\right)+m_1^2$}\\
       \hline
       $(2\,1\,4\,3)$&$(\widetilde{\sigma}_1^{(1)}-m_1)(e_1(\widetilde{\sigma}^{(2)}) - e_1^3(m))$ & $(\sigma_1-m_1)(e_1^3(\sigma) - e_1^3(m))$ \\
       \hline
       $(3\,1\,2\,4)$ &$(\widetilde{\sigma}_1^{(1)}-m_1)(\widetilde{\sigma}_1^{(1)}-m_2)$&$(\sigma_1-m_1)(\sigma_1-m_2)$\\
       \hline
       $(2\,3\,4\,1)$ &$\prod_{a=1}^3 (\widetilde{\sigma}_a^{(2)}-m_1)$&$\prod_{a=1}^3 (\sigma_a-m_1) +\qcoh_1$ \\
        \hline
        $(3\,1\,4\,2)$&\parbox{7cm}{\centering $m_1^2 (\widetilde{\sigma}_1^{(1)}-m_2)-m_1 (\widetilde{\sigma}_1^{(1)}-m_2) (\widetilde{\sigma}_1^{(2)}+\widetilde{\sigma}_2^{(2)}+\widetilde{\sigma}_3^{(2)}-m_2)+\widetilde{\sigma}_1^{(1)} (\widetilde{\sigma}_1^{(2)}-m_2) (\widetilde{\sigma}_2^{(2)}-m_2)+\widetilde{\sigma}_3^{(2)} (\widetilde{\sigma}_1^{(1)} (\widetilde{\sigma}_1^{(2)}-m_2)+(\widetilde{\sigma}_1^{(1)}-\widetilde{\sigma}_1^{(2)}) \widetilde{\sigma}_2^{(2)})$}&\parbox{6.5cm}{\centering $(\sigma _1-m_1) \left(\sigma _1-m_2\right) (\sigma _2+\sigma _3-m_1-m_2)-\qcoh_1$}\\
        \hline
        $(4\,1\,2\,3)$ &\parbox{7cm}{\centering $m_1 (m_2-\widetilde{\sigma}_1^{(1)}) (\widetilde{\sigma}_1^{(1)}-m_3)+\widetilde{\sigma}_1^{(1)}(m_2 (m_3-\widetilde{\sigma}_1^{(1)})+\widetilde{\sigma}_1^{(1)} (\widetilde{\sigma}_1^{(2)}-m_3)+(\widetilde{\sigma}_1^{(1)}-\widetilde{\sigma}_1^{(2)}) \widetilde{\sigma}_2^{(2)})+(\widetilde{\sigma}_1^{(1)}-\widetilde{\sigma}_1^{(2)}) (\widetilde{\sigma}_1^{(1)}-\widetilde{\sigma}_2^{(2)}) \widetilde{\sigma}_3^{(2)}$}&$(\sigma _1-m_1) (\sigma _1-m_2) (\sigma _1-m_3)+\qcoh_1$\\
         \hline
        $(3\,2\,4\,1)$ &$(\widetilde{\sigma}_1^{(1)}-m_2) (\widetilde{\sigma}_1^{(2)}-m_1) (\widetilde{\sigma}_2^{(2)}-m_1) (\widetilde{\sigma}_3^{(2)}-m_1)$ &$(\sigma _1-m_1) (\sigma _1-m_2) (\sigma _1-m_3)+\qcoh_1$\\
        \hline
        $(4\,1\,3\,2)$&\parbox{7cm}{\centering $(\widetilde{\sigma}_1^{(1)}-m_3) (m_1^2 (\widetilde{\sigma}_1^{(1)}-m_2)-m_1 (\widetilde{\sigma}_1^{(1)}-m_2) (\widetilde{\sigma}_1^{(2)}+\widetilde{\sigma}_2^{(2)}+\widetilde{\sigma}_3^{(2)}-m_2)+\widetilde{\sigma} _1^{(1)} (\widetilde{\sigma}_1^{(2)}-m_2)
   (\widetilde{\sigma}_2^{(2)}-m_2)-\widetilde{\sigma}_3^{(2)} (m_2 \widetilde{\sigma}_1^{(1)}-(\widetilde{\sigma}_1^{(2)}+\widetilde{\sigma}_2^{(2)}) \widetilde{\sigma}_1^{(1)}+\widetilde{\sigma}_1^{(2)} \widetilde{\sigma}_2^{(2)}))$}&\parbox{6.5cm}{\centering $(\sigma _1-m_3) ((\sigma _1-m_1) (\sigma _1-m_2) (+\sigma _2+\sigma _3-m_1-m_2)-\qcoh_1)$}\\
        \hline
        $(4\,2\,3\,1)$ &$(\widetilde{\sigma}_1^{(1)}-m_2)(\widetilde{\sigma}_1^{(1)}-m_3)\prod_{a=1}^3 (\widetilde{\sigma}_a^{(2)}-m_1)$&$(\sigma_1-m_2)(\sigma_1-m_3)\left[\prod_{a=1}^3 (\sigma_a-m_1)+\qcoh_1\right] $\\
        \hline
\end{longtable}
\end{center}
\FloatBarrier

\newpage 
\section{Schubert defects in Fl\texorpdfstring{$(2,3;4)$}{234}}\label{app:234}
\subsection*{All possible Schubert varieties}
In this partial flag, we have 12 possible Schubert varieties indexed by the equivalence classes in ${\rm W}^{(2,3;4)}$. These classes and their corresponding dimensions are summarized in the following table:
\begin{table}[!htbp]
\centering
\begin{subtable}{0.45\textwidth}
\centering
\begin{equation*}
        \begin{array}{|c||c|c|}
    \hline
        w&{\ell(w)}&{\dim X_w}  \\
        \hline
        \hline
        {(1\,2\,3\,4)} & 0&5 \\
        \hline
        {(1\,2\,4\,3)} &  1 &4\\
        \hline
         {(1\,3\,2\,4)} &  1&4  \\
                 \hline
         {(1\,3\,4\,2)} & 2&3  \\
        \hline
        {(1\,4\,2\,3)} &  2 &3\\
             \hline
           {(2\,3\,1\,4)} & 2&3\\
        \hline
        \end{array}
    \end{equation*}
\end{subtable}
\hfill
\begin{subtable}{0.45\textwidth}
\centering
\begin{equation*}
        \begin{array}{|c||c|c|}
    \hline
         w&{\ell(w)}&{\dim X_w}  \\
        \hline
          \hline
        {(1\,4\,3\,2)} &  3&2\\
        \hline
        {(2\,3\,4\,1)} &  3&2\\
        \hline
        {(2\,4\,1\,3)} &  3&2\\
        \hline
        {(2\,4\,3\,1)} &  4&1 \\
        \hline
        {(3\,4\,1\,2)} &  4&1\\
        \hline
        {(3\,4\,2\,1)} & 5 &0\\
        \hline
        \end{array}
    \end{equation*}
\end{subtable}
\end{table}
\FloatBarrier

\noindent
These Schubert varieties are connected via the following Hasse diagram:
\begin{figure}[h!]
    \centering
    \includegraphics[width=0.70\linewidth]{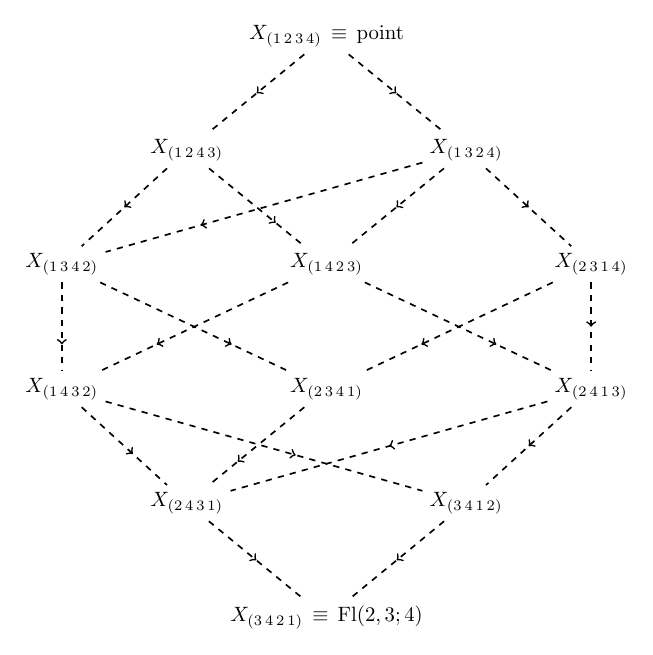}
\end{figure}

\newpage
\newpage
\subsection*{Coupled systems defining the Schubert defects}\label{subsec:defects234}
For each one of the 12 permutation classes $[w]\in {\rm W}^{(2,3;4)}$, let us now write down the 1d-3d (0d-2d) coupled system defining the Schubert line (point) defect. Here we follow the conventions of figure \ref{fig:GenProposal} for the general proposal. Viewed as Schubert line defects, in the last column, we also include the reduced form of the quiver after applying the duality moves reviewed in section 2 of \cite{Closset:2025cfm}.
\begin{table}[!h]
\centering
\begin{subtable}{0.45\textwidth}
\centering
\begin{equation*}
        \begin{array}{|c||c|c|c|}
    \hline
        w&{\rm General\,\,proposal} & {\rm Reduced \,\,quiver}\\
        \hline
        \hline
         \raisebox{9ex}{(1\,2\,3\,4)}   & \includegraphics[scale=0.3]{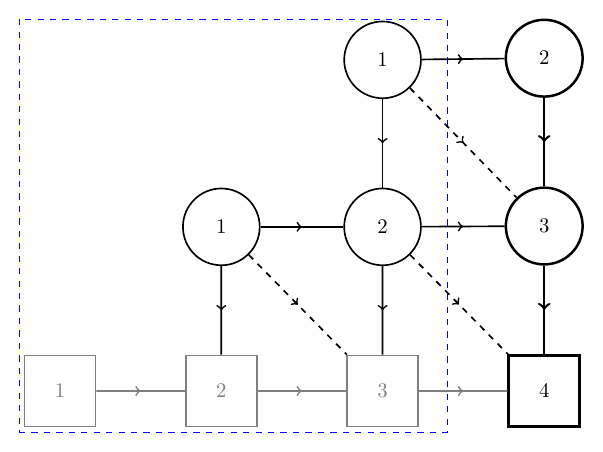}&\includegraphics[scale=0.3]{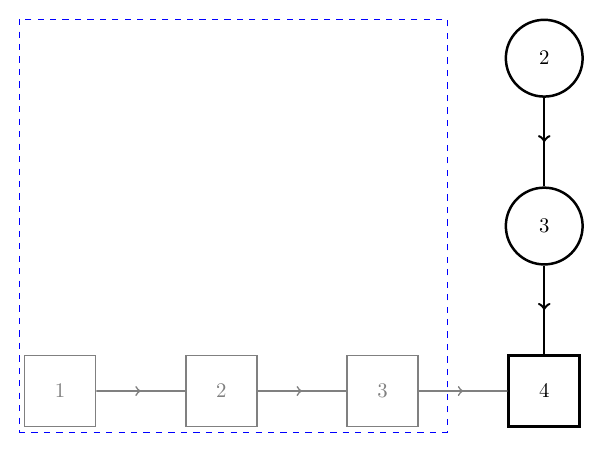}\\
           \hline
                  \raisebox{9ex}{(1\,2\,4\,3)}   & \includegraphics[scale=0.3]{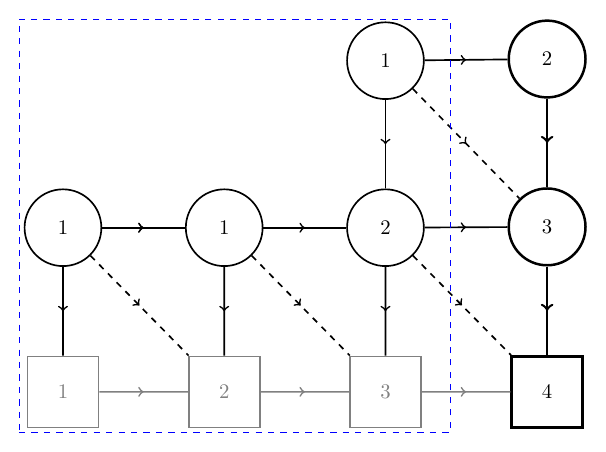}&\includegraphics[scale=0.3]{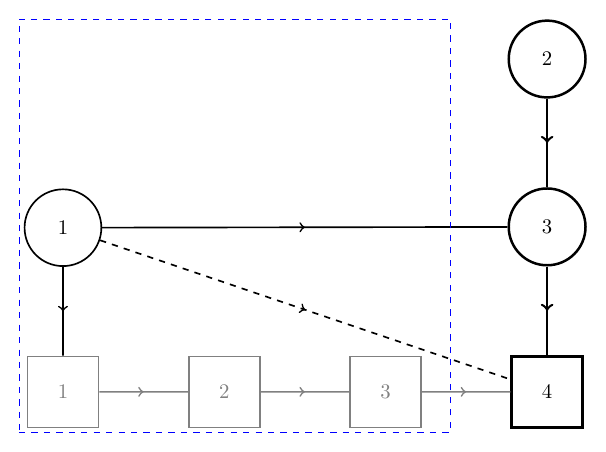}\\
                    \hline
           \raisebox{9ex}{(1\,3\,2\,4)} &\includegraphics[scale=0.3]{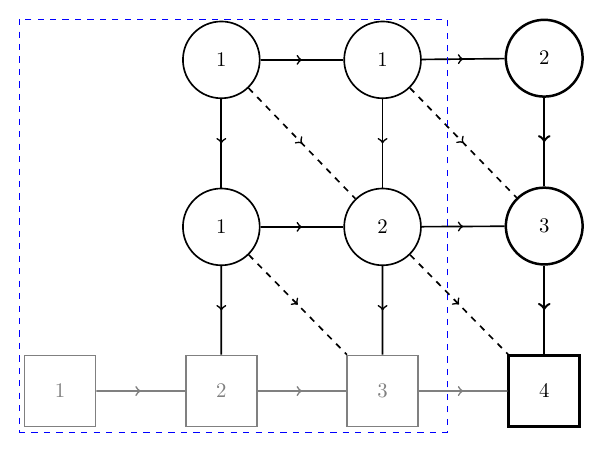}&\includegraphics[scale=0.3]{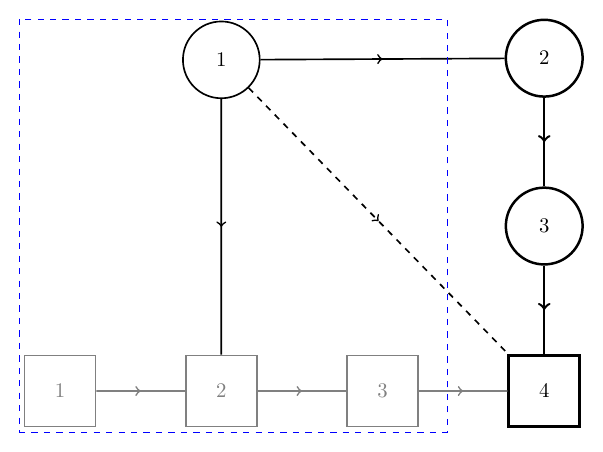}\\
              \hline
           \raisebox{9ex}{(1\,3\,4\,2)} &\includegraphics[scale=0.3]{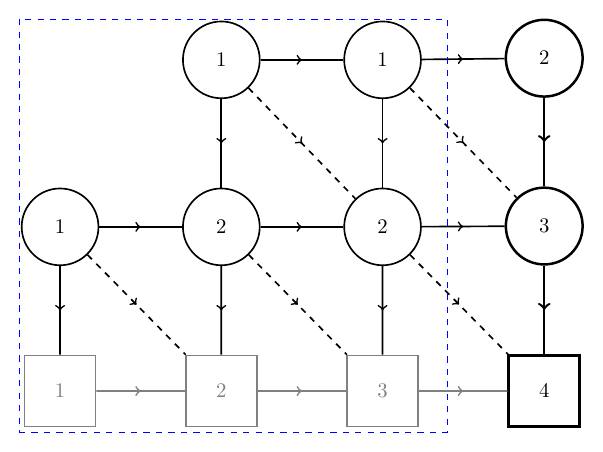}&\includegraphics[scale=0.3]{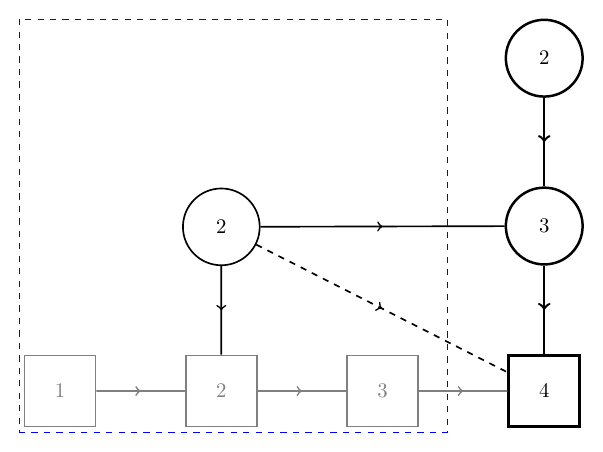}\\
           \hline
        \raisebox{9ex}{(1\,4\,2\,3)}   &\includegraphics[scale=0.3]{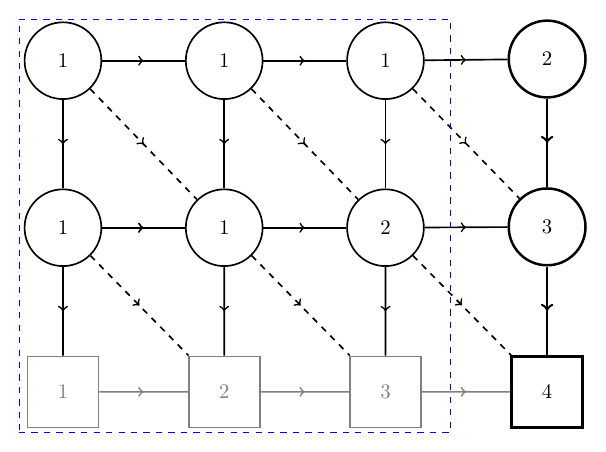} &\includegraphics[scale=0.3]{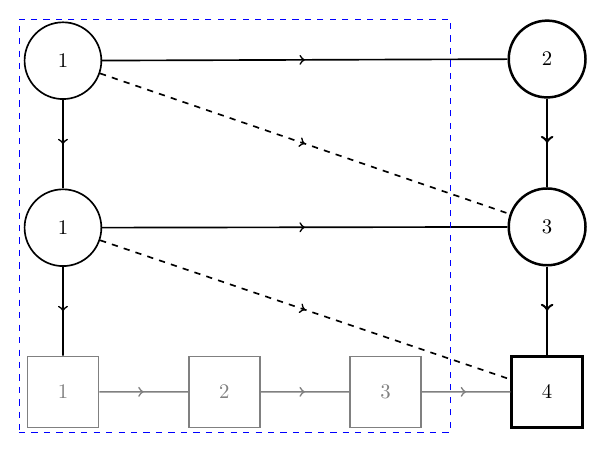} \\
               \hline
        \raisebox{9ex}{(2\,3\,1\,4)} &\includegraphics[scale=0.3]{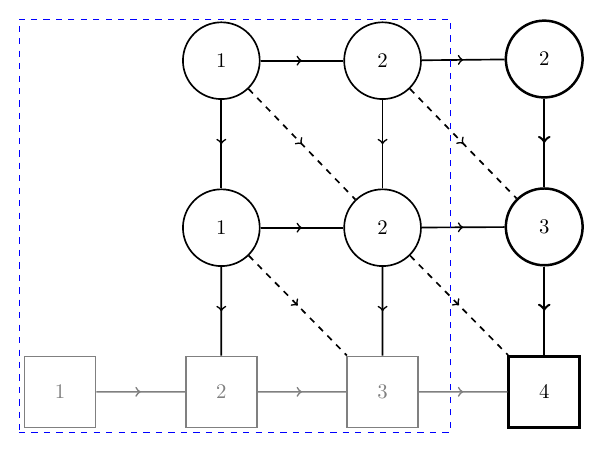}& \includegraphics[scale=0.3]{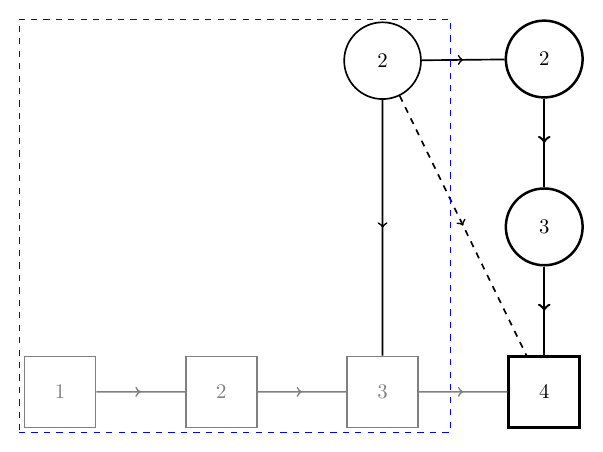}\\
           \hline
        \end{array}
    \end{equation*}
\end{subtable}
\hfill
\begin{subtable}{0.45\textwidth}
\centering
\begin{equation*}
        \begin{array}{|c||c|c|c|}
    \hline
        w&{\rm General\,\,proposal}&  {\rm Reduced \,\,quiver}\\
        \hline
           \hline
             \raisebox{9ex}{(1\,4\,3\,2)} &\includegraphics[scale=0.3]{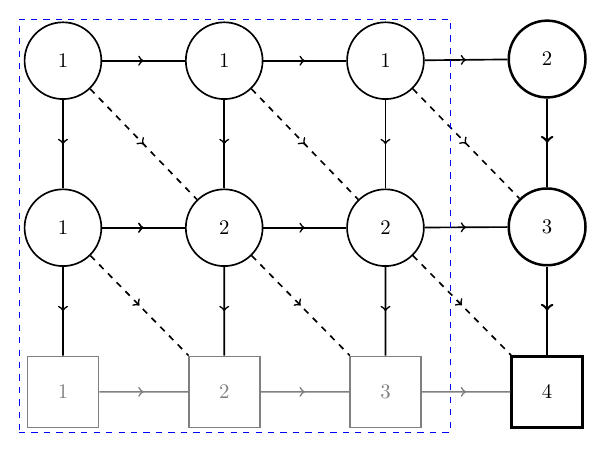}& \includegraphics[scale=0.3]{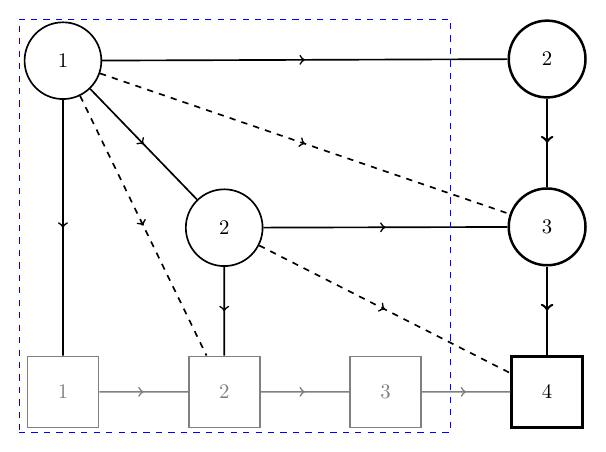}\\
           \hline      
            \raisebox{9ex}{(2\,3\,4\,1)}   & \includegraphics[scale=0.3]{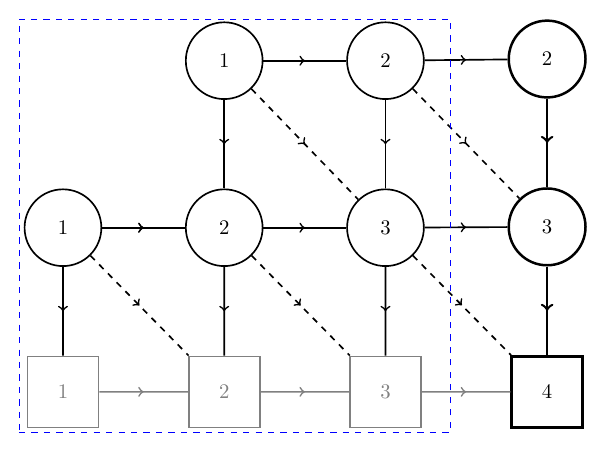}&\includegraphics[scale=0.3]{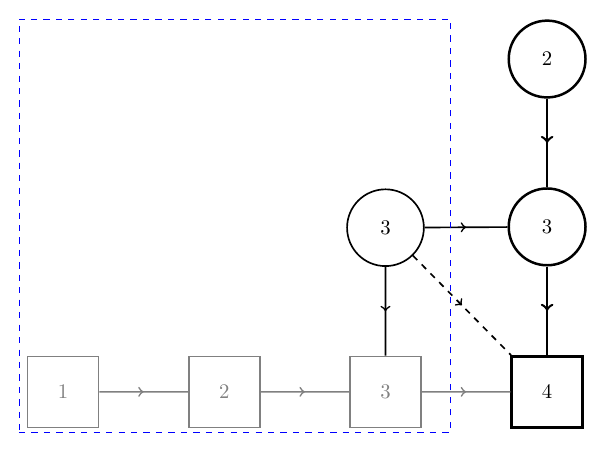}\\
           \hline
           \raisebox{9ex}{(2\,4\,1\,3)} &\includegraphics[scale=0.3]{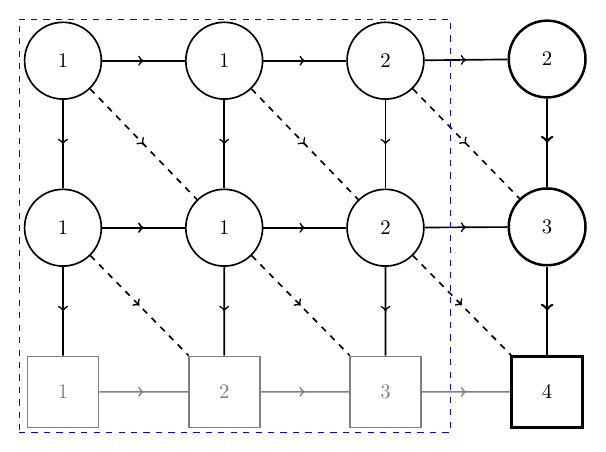}&\includegraphics[scale=0.3]{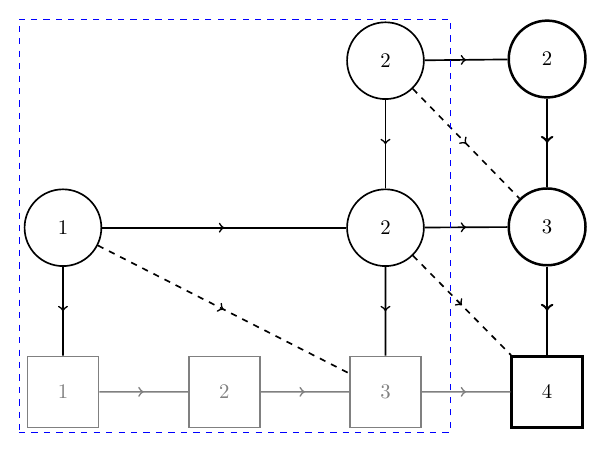}\\
           \hline
           \raisebox{9ex}{(2\,4\,3\,1)} &\includegraphics[scale=0.3]{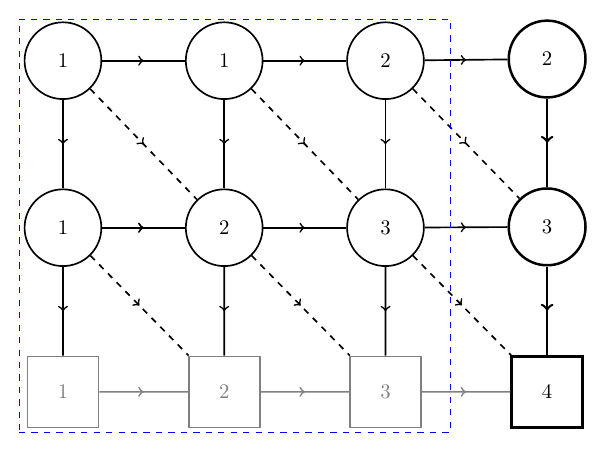}& \includegraphics[scale=0.3]{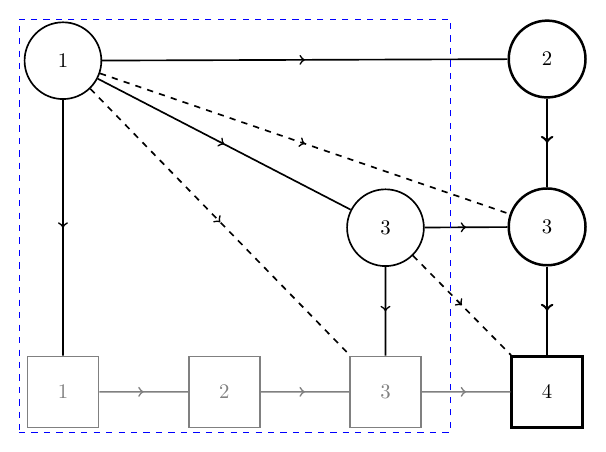}\\
           \hline
             \raisebox{9ex}{(3\,4\,1\,2)} &\includegraphics[scale=0.3]{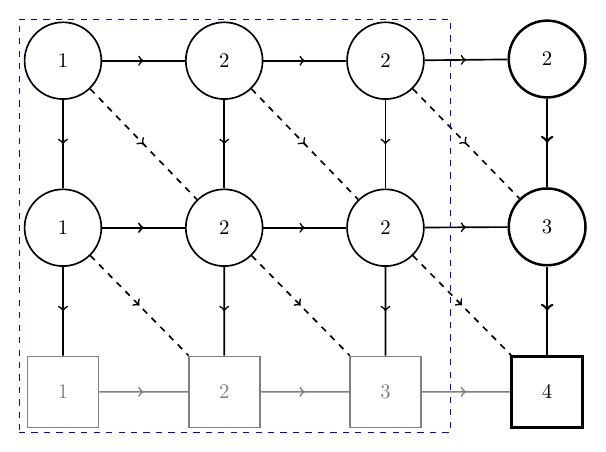}&\includegraphics[scale=0.3]{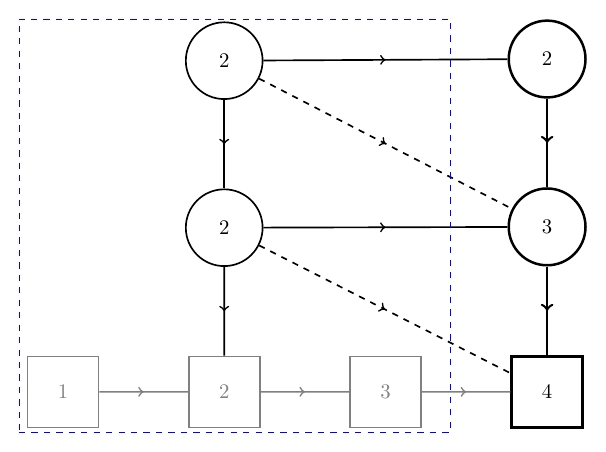}\\
           \hline
           \raisebox{9ex}{(3\,4\,2\,1)} &\includegraphics[scale=0.3]{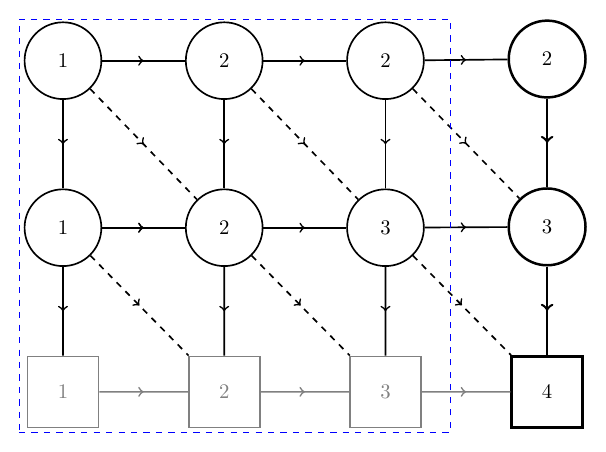}&\includegraphics[scale=0.3]{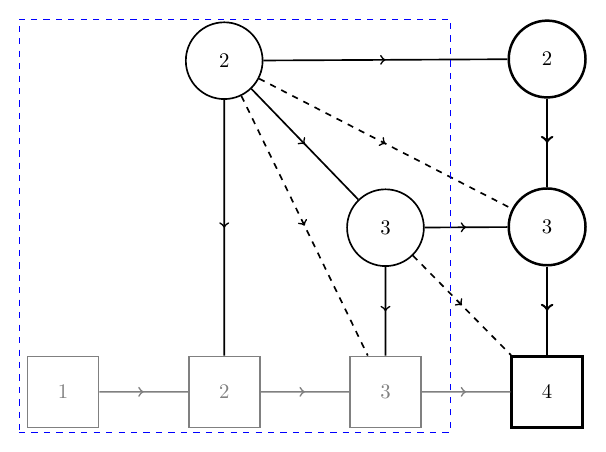}\\
           \hline
        \end{array}
    \end{equation*}
\end{subtable}
\end{table}

\newpage
\subsection*{1d indices and parabolic quantum Grothendieck polynomials}
\begin{center}
\begin{longtable}[!h]{|c||c|c|}
        \hline
     $w$&$\mathcal{I}_w^{(\rm 1d)}(x,y)$&$\mathcal{I}_w^{(\rm 1d)}{\scriptsize\begin{bmatrix}
        \qk_1&\qk_2\\
        2&3
    \end{bmatrix}}(x,y)$\\
    \hline
    \hline
     $(1\,2\,4\,3) $ & \parbox{7cm}{\centering $1-\frac{x_1^{(2)}x_2^{(2)}x_3^{(2)}}{y_1y_2y_3}$}& \parbox{7cm}{\centering $1-\frac{x_1x_2x_3}{(1-q_1)\,y_1y_2y_3}$} \\
      \hline
        $ (1\,3\,2\,4)$ & \parbox{7cm}{\centering $1-\frac{x_1^{(1)}x_2^{(1)}}{y_1y_2}$}& \parbox{7cm}{\centering$1-\frac{x_1x_2}{y_1y_2}$} \\
        \hline
         $(1\,3\,4\,2)$ &\parbox{7cm}{\centering $1-\frac{x_1^{(2)} x_2^{(2)}}{y_1 y_2}-\frac{x_2^{(2)}x_3^{(2)} }{y_1 y_2}-\frac{x_1^{(2)} x_3^{(2)}}{y_1 y_2}+\frac{x_1^{(2)}x_2^{(2)} x_3^{(2)} }{y_1^2 y_2}+\frac{x_1^{(2)}x_2^{(2)} x_3^{(2)} }{y_1
   y_2^2}$} &\parbox{7cm}{\centering $1-\frac{x_1 x_2}{y_1 y_2}-\frac{x_2x_3 }{(1-\qk_1)\,y_1 y_2}-\frac{x_1 x_3}{(1-\qk_1)\,y_1 y_2}+\frac{x_1x_2 x_3 }{(1-\qk_1)\,y_1^2 y_2}+\frac{x_1x_2 x_3}{(1-\qk_1)\,y_1
   y_2^2}$}  \\
        \hline
         $(1\,4\,2\,3)$ &\parbox{7cm}{\centering $1-\frac{x_1^{(1)} x_2^{(1)}}{y_1 y_2}-\frac{x_1^{(1)} x_2^{(1)}}{y_1 y_3}-\frac{x_1^{(1)} x_2^{(1)}}{y_2 y_3}+\frac{x_1^{(1)}x_2^{(1)} x_1^{(2)} }{y_1 y_2
   y_3}+\frac{x_1^{(1)} x_2^{(1)}x_2^{(2)} }{y_1 y_2 y_3}+\frac{x_1^{(1)}x_2^{(1)} x_3^{(2)} }{y_1 y_2 y_3}-\frac{x_1^{(2)} x_2^{(2)} x_3^{(2)}}{y_1 y_2 y_3}$} &\parbox{7cm}{\centering $1-\frac{x_1 x_2}{y_1 y_2}-\frac{x_1 x_2}{y_1 y_3}-\frac{x_1 x_2}{y_2 y_3}+\frac{x_1^2x_2  }{y_1 y_2
   y_3}+\frac{x_1 x_2^2 }{y_1 y_2 y_3}-\frac{\qk_1 x_1 x_2 x_3}{(1-\qk_1)y_1 y_2 y_3}$} \\
        \hline
         $(2\,3\,1\,4)$ & \parbox{7cm}{\centering $1-\frac{x^{(1)}_1}{y_1}-\frac{x^{(1)}_2}{y_1}+\frac{x^{(1)}_1 x^{(1)}_2}{y_1^2}$}&\parbox{7cm}{\centering $1-\frac{x_1}{y_1}-\frac{x_2}{y_1}+\frac{x_1 x_2}{y_1^2}$} \\
        \hline
         $(1\,4\,3\,2)$ &\parbox{7cm}{\centering $1-\frac{x_1^{(1)} x^{(1)}_2}{y_1 y_3}-\frac{x^{(1)}_1 x^{(1)}_2}{y_2 y_3}-\frac{x^{(2)}_1 x^{(2)}_2}{y_1
   y_2}-\frac{x^{(2)}_1 x^{(2)}_3}{y_1 y_2}-\frac{x^{(2)}_2 x_3^{(2)}}{y_1 y_2}+\frac{x^{(1)}_1 x^{(1)}_2 x^{(2)}_1 }{y_1 y_2 y_3}+\frac{x^{(1)}_1 x_2^{(1)}x_2^{(2)} }{y_1 y_2
   y_3}+\frac{x^{(1)}_1 x^{(1)}_2x^{(2)}_3 }{y_1 y_2 y_3}+\frac{x_1^{(2)} x_2^{(2)} x_3^{(2)}}{y_1^2 y_2}+\frac{x^{(2)}_1 x^{(2)}_2 x^{(2)}_3}{y_1
   y_2^2}-\frac{x^{(1)}_1 x^{(1)}_2 x^{(2)}_1 x^{(2)}_2 x^{(2)}_3 }{y_1^2 y_2^2 y_3}$} &\parbox{7cm}{\centering $1-\frac{x_1 x_2}{y_1 y_3}-\frac{x_1 x_2}{y_2 y_3}-\frac{x_1 x_2}{y_1
   y_2}-\frac{x_1 x_3}{(1-\qk_1)\, y_1 y_2}-\frac{x_2 x_3}{(1-\qk_1)\,y_1 y_2}+\frac{x^2_1 x_2 }{y_1 y_2 y_3}+\frac{x_1 x_2^{2} }{y_1 y_2
   y_3}+\frac{x_1 x_2x_3 }{y_1 y_2 y_3}+\frac{x_1 x_2 x_3}{(1-\qk_1)\,y_1^2 y_2}+\frac{x_1 x_2 x_3}{(1-\qk_1)\,y_1
   y_2^2}-\frac{x^{2}_1 x^{2}_2 x_3 }{(1-\qk_1)\,y_1^2 y_2^2 y_3}$}  \\
        \hline
   $ (2\,3\,4\,1)$ & \parbox{7cm}{\centering $1-\frac{x^{(2)}_1}{y_1}-\frac{x^{(2)}_2}{y_1}-\frac{x^{(2)}_3}{y_1}+\frac{x^{(2)}_1x^{(2)}_2}{y_1^2}+\frac{x^{(2)}_1x^{(2)}_3}{y_1^2}+\frac{x^{(2)}_2x^{(2)}_3}{y_1^2}-\frac{x^{(2)}_1x^{(2)}_2x^{(2)}_3}{y_1^3}$} & \parbox{7cm}{\centering $1-\frac{x_1}{y_1}-\frac{x_2}{y_1}-\frac{x_3}{y_1}+\frac{x_1x_2}{y_1^2} + \frac{x_1x_3+x_2 x_3}{(1-q_1)y_1^2} - \frac{x_1 x_2 x_3}{(1-q_1)y_1^3} $}\\
   \hline
   $ (2\,4\,1\,3)$ & \parbox{7cm}{\centering $1-\frac{x^{(1)}_1}{y_1}-\frac{x^{(1)}_2}{y_1}+\frac{x^{(1)}_1x^{(1)}_2}{y_1^2}-\frac{x^{(1)}_1x^{(1)}_2}{y_2y_3}+\frac{x^{(1)}_1x^{(1)}_2x^{(2)}_1}{y_1y_2y_3}+\frac{x^{(1)}_1x^{(1)}_2x^{(2)}_2}{y_1y_2y_3}+\frac{x^{(1)}_1x^{(1)}_2x^{(2)}_3}{y_1y_2y_3}-\frac{x^{(2)}_1x^{(2)}_2x^{(2)}_3}{y_1y_2y_3}-\frac{x^{(1)}_1x^{(1)}_2x^{(2)}_1x^{(2)}_2}{y_1^2y_2y_3}-\frac{x^{(1)}_1x^{(1)}_2x^{(2)}_1x^{(2)}_3}{y_1^2y_2y_3}-\frac{x^{(1)}_1x^{(1)}_2x^{(2)}_2x^{(2)}_3}{y_1^2y_2y_3}+\frac{x^{(1)}_1x^{(2)}_1x^{(2)}_2x^{(2)}_3}{y_1^2y_2y_3}+\frac{x^{(1)}_2x^{(2)}_1x^{(2)}_2x^{(2)}_3}{y_1^2y_2y_3}$} & \parbox{7cm}{\centering $1-\frac{x_1}{y_1}-\frac{x_2}{y_1} + \frac{x_1x_2}{y_1^2} - \frac{x_1x_2}{y_2y_3}+\frac{x_1^2x_2 + x_1x_2^2}{y_1y_2y_3} - \frac{q x_1x_2x_3}{(1-q_1)y_1y_2y_3 } - \frac{x_1^2x_2^2}{y_1^2 y_2 y_3}$ }\\
    \hline
   $ (2\,4\,3\,1)$ & \parbox{7cm}{\centering $1-\frac{x^{(2)}_1}{y_1}-\frac{x^{(2)}_2}{y_1}-\frac{x^{(2)}_3}{y_1} + \frac{x^{(2)}_1x^{(2)}_2}{y_1^2}+\frac{x^{(2)}_1x^{(2)}_3}{y_1^2}+\frac{x^{(2)}_2x^{(2)}_3}{y_1^2} - \frac{x^{(1)}_1x^{(1)}_2}{y_2 y_3} - \frac{x^{(2)}_1x^{(2)}_2x^{(2)}_3}{y_1^3} + \frac{x^{(1)}_1x^{(1)}_2x^{(2)}_1}{y_1y_2y_3}+\frac{x^{(1)}_1x^{(1)}_2x^{(2)}_2}{y_1y_2y_3}+\frac{x^{(1)}_1x^{(1)}_2x^{(2)}_3}{y_1y_2y_3} -\frac{x^{(1)}_1x^{(1)}_2x^{(2)}_1x^{(2)}_2}{y_1^2 y_2 y_3}-\frac{x^{(1)}_1x^{(1)}_2x^{(2)}_1x^{(2)}_3}{y_1^2 y_2 y_3}-\frac{x^{(1)}_1x^{(1)}_2x^{(2)}_2x^{(2)}_3}{y_1^2 y_2 y_3}+\frac{x^{(1)}_1x^{(1)}_2x^{(2)}_1x^{(2)}_2x^{(2)}_3}{y_1^3 y_2 y_3}$ } & \parbox{7cm}{\centering $1-\frac{x_1}{y_1}-\frac{x_2}{y_1}-\frac{x_3}{y_1} + \frac{x_1 x_2}{y_1^2}+\frac{x_1 x_3+x_2x_3}{(1-q_1)y_1^2} -\frac{x_1 x_2}{y_2 y_3} - \frac{x_1 x_2 x_3 }{(1-q_1)y_1^3} +\frac{x_1x_2(x_1+x_2+x_3)}{y_1y_2y_3}-\frac{x_1^2x_2^2}{y_1^2y_2y_3}-\frac{x_1^2x_2x_3}{(1-q_1)y_1^2y_2y_3}-\frac{x_1x_2^2x_3}{(1-q_1)y_1^2y_2y_3}+\frac{x_1^2x_2^2x_3}{(1-q_1)y_1^3y_2y_3} $  }\\
   \hline
   $ (3\,4\,1\,2)$ & \parbox{7cm}{\centering  $1-\frac{x^{(1)}_1}{y_1}-\frac{x^{(1)}_2}{y_1}-\frac{x^{(1)}_1}{y_2}-\frac{x^{(1)}_2}{y_2}+\frac{x^{(1)}_1x^{(1)}_2}{y_1^2}+\frac{x^{(1)}_1x^{(1)}_2}{y_2^2} + \frac{x^{(1)}_1x^{(1)}_2}{y_1 y_2} + \frac{x^{(1)}_1x^{(2)}_1}{y_1 y_2}+\frac{x^{(1)}_2x^{(2)}_1}{y_1 y_2}+\frac{x^{(1)}_1x^{(2)}_2}{y_1 y_2}+\frac{x^{(1)}_2x^{(2)}_2}{y_1 y_2} + \frac{x^{(1)}_1 x^{(2)}_3}{y_1y_2}+ \frac{x^{(1)}_2 x^{(2)}_3}{y_1y_2} - \frac{x^{(2)}_1 x^{(2)}_2}{y_1y_2} - \frac{x^{(2)}_1 x^{(2)}_3}{y_1y_2}- \frac{x^{(2)}_2 x^{(2)}_3}{y_1y_2} - \frac{x^{(1)}_1x^{(1)}_2x^{(2)}_1}{y_1y_2^2} - \frac{x^{(1)}_1x^{(1)}_2x^{(2)}_2}{y_1y_2^2} - \frac{x^{(1)}_1x^{(1)}_2x^{(2)}_3}{y_1y_2^2} + \frac{x^{(2)}_1x^{(2)}_2x^{(2)}_3}{y_1y_2^2} - \frac{x^{(1)}_1x^{(1)}_2x^{(2)}_1}{y_1^2y_2} - \frac{x^{(1)}_1x^{(1)}_2x^{(2)}_2}{y_1^2y_2} - \frac{x^{(1)}_1x^{(1)}_2x^{(2)}_3}{y_1^2y_2} + \frac{x^{(2)}_1x^{(2)}_2x^{(2)}_3}{y_1^2y_2} + \frac{x^{(1)}_1x^{(1)}_2x^{(2)}_1x^{(2)}_2}{y_1^2y_2^2} + \frac{x^{(1)}_1x^{(1)}_2x^{(2)}_1x^{(2)}_3}{y_1^2y_2^2} + \frac{x^{(1)}_1x^{(1)}_2x^{(2)}_2x^{(2)}_3}{y_1^2y_2^2} - \frac{x^{(1)}_1x^{(2)}_1x^{(2)}_2x^{(2)}_3}{y_1^2y_2^2} - \frac{x^{(1)}_2x^{(2)}_1x^{(2)}_2x^{(2)}_3}{y_1^2y_2^2} $} & \parbox{7cm}{\centering $ 1-\frac{x_1}{y_1}-\frac{x_2}{y_1}-\frac{x_1}{y_2}-\frac{x_2}{y_2} + \frac{2x_1x_2}{y_1y_2} +\frac{x_1x_2}{y_1^2}+\frac{x_1x_2}{y_2^2}-\frac{q_1(x_1+x_2)x_3}{(1-q_1)y_1y_2}+\frac{x_1^2+x_2^2}{y_1y_2} -\frac{x_1^2x_2}{y_1y_2^2}-\frac{x_1x_2^2}{y_1y_2^2}-\frac{x_1^2x_2}{y_1^2y_2}-\frac{x_1x_2^2}{y_1^2y_2}+\frac{q_1x_1x_2x_3}{(1-q_1)y_1^2y_2}+\frac{q_1x_1x_2x_3}{(1-q_1)y_1y_2^2}+\frac{x_1^2x_2^2}{y_1^2y_2^2}$ }\\
    \hline
   $ (3\,4\,2\,1)$ & \parbox{7cm}{\centering $1 -\frac{x^{(1)}_1}{y_2} -\frac{x^{(1)}_2}{y_2}-\frac{x^{(2)}_1}{y_1}-\frac{x^{(2)}_2}{y_1} -\frac{x^{(2)}_3}{y_1} +\frac{x^{(2)}_1x^{(2)}_2}{y_1^2}  +\frac{x^{(2)}_1x^{(2)}_3}{y_1^2} +\frac{x^{(2)}_2 x^{(2)}_3}{y_1^2} +\frac{x^{(1)}_1x^{(1)}_2}{y_2^2} + \frac{x^{(1)}_1x^{(2)}_1}{y_1y_2} + \frac{x^{(1)}_2x^{(2)}_1}{y_1y_2} + \frac{x^{(1)}_1x^{(2)}_2}{y_1y_2} + \frac{x^{(1)}_2x^{(2)}_2}{y_1y_2} + \frac{x^{(1)}_1x^{(2)}_3}{y_1y_2} + \frac{x^{(1)}_2x^{(2)}_3}{y_1y_2} -\frac{x^{(2)}_1 x^{(2)}_2 x^{(2)}_3}{y_1^3} -\frac{x^{(1)}_1 x^{(2)}_1 x^{(2)}_2}{y_1^2 y_2} -\frac{x^{(1)}_2 x^{(2)}_1 x^{(2)}_2}{y_1^2 y_2} -\frac{x^{(1)}_1 x^{(2)}_1 x^{(2)}_3}{y_1^2 y_2} -\frac{x^{(1)}_2 x^{(2)}_1 x^{(2)}_3}{y_1^2 y_2} -\frac{x^{(1)}_1 x^{(2)}_2 x^{(2)}_3}{y_1^2 y_2} -\frac{x^{(1)}_2 x^{(2)}_2 x^{(2)}_3}{y_1^2 y_2}     -\frac{x^{(1)}_1 x^{(1)}_2 x^{(2)}_1}{y_1 y_2^2} -\frac{x^{(1)}_1 x^{(1)}_2 x^{(2)}_2}{y_1 y_2^2} -\frac{x^{(1)}_1 x^{(1)}_2 x^{(2)}_3}{y_1 y_2^2} + \frac{x^{(1)}_1x^{(1)}_2x^{(2)}_1x^{(2)}_2}{y_1^2 y_2^2} + \frac{x^{(1)}_1x^{(1)}_2x^{(2)}_1x^{(2)}_3}{y_1^2 y_2^2} + \frac{x^{(1)}_1x^{(1)}_2x^{(2)}_2x^{(2)}_3}{y_1^2 y_2^2} + \frac{x^{(1)}_1x^{(2)}_1x^{(2)}_2x^{(2)}_3}{y_1^3 y_2} + \frac{x^{(1)}_2x^{(2)}_1x^{(2)}_2x^{(2)}_3}{y_1^3 y_2} - \frac{x^{(1)}_1x^{(1)}_2x^{(2)}_1x^{(2)}_2x^{(2)}_3}{y_1^3 y_2^2} $}  & \parbox{7cm}{\centering $1 - \frac{x_1}{y_1} - \frac{x_2}{y_1} - \frac{x_3}{y_1} - \frac{x_1}{y_2} - \frac{x_2}{y_2} + \frac{x_1 x_2}{y_1^2} + \frac{x_1 x_3+x_2 x_3}{(1-q_1)y_1^2} + \frac{x_1x_2}{y_2^2} + \frac{x_1 x_3 + x_2 x_3}{y_1 y_2} + \frac{(x_1+x_2)^2}{y_1 y_2} - \frac{2 x_1x_2x_3}{(1-q_1)y_1^2y_2} -\frac{x_1x_2x_3}{(1-q_1)y_1^2 y_2} - \frac{x_1x_2x_3}{(1-q_1) y_1^3} - \frac{x_1x_2x_3}{y_1y_2^2} - \frac{x_1^2x_2}{y_1y_2^2} - \frac{x_1x_2^2}{y_1y_2^2} - \frac{x_1^2x_2}{y_1^2y_2} - \frac{x_1x_2^2}{y_1^2y_2} - \frac{x_1^2x_3+x_2^2 x_3}{(1-q_1)y_1^2y_2} + \frac{x_1^2x_2^2}{y_1^2y_2^2}+\frac{x_1x_2(x_1+x_2)x_3}{(1-q_1)y_1^2y_2^2} + \frac{x_1x_2(x_1+x_2)x_3}{(1-q_1)y_1^3y_2}-\frac{x_1^2x_2^2x_3}{(1-q_1)y_1^3 y_2^2}$ }\\
    \hline
\end{longtable}
\end{center}

\subsection*{0d partition functions and parabolic quantum Schubert polynomials}
\begin{center}
    \renewcommand{\arraystretch}{1.4}
    \begin{longtable}[!h]{|c||c|c|}
        \hline
     $w$&$\mathcal{I}_w^{(\rm 0d)}(\widetilde{\sigma},m)$&$\mathcal{I}_w^{(\rm 0d)}{\scriptsize\begin{bmatrix}
        \qcoh_1&\qcoh_2\\
        2&3
    \end{bmatrix}}(\sigma,m)$\\
    \hline
      \hline
    $(1\,2\,4\,3)$ & \parbox{7cm}{\centering $\widetilde{\sigma}^{(2)}_1+\widetilde{\sigma}^{(2)}_2+\widetilde{\sigma}^{(2)}_3-m_1-m_2-m_3$} & \parbox{7cm}{\centering $\sigma_1 + \sigma_2 + \sigma_3 - m_1 -m_2 -m_3$}\\
     \hline
   $ (1\,3\,2\,4)$ & \parbox{7cm}{\centering $\widetilde{\sigma}^{(1)}_1+\widetilde{\sigma}^{(1)}_2-m_1-m_2$} &\parbox{7cm}{\centering $\sigma_1+\sigma_2-m_1-m_2$}\\
         \hline
    $(1\,3\,4\,2)$ & \parbox{7cm}{\centering $e_2(\widetilde{\sigma}^{(2)}) - e_1(\widetilde{\sigma}^{(2)})(m_1+m_2)+m_1^2+m_1m_2+m_2^2$} &\parbox{7cm}{\centering $\sigma_1\sigma_2+\sigma_1\sigma_3+\sigma_2\sigma_3-(\sigma_1+\sigma_2+\sigma_3)(m_1+m_2)+m_1^2+m_1m_2+m_2^2$}\\
    \hline
    $(1\,4\,2\,3)$ & \parbox{7cm}{\centering $(\widetilde{\sigma}^{(1)}_1+\widetilde{\sigma}^{(1)}_2)(\widetilde{\sigma}^{(2)}_1+\widetilde{\sigma}^{(2)}_2+\widetilde{\sigma}^{(2)}_3)-\widetilde{\sigma}^{(2)}_1\widetilde{\sigma}^{(2)}_2-\widetilde{\sigma}^{(2)}_1\widetilde{\sigma}^{(2)}_3-\widetilde{\sigma}^{(2)}_2\widetilde{\sigma}^{(2)}_3 - (\widetilde{\sigma}^{(1)}_1+\widetilde{\sigma}^{(1)}_2 )(m_1+m_2+m_3) + m_1m_2+m_1m_3+m_2m_3$} & \parbox{7cm}{\centering $\sigma_1^2+\sigma_1\sigma_2+\sigma_2^2 - (\sigma_1+\sigma_2)(m_1+m_2+m_3)+ m_1m_2+m_1m_3+m_2m_3$}\\
          \hline
    $(2\,3\,1\,4)$ & \parbox{7cm}{\centering $(\widetilde{\sigma}^{(1)}_1-m_1)(\widetilde{\sigma}^{(1)}_2-m_1)$} & \parbox{7cm}{\centering $(\sigma_1-m_1)(\sigma_2-m_1)$}\\
    \hline
    $(1\,4\,3\,2) $& \parbox{7cm}{\centering $e_1(\widetilde{\sigma}^{(1)}) e_2(\widetilde{\sigma}^{(2)})-e_3(\widetilde{\sigma}^{(2)}) - e_1(\widetilde{\sigma}^{(1)})e_1(\widetilde{\sigma}^{(2)})(m_1+m_2) - e_2(\widetilde{\sigma}^{(2)})m_3+e_1(\widetilde{\sigma}^{(2)})(m_1m_2+m_1m_3+m_2m_3)+e_1(\widetilde{\sigma}^{(1)})(m_1^2+m_1m_2+m_2^2)-m_1^2m_2-m_1m_2^2-m_1^2m_3-m_2^2m_3-m_1m_2m_3$} & \parbox{7cm}{\centering $(\sigma_1^2\sigma_2+\sigma_1\sigma_2^2+\sigma_1^2\sigma_3+\sigma_2^2\sigma_3+\sigma_1\sigma_2\sigma_3) - (\sigma_1\sigma_2+\sigma_2\sigma_3+\sigma_1\sigma_3)(m_1+m_2+m_3)-(\sigma_1^2+\sigma_1\sigma_2+\sigma_2^2)(m_1+m_2)+(\sigma_1+\sigma_2+\sigma_3)(m_1m_2+m_1m_3+m_2m_3)+(\sigma_1+\sigma_2)(m_1^2+m_1m_2+m_2^2)-(m_1^2m_2+m_1m_2^2 + m_1^2m_3+m_2^2 m_3 + m_1m_2m_3) + \qcoh_1$}\\
    \hline
   $ (2\,3\,4\,1)$ & \parbox{7cm}{\centering $\prod_{a=1}^3(\widetilde{\sigma}^{(2)}_a - m_1)$} &\parbox{7cm}{\centering $(\sigma_1-m_1)(\sigma_2-m_1)(\sigma_3-m_1)-\qcoh_1$}\\
    \hline
    $(2\,4\,1\,3)$ & \parbox{7cm}{\centering $e_2(\widetilde{\sigma}^{(1)})e_1(\widetilde{\sigma}^{(2)}) - e_3(\widetilde{\sigma}^{(2)}) + e_2(\widetilde{\sigma}^{(2)})m_1-e_1(\widetilde{\sigma}^{(1)})e_1(\widetilde{\sigma}^{(2)})m_1 -e_2(\widetilde{\sigma}^{(1)}) (m_1+m_2+m_3) + e_1(\widetilde{\sigma}^{(1)})(m_1m_2+m_1m_3+m_2m_3) - m_1^2m_2-m_1^2 m_3$} &\parbox{7cm}{\centering $(\sigma_1-m_1)(\sigma_2-m_1)(\sigma_1+\sigma_2-m_2-m_3)+\qcoh_1$}\\
    \hline
    $(2\,4\,3\,1)$ & \parbox{7cm}{\centering $(\widetilde{\sigma}^{(1)}_1+\widetilde{\sigma}^{(1)}_2-m_2-m_3)(\widetilde{\sigma}^{(2)}_1-m_1)(\widetilde{\sigma}^{(2)}_2-m_1)(\widetilde{\sigma}^{(2)}_3-m_1)$} &\parbox{7cm}{\centering $(\sigma_1+\sigma_2-m_2-m_3)((\sigma_1-m_1)(\sigma_2-m_1)(\sigma_3-m_1)-\qcoh_1)$}\\
    \hline
    $(3\,4\,1\,2) $& \parbox{7cm}{\centering $e_2(\widetilde{\sigma}^{(1)})e_2(\widetilde{\sigma}^{(2)})- e_1(\widetilde{\sigma}^{(1)})e_3(\widetilde{\sigma}^{(2)}) + (e_3(\widetilde{\sigma}^{(2)}) - e_2(\widetilde{\sigma}^{(1)})e_1(\widetilde{\sigma}^{(2)}))(m_1+m_2) + e_2(\widetilde{\sigma}^{(1)}) m_1 m_2 +e_1(\widetilde{\sigma}^{(1)})e_1(\widetilde{\sigma}^{(2)})m_1m_2 - e_2(\widetilde{\sigma}^{(2)})m_1m_2 + e_2(\widetilde{\sigma}^{(1)})(m_1^2+m_2^2) - e_1(\widetilde{\sigma}^{(1)})(m_1^2m_2+m_1m_2^2) + m_1^2m_2^2 $} &\parbox{7cm}{\centering $\sigma_1^2\sigma_2^2-(\sigma_1^2\sigma_2+\sigma_1\sigma_2^2)(m_1+m_2)+\sigma_1\sigma_2(m_1+m_2)^2+(\sigma_1^2+\sigma_2^2)m_1m_2-(\sigma_1+\sigma_2)(m_1^2m_2+m_1m_2^2-\qcoh_1) - (m_1+m_2)\qcoh_1 +m_1^2m_2^2$}\\
    \hline
    $(3\,4\,2\,1)$ & \parbox{7cm}{\centering $(\widetilde{\sigma}^{(1)}_1-m_2)(\widetilde{\sigma}^{(1)}_2-m_2)(\widetilde{\sigma}^{(2)}_1-m_1)(\widetilde{\sigma}^{(2)}_2-m_1)(\widetilde{\sigma}^{(2)}_3-m_1)$} &\parbox{7cm}{\centering $(\sigma_1-m_2)(\sigma_2-m_2)((\sigma_1-m_1)(\sigma_2-m_1)(\sigma_3-m_1)-\qcoh_1)$}\\
    \hline
\end{longtable}
\end{center}
\section{Geometry of examples of Schubert varieties}\label{app:geometry}

\label{sec:geometry}

In this appendix, we will explicitly present geometries of a few examples of Schubert varieties.
We will illustrate in a few examples that our proposed quivers in figure \ref{fig:GenProposal} describe resolutions of Schubert varieties.  In the special cases of Gr$(1,n)$ and Gr$(2,4)$, we will describe the Schubert varieties explicitly and also discuss how the constructions discussed here and elsewhere provide resolutions.
We also discuss Schubert varieties in Fl$(1,n-1;n)$.  Recall that, Gr$(1,n) \equiv {\mathbb P}^{n-1}$ and all its Schubert varieties are smaller-dimensional projective spaces, and both Gr$(2,4)$ and Fl$(1,n-1;n)$ can be represented as hypersurfaces in the images of their Pl\"ucker embeddings, which makes the analysis reasonably concise.

\subsection{The projective space \texorpdfstring{$\mathbb{P}^{n-1}$}{Pn-1}}

Consider $X = {\mathbb P}^{n-1} = {\rm Gr}(1,n)$.  
Now, $X = {\mathbb C}^n \sslash {\mathbb C}^{\times}$, where the
${\mathbb C}^n$ is a $1 \times n$ matrix:
\begin{equation}
(\phi_i)_{i=1,\cdots,n}~,
\end{equation}
interpreted as homogeneous coordinates on ${\mathbb P}^{n-1}$. Consider a Schubert variety corresponding to a partition $(\lambda)$, i.e.,
a row of $\lambda$ boxes.  This is described by the $1 \times n$
matrix obeying:
\begin{equation}
\phi_n \: = \: 0 \: = \: \phi_{n-1} \: = \: \cdots \: = \:
\phi_{n-\lambda+1}~,
\end{equation}
with the first $n-\lambda$ $\phi_i$'s left unspecified.  This is just a complete
intersection of $\lambda$ hyperplanes, meaning"
\begin{equation}
\Omega_{(\lambda)} \: = \: {\mathbb P}^{n-1-\lambda}~.
\end{equation}
In particular, $\Omega_{(\lambda)}$ is smooth.

Now, consider the Kempf--Laksov resolution \cite{kempf-laskov}. This is defined by adding one field along the defect, call it $z$,
and a $U(1)$ gauge group along the defect, for which $z$ is a bifundamental between the bulk and defect $U(1)$'s, together with constraints of the form:
\begin{equation}
z \,\phi_n \: = \: 0 \: = \: z\, \phi_{n-1} \: = \:
\cdots \: = \: z \,\phi_{n-\lambda+1}~.
\end{equation}
The $D$-term equations of the 1d SQM force $z \neq 0$, so the equations above reproduce the
Schubert variety.  Furthermore, since there is only one field ($z$)
charged under the $U(1)$, altogether it contributes:
\begin{equation}
[ {\mathbb C}/ {\mathbb C}^{\times} ] \: = \:
\frac{ \{ |z|^2\, =\, r \} }{ U(1) } \: = \: {\rm point}~.
\end{equation}
Therefore, the Kempf--Laksov `resolution' of these smooth Schubert varieties is identical to those Schubert varieties, as expected.

\subsection{The Grassmannian manifold \texorpdfstring{${\rm Gr}(2,4)$}{Gr24}}

Next, we consider $X = {\rm Gr}(2,4)$.
Now, $X = {\mathbb C}^8 \sslash GL(2)$, where the ${\mathbb C}^8$
is described as $2 \times 4$ matrices:
\begin{equation}
\left( \phi_i^a \right)^{a=1,2}_{i=1\cdots 4}~.
\end{equation}
We define Pl\"ucker embedding coordinates:
\begin{equation}
p_{ij} \: := \: \epsilon_{ab}\, \phi^a_i\, \phi^b_j~,
\end{equation}
where it is understood that repeated indices are summed over. In this embedding, this Grassmannian can be described as a quadric
hypersurface:
\begin{equation}\label{Gr24pluker}
{\rm Gr}(2,4) \: = \: \{ p_{12}\, p_{34}\, -\, p_{13}\, p_{24}\, +\, p_{14}\, p_{23}\, = \,0 \}
\: \subset \: {\mathbb P}^5~.
\end{equation}

As we reviewed earlier in section \ref{sec:Schubert in partial}, Schubert varieties in this case are indexed by 2-partitions $\lambda = [\lambda_1, \lambda_2]$ with $0\,\leq\,\lambda_2\,\leq\,\lambda_1\leq 2$. In what follows, we will consider the two cases with $\lambda = [1,0],[2,0]$.

\subsubsection{The Schubert variety with \texorpdfstring{$\lambda \,=\, [1,0]$}{lambda(1)}}
Next, specialize to the Schubert variety $X_{\tiny\yng(1)}$
corresponding to $\tiny\yng(1)$.
This can be described as $2 \times 4$ matrices:
\begin{equation}
\left( \phi^a_i \right) \: = \: \left[ \begin{array}{cccc}
* & * & * & * \\
* & * & 0 & 0
\end{array} \right]~.
\end{equation}
From \eqref{Gr24pluker}, this is precisely the hypersurface $\{ p_{34} = 0 \}$.
(As a consistency check, $X_{\tiny\yng(1)}$ should be codimension 1 in $X = {\rm Gr}(2,4)$.)

\medskip
\noindent
{\bf The singularity.} For simplicity, specialize to the patch $\{ p_{12} = 1 \}$ on ${\mathbb P}^5$.
In that patch, the hypersurface describing the Grassmannian Gr$(2,4)$ is:
\begin{equation}
f(p_{ij}) \: = \: 
p_{34} \: - \: p_{13}\, p_{24} \: + \: p_{14}\, p_{23} \: = \: 0~.
\end{equation}
This hypersurface is smooth:  because of the presence of the first
term, there are no solutions to $f = df  = 0$.

Now, note that along $X_{\tiny\yng(1)}$, where $p_{34} = 0$,
\begin{equation}  \label{eq:conifold}
f|_{X_{{\tiny\yng(1)}}} \: = \: - p_{13}\, p_{24} \: + \: p_{14}\, p_{23} \: = \: 0~.
\end{equation}
This is a hypersurface in 
\begin{equation}
{\mathbb C}^4 \: = \: {\rm Spec} \, {\mathbb C}[p_{13}, p_{14},
p_{23}, p_{24} ]~.
\end{equation}
This hypersurface is singular -- it's the canonical example of a 
rational double point, the conifold singularity. Hence, we see that the Schubert variety $X_{\tiny\yng(1)}$ is singular.

\medskip
\noindent
{\bf Kempf--Laksov resolution.} In the Kempf--Laksov resolution of $X_{\tiny\yng(1)}$, we add new fields $(z_a) = \{z_1, z_2\}$, which have charge 1 under
a $U(1)$, i.e., related by a rescaling $[z_1, z_2] \sim
[\lambda z_1, \lambda z_2]$ for $\lambda \in {\mathbb C}^{\times}$.
In addition, we also impose two constraints:
\begin{equation}
\sum_a \,z_a \,\phi^a_3 \: = \: 0 \: = \:\sum_a\, z_a\, \phi^a_4~.
\end{equation}

Now, define:
\begin{equation}
u \: := \: z_1\, \phi^1_3, \: \: \:
v \: := \: z_2 \,\phi^2_4~.
\end{equation}
Then, it is straightforward to check that:
\begin{eqnarray}
p_{14}\, u \: + \: p_{13}\, v & \,=\, & 0~,
\\
p_{24}\, u \: + \: p_{23}\, v &\, = \,& 0~,
\end{eqnarray}
or more simply,
\begin{equation}
\left[ \begin{array}{cc}
p_{14} & p_{13} \\
p_{24} & p_{23} \end{array} \right]
\left[ \begin{array}{c} u \\ v \end{array} \right]
\: = \: 0~.
\end{equation}

This defines a small resolution of the conifold singularity~(\ref{eq:conifold}).
(The form of this resolution is well-known; as a check, this implies that the
$2 \times 2$ matrix of $p$'s has a zero eigenvalue, and the
conifold equation~(\ref{eq:conifold}) simply states that the determinant
vanishes.) Hence, the Kempf-Laksov resolution really does resolve
$X_{\tiny\yng(1)}$.

Physically, the defect is supported on $X_{\tiny\yng(1)}$,
but sees itself propagating on 
the small resolution $\widetilde{X}_{\tiny\yng(1)}$.
(This is analogous to D-branes on orbifolds; although the orbifold
itself is a stack, D-branes often perceive a resolution of any
orbifold singularity \cite{Douglas:1997de}, a fact which in those circumstances
was explained in terms of nonreduced scheme structures in \cite{Donagi:2003hh}.)

\subsubsection{The Schubert variety with \texorpdfstring{$\lambda \,=\,[2,0]$}{lambda(2)}}

Consider the Schubert variety $X_{\tiny\yng(2)} \subset {\rm Gr}(2,4)$,
given by matrices of the form:
\begin{equation}
\left[ \begin{array}{cccc}
* & * & * & * \\
* & 0 & 0 & 0 \end{array} \right]~.
\end{equation}

This is the subvariety $p_{23} = 0 = p_{24} = p_{34}$.
It is the subvariety of the conifold $X_{\tiny\yng(1)}$,
which is itself given by $\{ p_{34} = 0 \}$.
Here, that means it is the subvariety of:
\begin{equation}
\{ p_{13}\, p_{24}\, = \,p_{14}\, p_{23} \}~,
\end{equation}
given by $p_{23} = p_{24} = 0$. This is codimension one -- a Weil divisor, but not a Cartier divisor
(as it is codimension one but is not expressed holomorphically as
$\{f = 0 \}$ for a single function $f$). Explicitly, this divisor is smooth, given by
\begin{equation}
{\mathbb C}^2 \: = \: {\rm Spec} \, {\mathbb C}[p_{13}, p_{14}]~.
\end{equation}

\medskip
\noindent
\textbf{Kempf–Laksov resolution.} In the Kempf--Laksov resolution, we add a single $U(1)$ along
the defect plus a chiral superfield $z$ in the bifundamental between
the defect $U(1)$ and the bulk $U(2)$, together with the constraints:
\begin{equation}
z_a \,\phi^a_2 \: = \: 0 \: = \: z_a \,\phi^a_3 \: = \: z_a \,\phi^a_4~,
\end{equation}
which reproduces the structure of the matrix above.

Relative to the Kempf--Laksov resolution of $X_{\tiny\yng(1)}$,
this has the same fields but a new constraint, namely:
\begin{equation}  \label{eq:g24:2:newconstr}
z_a \,\phi^a_2 \: = \: 0~.
\end{equation}
If we work in the coordinate patch $p_{12} \neq 0$, as for
$X_{\tiny\yng(1)}$, then $\{ \phi^1_2, \phi^2_2 \}$ are not
both zero, so the constraint~(\ref{eq:g24:2:newconstr}) can be used
to determine a unique element of $[z_1, z_2]$.  For example, 
if $\phi^1_2 \neq 0$, then:
\begin{equation}
z_1 \: = \: -\, z_2 \,\frac{ \phi^2_2 }{ \phi_2^1 }~.
\end{equation}

In the case of the small resolution $\widetilde{X}_{\tiny\yng(1)}$ discussed above,
the $[z_1, z_2]$ acted as homogeneous coordinates on the small resolution; away from the singularity, their ratio was determined, but at the singularity, they were unconstrained.  Here, by contrast, we see that their ratio is uniquely determined everywhere, including at the singularity.

Thus, the Kempf--Laksov resolution returns $X_{\tiny\yng(2)}$ itself, as expected since $X_{\tiny\yng(2)}$ is smooth. Phrased another way, we see that:
\begin{equation}
X_{\tiny\yng(2)} \: = \: \{ z_a\, \phi^a_2 \,= \,0 \} \subset
\widetilde{X}_{\tiny\yng(1)}~.
\end{equation}

\subsection{The incidence variety}

Recall from our discussion in section \ref{sec: GLSM to partial flag}, an incidence variety is a partial flag manifold of the form Fl$(1,n-1;n)$.
These can be described as hypersurfaces of degree $(1,1)$ in 
${\mathbb P}^{n-1} \times {\mathbb P}^{n-1}$,
see e.g.~\cite[section 2.3]{Donagi:2007hi}.

We can understand this explicitly as follows. Describe Fl$(1,n-1;n)$ in terms of a bifundamental $\Phi$ of $U(1) \times U(n-1)$,
and a bifundamental $\Psi$ of $U(n-1) \times U(n)$.\footnote{Note the change in our notation here. Instead of denoting these fields by $\phi_1^2$ and $\phi_2^3$ as in figure \ref{fig:partial Flag quiver}, we denote them by $\Phi$ and $\Psi$, respectively.} The Pl\"ucker coordinates on this flag manifold are given by:
\begin{eqnarray}
    p(\Psi\, \Phi)_{i} &\, =\, & \Psi^a_i \,\Phi_a~, \\
    p(\Psi)_{i_1 \cdots i_{n-1}} &\, =\, &
    \epsilon_{a_1 \cdots a_{n-1}}\, \Psi^{a_1}_{i_1}\, \cdots \,\Psi^{a_{n-1}}_{i_{n-1}}~.
\end{eqnarray}
Then, $\{ p(\Phi\, \Psi) \}$ and $\{ p(\Psi) \}$ both separately define maps
Fl$(1,n-1;n) \,\rightarrow \,{\mathbb P}^{n-1}$, where the $p$'s are interpreted as homogeneous coordinates, so that altogether we have a map:
\begin{equation}
    {\rm Fl}(1,n-1;n) \: \longrightarrow \: {\mathbb P}^{n-1} \,\times\, {\mathbb P}^{n-1}~.
\end{equation}
Furthermore, it is straightforward to check, after unwinding definitions, that
the Pl\"ucker coordinates satisfy the following relation:
\begin{equation}  \label{eq:inc:hyp}
    \epsilon^{i_1 \cdots i_n} \, p(\Psi \,\Phi)_{i_1} \,
    p(\Psi)_{i_2 \cdots i_n} \: = \: 0~.
\end{equation}
It can be shown that 
this relation completely specifies Fl$(1,n-1;n)$, hence we can understand Fl$(1,n-1;n)$ as a degree $(1,1)$ hypersurface in ${\mathbb P}^{n-1} \,\times\, {\mathbb P}^{n-1}$.

In this section, we will verify in examples that the GLSM defect construction given in figure \ref{fig:GenProposal} correctly localizes on the claimed corresponding Schubert varieties.

\medskip
\noindent
\textbf{Singular Schubert varieties in Fl$(1,n-1;n)$.} Following \cite[examples 1.2.3.(5)]{brion-lec},
the Schubert varieties in Fl$(1,n-1;n)$ are labeled as $X_{i,j}$ ($i, j \in \{1, \cdots, n\}$,
$i \neq j$) and are given by:
\begin{equation}
    X_{i,j} \: = \: \{ (V_1, V_{n-1}) \in {\rm Fl}(1,n-1;n) \, | \,
    V_1 \,\subseteq\, \widetilde{E}_i \,\mbox{ and }\,
    V_{n-1}\, \supseteq \,\widetilde{E}_{j-1}\, \}~.
\end{equation}
Where $\widetilde{E}_i\, :=\,{\rm Span}_\C\{{\rm e}_1, \cdots, {\rm e}_i\} $. In terms of $\Phi$ and $\Psi$, the $1 \times n$ matrix $\Psi\, \Phi$ is a generator of $V_1 \subset {\mathbb C}^n$,
and the $(n-1) \times n$ matrix $\Psi$ encodes $n-1$ generators of $V_{n-1} \subset {\mathbb C}^n$.
From the description of $X_{i,j}$, we see immediately that: 
\begin{equation}
    p(\Psi \,\Phi)_k \: = \: 0~,
\end{equation}
for $k > i$, as $V_1$ is in the span of $e_1, \cdots, e_i$. Similarly, since $V_{n-1}$ includes the span of $e_1, \cdots, e_{j-1}$,  we expect that:
\begin{equation}
    p(\Psi)_{i_1 \cdots i_{n-1}} \: = \: 0
\end{equation}
if any of $1, \cdots, j-1 \not\in \{i_1, i_2, \cdots, i_n\}$.  To make this clearer,
we can use linear transformations to rotate $\Psi$ to a matrix of the form:
\begin{equation}
    \left[ \begin{array}{ccccccc} 
    1 & 0 & \cdots & 0 & 0 & \cdots & 0 \\
    0 & 1 & \cdots & 0 & 0 & \cdots & 0 \\
      &   & \vdots &   & 0 & \cdots & 0 \\
    0 & 0 & \cdots & 1 & 0 & \cdots & 0 \\
    0 & 0 & \cdots & 0 & * & \cdots & * \\
      &   & \vdots &   & * & \cdots & * 
      \end{array}
    \right]~,
\end{equation}  
where $*$ denotes an undetermined entry.  If any of the first $j-1$ columns are omitted, the resulting matrix has a row that is zero, hence its determinant vanishes.

Following  \cite[examples 1.2.3.(5)]{brion-lec},
define:
\begin{equation}
    x_k \: = \: p(\Psi\, \Phi)_k~, \qquad
    y_k \: = \: \epsilon^{k i_1 \cdots i_{n-1} } p(\Psi)_{i_1 \cdots i){n-1}}~.
\end{equation}
Then, the Schubert variety $X_{i,j}$ is given by:
\begin{equation}
    x_{i+1} \: = \: \cdots \: = \: x_n \: = \: 0
    \: = \: y_1 \: = \: \cdots \: = \: y_{j-1}~.
\end{equation}
From  \cite[examples 1.2.3.(5)]{brion-lec},
the Schubert variety $X_{i,j}$ will be singular if $1 \,<\, j\, <\, i\, <\, n$,
with singular locus $X_{j-1,i+1}$.  For other values of $i, j$, $X_{i,j}$ is smooth. In passing, note that for $n=3$, the full flag manifold Fl$(3)$, these conditions cannot be satisfied, so all its Schubert varieties are smooth.

\medskip
\noindent
\textbf{Explicit examples in Fl$(1,3;4)$.} Consider for example the case $n=4$.  In this case, the only singular Schubert variety is:
\begin{equation}
    X_{3,2} \: = \: \{ x_4 \: = \: 0 \: = \: y_1 \}~.
\end{equation}
Then, the equation for the incidence variety as a hypersurface~(\ref{eq:inc:hyp}) reduces to:
\begin{equation}
    x_2 \,y_2 \: + \: x_3 \,y_3 \: = \: 0~,
\end{equation}
which has a conifold singularity at the origin.
Note that the origin is given as:
\begin{equation}
    \{ x_2 \: = \: x_3 \: = \: x_4 \: = \: 0 \: = \: y_1 \: = \: y_2 \: = \: y_3 \}~,
\end{equation}
which coincides with the Schubert variety $X_{2-1, 3+1} = X_{1,4}$, as expected.

Now, let us compare this to the physics proposal of section~\ref{sect:proposal defect}.
The Schubert variety $X_{3,2}$ (or rather its resolution), corresponding to the permutation $w = (2\,1\,4\,3)$, should be described by the quiver:
\begin{center}
    \begin{tikzpicture}[->-/.style={decoration={
  markings,
  mark=at position #1 with {\arrow{>}}},postaction={decorate}}]
        \node[circle,draw] (r21) {\parbox[c][0.3cm][c]{0.3cm}{\centering $1$}};
        \node[circle,draw,right=1cm of r21] (r22) {\parbox[c][0.3cm][c]{0.3cm}{\centering $1$}};
        \node[circle,draw,right=1cm of r22] (r23) {\parbox[c][0.3cm][c]{0.3cm}{\centering $2$}};
        \node[circle,draw,right=1cm of r23] (r24) {\parbox[c][0.3cm][c]{0.3cm}{\centering $3$}};
        \node[rectangle,draw,gray, below=1cm of r21] (r31) {\parbox[c][0.3cm][c]{0.3cm}{\centering $1$}};
        \node[rectangle,draw,gray,below=1cm of r22] (r32) {\parbox[c][0.3cm][c]{0.3cm}{\centering $2$}};
        \node[rectangle,draw,gray,below=1cm of r23] (r33) {\parbox[c][0.3cm][c]{0.3cm}{\centering $3$}};
        \node[rectangle,draw,below=1cm of r24] (r34) {\parbox[c][0.3cm][c]{0.3cm}{\centering $4$}};
        \node[circle,draw,above=1cm of r23] (r13) {\parbox[c][0.3cm][c]{0.3cm}{\centering $1$}};
        \node[circle,draw,above=1cm of r24] (r14) {\parbox[c][0.3cm][c]{0.3cm}{\centering $1$}};
        \draw[->-=0.6] (r14) -- (r24) node[midway,right] {$\Phi$};  
        \draw[->-=0.6] (r24) -- (r34) node[midway,right] {$\Psi$};
        \draw[->-=0.6] (r13) -- (r14) node[midway,above] {$\phi_{31}$};
        \draw[->-=0.6] (r13) -- (r23) node[midway,left] {$\varphi_{31}$};  
        \draw[->-=0.6,dashed] (r13) -- (r24);
        \draw[->-=0.6] (r21) -- (r22) node[midway,above] {$\phi_{13}$}; 
        \draw[->-=0.6] (r22) -- (r23) node[midway,above] {$\phi_{23}$}; 
        \draw[->-=0.6] (r23) -- (r24) node[midway,above] {$\phi_{33}$};
        \draw[->-=0.6] (r21) -- (r31) node[midway,left] {$\varphi_{13}$}; 
        \draw[->-=0.6] (r22) -- (r32) node[midway,left] {$\varphi_{23}$}; 
        \draw[->-=0.6] (r23) -- (r33) node[midway,left] {$\varphi_{33}$};
        %
        \draw[->-=0.6,gray] (r31) -- (r32) node[midway,above] {$\iota$}; 
        \draw[->-=0.6,gray] (r32) -- (r33) node[midway,above] {$\iota$}; 
        \draw[->-=0.6,gray] (r33) -- (r34) node[midway,above] {$\iota$};;
        \draw[->-=0.6,dashed] (r23) -- (r34); \draw[->,dashed] (r22) -- (r33); 
        \draw[->-=0.6,dashed] (r21) -- (r32);
        \node[fit=(r32)(r33), dashed, blue, draw, inner sep=8pt, minimum width=5.6cm,minimum height=4.5cm, shift={(-0.7cm,1.8cm)}] () {};
    \end{tikzpicture}
\end{center}
where the dashed blue box encloses data lying along the defect itself, the
$\iota$'s denote canonical inclusions, and horizontal and vertical bifundamentals are denoted $\phi$, $\varphi$, respectively.\footnote{Again, note here the change in our notation for the 1d chiral fields. This is just for compactness.}

The $E$-term conditions arising from the diagonal Fermi multiplets are:
\begin{eqnarray}
    \Phi \circ \phi_{31} & = & \phi_{33} \circ \varphi_{31}~,  \label{eq:f134:s32:A}
    \\
    \Psi \circ \phi_{33} & = & \iota \circ \varphi_{33}~,
    \label{eq:f134:s32:B}
    \\
    \varphi_{33} \circ \phi_{23} & = & \iota \circ \varphi_{23}~, \label{eq:f134:s32:C}
    \\
    \varphi_{23} \circ \phi_{13} & = & \iota \circ \varphi_{13}~.  \label{eq:f134:s32:D}
\end{eqnarray}
From the $D$-term conditions, $\phi_{31}$ is an invertible scalar, so from~(\ref{eq:f134:s32:A}) we can write:
\begin{equation}
    \Phi_a \: = \: \frac{ (\phi_{33})^{\alpha}_a \,(\varphi_{31})_{\alpha} }{\phi_{31}}~, 
\end{equation}
and combining with~(\ref{eq:f134:s32:B}) yields:
\begin{eqnarray}
    \Psi_i^a \,\Phi_a & = &
    \frac{ \Psi^a_i \,(\phi_{33})^{\alpha}_a\, (\varphi_{31})_{\alpha} }{\phi_{31}}~, 
    \\
    & = &
    \frac{ (\iota \,\circ\, \varphi_{33})^{\alpha}_i \,(\varphi_{31})_{\alpha} }{\phi_{31}}~, 
\end{eqnarray}
which vanishes for $i=4$ as this is outside the range of $\iota$.
In particular, we see that:
\begin{equation}
    x_4 \: = \: p(\Psi \,\Phi)_4 \: = \: 0~,
\end{equation}
which is one of the equations defining the Schubert variety $X_{3,2}$.

Similarly, combining~(\ref{eq:f134:s32:B}), (\ref{eq:f134:s32:C}), (\ref{eq:f134:s32:D}),
we have that:
\begin{equation}
    \phi_{13} \,\left( \Psi\, \circ\, \phi_{33} \,\circ\, \phi_{23} \right)_{2,3,4} \: = \: 0~.
\end{equation}
Write $A \equiv \phi_{33} \circ \phi_{23}$, then in components we can write this as:
\begin{equation}
    A_1 \,\Psi^1_i \: + \: A_2\, \Psi^2_i \: + \: A_3 \,\Psi^3_i \: = \: 0~,
\end{equation}
for $i= 2, 3, 4$, or in terms of matrices:
\begin{equation}
    \left[ A_1, A_2, A_3 \right]\, \left[ \begin{array}{ccc}
    \Psi^1_2 & \Psi^1_3 & \Psi^1_4 \\
    \Psi^2_2 & \Psi^2_3 & \Psi^2_4 \\
    \Psi^3_2 & \Psi^3_3 & \Psi^3_4 \end{array} \right] \: = \: 0~.
\end{equation}
However, $A$ is a nonzero matrix, hence the determinant of the $3 \times 3$ matrix of $\Psi$ components
must vanish, which implies:
\begin{equation}
    y_1 \: = \: p(\Psi)_1 \: = \: 0~.
\end{equation}

To study $p(\Psi)$, we can also focus on the lower portion of the quiver:
\begin{center}
    \begin{tikzpicture}[->-/.style={decoration={
  markings,
  mark=at position #1 with {\arrow{>}}},postaction={decorate}}]
        \node[circle,draw] (r21) {\parbox[c][0.3cm][c]{0.3cm}{\centering $1$}};
        \node[circle,draw,right=1cm of r21] (r22) {\parbox[c][0.3cm][c]{0.3cm}{\centering $1$}};
        \node[circle,draw,right=1cm of r22] (r23) {\parbox[c][0.3cm][c]{0.3cm}{\centering $2$}};
        \node[circle,draw,right=1cm of r23] (r24) {\parbox[c][0.3cm][c]{0.3cm}{\centering $3$}};
        \node[rectangle,draw,gray,below=1cm of r21] (r31) {\parbox[c][0.3cm][c]{0.3cm}{\centering $1$}};
        \node[rectangle,draw,gray,below=1cm of r22] (r32) {\parbox[c][0.3cm][c]{0.3cm}{\centering $2$}};
        \node[rectangle,draw,gray,below=1cm of r23] (r33) {\parbox[c][0.3cm][c]{0.3cm}{\centering $3$}};
        \node[rectangle,draw,below=1cm of r24] (r34) {\parbox[c][0.3cm][c]{0.3cm}{\centering $4$}};
        \draw[->-=0.6] (r24) -- (r34) node[midway,right] {$\Psi$};
        \draw[->-=0.6] (r21) -- (r22) node[midway,above] {$\phi_{13}$}; 
        \draw[->-=0.6] (r22) -- (r23) node[midway,above] {$\phi_{23}$}; 
        \draw[->-=0.6] (r23) -- (r24) node[midway,above] {$\phi_{33}$};
        \draw[->-=0.6] (r21) -- (r31) node[midway,left] {$\varphi_{13}$}; 
        \draw[->-=0.6] (r22) -- (r32) node[midway,left] {$\varphi_{23}$}; 
        \draw[->-=0.6] (r23) -- (r33) node[midway,left] {$\varphi_{33}$};
        %
        \draw[->-=0.6,gray] (r31) -- (r32) node[midway,above] {$\iota$}; 
        \draw[->-=0.6,gray] (r32) -- (r33) node[midway,above] {$\iota$}; 
        \draw[->-=0.6,gray] (r33) -- (r34) node[midway,above] {$\iota$};;
        \draw[->-=0.6,dashed] (r23) -- (r34); 
        \draw[->-=0.6,dashed] (r22) -- (r33); 
        \draw[->-=0.6,dashed] (r21) -- (r32);
    \end{tikzpicture}
\end{center}
This describes a Schubert variety in Gr$(3,4) \cong {\rm Gr}(1,4) \equiv {\mathbb P}^3$.
To understand which Schubert variety, we refer to the definition \eqref{Schubert in Grass defn}:
\begin{equation}
    \dim \left( \Sigma \,\cap \,F_{n-k+i-\lambda_i} \right) \: \geq \: i~.
\end{equation}
Here, $k=3$, $n=4$, so $n-k=1$.
From the column:
\begin{center}
    \begin{tikzpicture}[->-/.style={decoration={
  markings,
  mark=at position #1 with {\arrow{>}}},postaction={decorate}}]
        \node[circle,draw] (r1) {\parbox[c][0.3cm][c]{0.3cm}{\centering $2$}};
        \node[rectangle,draw,gray,below=1cm of r1] (r2) {\parbox[c][0.3cm][c]{0.3cm}{\centering $3$}};
        \draw[->-=0.5] (r1) -- (r2);
    \end{tikzpicture}
\end{center}
we find for $i=2$ that:
\begin{equation}
    1\, +\, i\, -\, \lambda_i \: = \: 3~,
\end{equation}
hence $\lambda_2 = 0$.
From the column:
\begin{center}
    \begin{tikzpicture}[->-/.style={decoration={
  markings,
  mark=at position #1 with {\arrow{>}}},postaction={decorate}}]
        \node[circle,draw] (r1) {\parbox[c][0.3cm][c]{0.3cm}{\centering $1$}};
        \node[rectangle,draw,gray,below=1cm of r1] (r2) {\parbox[c][0.3cm][c]{0.3cm}{\centering $2$}};
        \draw[->-=0.5] (r1) -- (r2);
    \end{tikzpicture}
\end{center}
we find for $i=1$ that
\begin{equation}
    2 - \lambda_1 \: = \: 2~,
\end{equation}
which would imply that $\lambda_1 = 0$, but is redundant. From the column:
\begin{center}
    \begin{tikzpicture}[->-/.style={decoration={
  markings,
  mark=at position #1 with {\arrow{>}}},postaction={decorate}}]
        \node[circle,draw] (r1) {\parbox[c][0.3cm][c]{0.3cm}{\centering $1$}};
        \node[rectangle,draw,gray,below=1cm of r1] (r2) {\parbox[c][0.3cm][c]{0.3cm}{\centering $1$}};
        \draw[->-=0.5] (r1) -- (r2);
    \end{tikzpicture}
\end{center}
we find for $i=1$ that $\lambda_1 = 1$. Altogether, we find that this describes the Schubert variety $X_{\tiny\yng(1)}$, which is what we need on this ${\mathbb P}^2$ factor to correctly describe the Schubert variety $X_{3,2} \subset {\rm Fl}(1,3;4)$.


\bibliography{flagGLSM} \addcontentsline{toc}{section}{References}
\bibliographystyle{JHEP} 

\end{document}